\documentclass[finals,epj,nopacs,twocolumn]{svjour}

\usepackage{latexsym}
\def\makeheadbox{\relax}

\usepackage{dcolumn}
\usepackage{graphics}
\usepackage{hyperref}
\usepackage{amsmath}
\usepackage{mathtools,cuted}

  \newcommand{\Qa}{\mathcal{Q}}
 \usepackage[latin1]{inputenc}
\usepackage{color}
\usepackage{url}
\usepackage{graphicx}
\hypersetup{linkcolor=red,citecolor=blue,filecolor=cyan,urlcolor=magenta}
\usepackage{float}
\usepackage{amsmath}
\usepackage{tikz}
\usepackage{graphicx}
\usepackage{latexsym,color}
\usepackage{amssymb}
\usepackage{charter}

\usepackage{xcolor}
\usepackage{amsfonts,dsfont}

\usepackage{txfonts}
\usepackage[T1]{fontenc}
\usepackage{ae,aecompl,latexsym}

\definecolor{lightgray}{rgb}{0.75,0.75,0.75}

\def\be{\begin{equation}}
\def\ee{\end{equation}}
\def\bea{\begin{eqnarray}}

\def\eea{\end{eqnarray}}

\newcommand{\la}{\mathcal{A}}
\newcommand{\Sa}{\mathcal{S}}

\usepackage{physics}

\newcommand{\il}{~}

\begin{document}
\title{Graph model overview,  events scales structure  and chains of events.}
\author{D. Pugliese 
}                     
%
%
\institute{Research Centre of Theoretical Physics and Astrophysics,
Institute of Physics,
  Silesian University in Opava,
 Bezru\v{c}ovo n\'{a}m\v{e}st\'{i} 13, CZ-74601 Opava, Czech Republic
}
\date{Received: date / Revised version: date}
%
\abstract{
We present a graph  model     for  a background independent, relational approach to spacetime emergence. The general idea and  the graph main features,   detailed   in \cite{start01}, are discussed. This is a combinatorial  (dynamical) metric graph,  colored on vertexes, endowed with  a classical distribution of colors probability  on the graph vertexes. The graph coloring determines the graph structure in  clusters of graph vertices (events)  that can be monochromatic (homogeneous loops) or polychromatic (inhomogeneous loops). The probability is conserved after the graph  conformal expansion  from an initial  seed graph state to higher (conformally expanded) graph states. The emerging  structure has self-similar characteristics on different scales (states).  From the coloring, different levels of vertices  and thus  graph levels arise as new aggregates of colored vertices. In this  second (derived)  graphs level, the derived graph  vertices correspond to  the polychromatic edges (with  differently colored vertices) of the initial graph.   Vertex aggregates are related, as some levels (graph states) to plexors and twistors (involving Clifford statistics).  Two metric levels are defined on the colored graph, the first level is   a natural metric defined on the graph, the second level  emerges from the first   and related, due to symmetries. Metric structure  reflects the graph colored structure under conformal transformations evolving with its states under conformal expansion. In some special
 cases vertices/events chains    could be related to  strings generalizations.
} 
%
\maketitle

\newcommand{\ti}[1]{\mbox{\tiny{#1}}}
\newcommand{\im}{\mathop{\mathrm{Im}}}
\def\be{\begin{equation}}
\def\ee{\end{equation}}
\def\bew{\begin{widetext}}
\def\eew{\end{widetext}}
\def\Rem{\textbf{Remark}}
\def\bea{\begin{eqnarray}}
\def\eea{\end{eqnarray}}
\newcommand{\tb}[1]{\textbf{\texttt{#1}}}
\newcommand{\oft}{\mathcal{o}_T}
\newcommand{\ofx}{\mathcal{o}_X}

\newcommand{\ttb}[1]{\textbf{#1}}
\newcommand{\ctimes}{\overset{\mathbf{\cdot}}{\times}}
\newcommand{\rtb}[1]{\textcolor[rgb]{1.00,0.00,0.00}{\tb{#1}}}
\newcommand{\gtb}[1]{\textcolor[rgb]{0.17,0.72,0.40}{\tb{#1}}}
\newcommand{\ptb}[1]{\textcolor[rgb]{0.77,0.04,0.95}{\tb{#1}}}
\newcommand{\btb}[1]{\textcolor[rgb]{0.00,0.00,1.00}{\tb{#1}}}
\newcommand{\otb}[1]{\textcolor[rgb]{1.00,0.50,0.25}{\tb{#1}}}
\newcommand{\non}[1]{{\LARGE{ \not{ }}}{#1}}
\newcommand{\nnon}[1]{{\Huge{\not}}{#1}}
\newcommand{\parr}[1]{\breve{\partial}}
\newcommand{\body}[1]{\|\mathbf{#1}\|}
\newcommand{\ff}[1]{\left\lfloor #1\right\rfloor}
\newcommand{\Hl}{{\large{ \mathrm{H}}}}
\newcommand{\srt}[1]{{\scriptsize{#1}}}

\newcommand{\bp}[1]{$\eth$}
\newcommand{\Ga}{\mathrm{G}}
\newcommand{\hh}[1]{\left\lceil #1\right\rceil}
\newcommand{\kets}[1]{\ket{#1}}
\newcommand{\kett}[1]{\kets_{#1}}
\newcommand{\ssp}{{\footnotesize{\textsf{\textbf{{S}}}}}}

\newcommand{\ddp}[1]{\vec{\dda}^{\scriptscriptstyle{#1}}}
\newcommand{\rc}{\rho_{\ti{C}}}
\newcommand{\ex}{\exists}
\newcommand{\vex}{\vec{\exists}}
\newcommand{\hex}{\hat{\exists}}
\newcommand{\oftt}{\mathcal{o}_{\widetilde{T}}}
\newcommand{\ofs}{\mathcal{o_S}}
\newcommand{\scr}{${\scriptsize{$\circ$}}$}

\newcommand{\Ca}{\mathcal{\mathbf{C}}}
\newcommand{\Ha}{\mathcal{H}}
\newcommand{\dda}{\mathcal{D}}
\newcommand{\1}{\mathds{1}}

\newcommand{\loops}{\mathbf{\mathrlap{\circlearrowleft}{\circlearrowright}}}

  \newcommand{\downmapsto}{\rotatebox[origin=c]{-90}{${\longmapsto}$}\mkern2mu}
   \newcommand{\vpoints}{\rotatebox[origin=c]{-90}{$...$}\mkern2mu}
  \newcommand{\upmapsto}{\rotatebox[origin=c]{90}{${\longmapsto}$}\mkern2mu}

\newcommand{\Tem}{T^{\rm{em}}}

\makeatletter
\newenvironment{sqcases}{%
  \matrix@check\sqcases\env@sqcases
}{%
  \endarray\right.%
}
\def\env@sqcases{%
  \let\@ifnextchar\new@ifnextchar
  \left\lbrack
  \def\arraystretch{1.2}%
  \array{@{}l@{\quad}l@{}}%
  }
\makeatother
\section{Introduction}
We introduce a (decomposition) graph for a background independent model of relational emergent spacetime in the frame of (dynamical)  {quantum graph},  here we summarize  some  ideas underling  the   model and we sketch the graph (lattice) main features, a more extensive discussion is   in \cite{start01}.  Eventually, with an  adaptation of the  metric structures (metric graph) to a {dynamical} quantum  graph one goal is to address  in a  combination of  top-down  and down-top strategy  the frame of spacetime emergence.
 The graph closes  the decompositions of the spacetime texture  in graph vertexes  (events) and edges,  the spacetime topological,  causal and  metric structure naturally  change in the graph frame.    The replacement of the spacetime (differential)  manifold  with a
graph grounds the  search for a common language for different scales,
for (assumed) different  events scales.
 Space-time  can be assumed to  emerge at  certain scales of the model, connected to   graph conformal transformations   in higher graph states (conformal expansion).   This brief note, discussing  the  decomposition graph   main properties and it can be considered as a model  overview  and a graph proposal.
 The spacetime emergence problem is translated  into events scale emergence and structures of a "crystalized" graph.
The (combinatorial) graph (close to  a Cayley graph) is colored and characterized by a classic definition of color probabilities (frequencies)  distribution.
 A relevant aspect of the analysis consists in the  construction of vertexes and clusters of vertexes  with inner colored structure (monochromatic-homogenous loops $\ell$, or polychromatic-inhomogeneous in colors- loops) that can be studied as emerging from the conformal  transformation of the colored graph, described through the concept of event algebras $\la$ and created through evolutive operators  (introducing a  vertex order relation) and shift operators (for color shift), related to graph edges, and  acting on vertexes.
 Lower graph states (scales)  admit fluctuations in their structure, understood,  in a first approximation, as variation of algebra distribution in the  adjacent chain vertexes (in the sense of the order relation in the chain)    for  color and for graph state and paths (chain). Therefore  algebra inhomogeneity (related  to a definition of graph radius and  inertia) leads to  inhomogeneous in algebra chain  with  valence two vertexes (algebraic valence ($\pm1$, 0)).
We could compare this  approach  with   theories with   graphs as an underlying  language: the   (causal)  loop quantum gravity (LQG), spin network  and causal set theories.
This model also
shares  some aspects with the analysis in combinatorial spacetime, stochastic metric and quantum computation spacetime and more generally with it-from (qu)bit ideas--\cite{Padma,Garay:1994en,Science,2015NJPh...17j2001A,Sorkingeven,DeHaro:2018rvw}. Chains of events  have  an event strings-formulation  as collateral:   chains might be  translated as   strings generalization.
 Our analysis    may  be also  read   into quantum graphity framework ({QGF}) as a graph  dynamical model \cite{Konopka:2008hp,Konopka:2006hu,Caravelli:2010xx,Oriti}.
Quantum graphity   may be defined as a model of emergent geometry (locality) in the context of quantum gravity theory, based on a background
independent formulation of condensed matter systems on graphs  that can be  coupled to a heat bath with temperature.
In this frame a quantum model  is attached as  an Hilbert space   to some graph elements, for example  the dynamical graph vertices and therefore
defining   system states which are ultimately given by bosonic (of fermionic) vertex degrees of freedom. An hamiltonian is constructed for the system.
The decomposition set, in {QGF} as  in our approach, constitutes a pre-geometric (quantum)  system  from which one can extract effective excitations  and from them geometry with matter, $\mu$matter, emerge. In our approach  matter would emerge from the graph coloring determining the  graph "dynamics" and  structure.
There are however   different versions of the {QGFs}.
Generally the approach followed in {QGF} is the graph dynamic one. In this work however we aim to  combine a dynamical and metric graph.
The dynamical setup is pursued on the  elements associated with parts of the graph which, according to the realization of the theory, are the graph edges (lines) or loops.
We start  from a vertex approach, the coloring is on the vertices, in this way it is immediate to define loops as clusters of vertices characterized by their coloring. Other levels of events/vertices are  associated with connections between differently colored vertices of a new different associated graph of higher level, \bp-events ("partial"-events), thus a  derived graph (composed by these elements) and  a derived metric structure follow.
The metric reflects this internal graph clustering providing, in first approximation, a loop measure.
 In {QGF}, in general,  matter would emerge only after the phase transition from the  (net) condensation of the graph as emergent symmetry in the
phase transition to the ordered phase graph.
At this point we can say that  the graph lives in two different regimes: a (final)  low energy and the (initial, cosmological) high energy regime.
At the high temperature, there is a
disordered phase where  notions of geometry, dimension and topology
drastically changes. In this phase  physics seems compelled to  be described  in  quantum mechanical terms while (a phase) transition occurs and the
low temperature phase, where the system is  ordered  and a background  and  fields  emerge, living on a low dimensional spacetime (low temperature and large volume) manifold with a metric, implying the geometry of general relativity (gravity geometrizations).
Therefore in {QGF} the high energy regime features a not  local theory where  the graph is   completely connected
    and invariant  under the full symmetric group acting on the vertices,  graph symmetries are   associated to the symmetries of permutation.
In the second phase of {QGF}, a  graph ground state   is defined, the system turns by this cooling  process ordered, local and, low-dimensional.
This dynamics breaks the permutation symmetry into  translations and rotations.
In this scenario   a U(1) gauge theory can emerge  in the ground state by a mechanism of  string-net condensation as  Levin-Wen mechanism \cite{Levin}.  Maintaining the discretization, the transition  should clearly leave a latex with a translation group.  Emergence of geometry by such a phase transition is indicated as geometrygenesis.
We retain  in our investigation this aspect developing a  combinatorial colored graph approach  although we do not foresee a development in  graph structure  evolution eras  with a cosmological geometro-genesis phase.

More specifically on regards on the graph construction, which is one main focus on this article, we note that
 in LQG, 
(space)
geometry  is on graphs,
with lines  "colored" with a half-integral
number associated to {SU(2)} representations. Thus, a spin network  represents the
quantum state of space at a certain point in time, more precisely the states of the
spin network basis are the eigenvectors of
operators  area (quanta on the links) and volume (quanta on the intersections) and are a basis
for the (kinematical) LQG states.
 Introduced by Penrose as a combinatorial foundation for Euclidean 3-space spin network was conceived mainly as a trivalent graphs with spins assigned edges,  with vertices labelled by intertwining operators\cite{P1,P2,P3}. LQG has been  formulated canonically in the frozen time formalism where a  state space is the
direct sum of Hilbert spaces  associated to the  graphs.
The configuration variable turned to be then  an SU(2)-connection (where LQG path integral can be realized   as a sequence of
spin-network states  (3-geometries) as  transition terms of
spin foam models). 
On the other hand, these analysis, featuring a combinatorial model,  consider cluster of colored vertices,
and it is clear that a Clifford algebra will derive, this should guarantee a spinorial representation also after conformal transformations and algebraic re-parametrization (as algebra homogenization i.e.  a graph and metric chains parametrization  in different events bases adapted to the coloring, it is then important to  consider  transformations between graphs states  with different bases.).
We here    question more generally  the graph model  underlying the  spin structure selection (emergence),   addressing  this problem   from general considerations on vertex structure and  the graph main properties; there is no (a-priori) imposition of  a specific emerging spinoral structure (a specific Clifford or Grassmann algebra),     arguing a "best-fit" description from general considerations   described through the graph language.
  A "spinorial" structure would  emerge from the dynamics of  colored graph elements and graph scales related  to the graph conformal transformations and associated to a first level (boson) structure, as a "super-symmetric" model where however the associated elements are not simply conserved for state and graph realization (chain) transitions.

Vertexes  and clusters could be considered as (geometry) fundamental elements, like in
 several  other approaches,  where the  fundamental  spacetime building blocks    are   considered as qubit: in these modes
space would be a collection of qubits, where the
empty space (therefore the vacuum)  would correspond to a  ground state of
qubits, and elementary particles as collective excitations  above the ground states.
One crucial  aspect of these approaches consists in the determination of  quibit (and multibit) organizations and, eventually, interaction (and dimensionality of associated Hilbert space).  That is  a  qubit might be a superposed quantum
state of  vertices or clusters in the graph,
multi-(qu)bit states can be  constructed considering aspects of  the combinatorial graphs and
  these should be built by graph generations associated with an initial seed graph and emerging after conformal graph expansion where concept of loops, algebra and inertia on the graph and events scale structure are introduced.
 The role of the graph conformal expansion is relevant as related to  the large scales and eventually has a role in the bit- to-qubit transition in this model.
We  implemented in \cite{start01} a simple logic signal representation, using mainly a 2-level logic signal representation (bichromatic graph) with a  set of   \bp-events of a "reference" (i.e. an "inner" parameter) chain. This analysis however has been applied including more basis (coloring), introducing the concept of complementary-chains, immediate in the signals representation   relating  the signal frequency to the vertex algebra in  a first approximation  of a bundle chain   (a path graph state formulations featuring all possible  {histories} at fixed graph state).
The special case of an equiprobable graph  i.e. with equiprobable colors distribution,  and  where there is a maximum in the colors probability distribution are investigated, these  aspects  regulate the graph vertex loops  and  their algebra variation after conformal transformation, the presence of a maximum decomposition in events of a vertex  sequence, and the   \bp-events emergence.
 Using such representations, events-aggregates, according to the permutations,  obey
Clifford statistics (vectors as  plexors-for permutation-and spinors-for rotation-in some way as  plexors  of  aggregates)\footnote{Clifford algebra can be  realized by gamma matrices satisfying  a set of canonical anti-commutation relations, the spinors are
the column vectors on which these matrices act. It might  be convenient
to decompose into a pair of so-called "half-spin" or Weyl representations if the dimension is even.
(Note the Weyl group is the symmetric group $S_n$, which is represented by signed
permutation matrices).  We note that  the  Schur-Weyl duality (the  theorem  relating  irreducible finite-dimensional
representations of the general linear group and  the symmetric group) connects the irreducible representations of the symmetric group to irreducible algebraic representations of the
general linear group of a complex vector space.} \cite{start01}.
Emergent symmetries are explored, using {{correspondent ``lightcone coordinates''}}, and metric transformations are investigated within certain approximations. The $(1+1)$  dimensions (coloring) case is particularly considered.
A doubled metric structure is adapted to this model.
The first metric level, $\sigma(\cdot,\cdot)$, is adapted to the edges of the graph acting on its vertexes and functions  of the loops number on vertex (cluster of homogeneous in color vertices attached to one event). This is an on-vertex approach, edges are "flat" and not-colored.
The second metric structure, $g(\cdot,\cdot)$, is a composition of the firsts and  can be derived from this.  We detail the  transformations  and symmetries of the colored  graph  assumed as reference, {(related to "world
sheet" inner coordinates}) binding the $g(\cdot,\cdot)$  to ${\sigma}(\cdot,\cdot)$ metric.
There are several graphs representations    of graph   states and states evolution.
Mainly we will use a graph with maximum valence 2 per vertex, which are called \emph{chains} $\Ca$ (if the events component are related by an "evolutive" operator $\dda$ establishing an order  relation between the events) or sequences  $\Sa$ (if components are related  by shift operator $\dda^*$ exchanging colors on the vertices but without order relation among vertices-  graph with not connected homogeneous in colors loops).
 At fixed graph state $\Ga_m$, each sequence $\Sa$ has a  chain  $\Ca$ degeneration, each sequence corresponds to a set of chains, we  detail in\cite{start01}.
  The  graph state is constituted by its sequences (and chains), and the set  of these is called the graph sequence  (chains)  decomposition (or structure).
  In \cite{start01} we study this inner structure, and the  decompositions under conformal transformation; the doubled graph metric structure is adapted  to the graph degeneracy in chains and sequences. We give the chain (and sequence) degeneration of the graph, the definition of associated entropy of the  graph state  $\Ga_m$ with this structure and  the transformation   laws   under  conformal graph expansion.
  In this set, it is clear the self-similar structure of the graph. Metrics  reflect this structure. The transformations (in fact at fixed graph  state $\Ga_m$ this is reduced to action of permutation matrices and for different states to  generalized permutation which can be matched with typical graph matrices) act on events/vertexes for the first metric level $\sigma(\cdot,\cdot)$, and on edges for the second metric $g(\cdot,\cdot)$.
 To highlight symmetries  and clarify  the  graph seeds replicas embedded in superior states (conformally  expanded graph), we  have decomposed the graph in chains with maximum valence $2$  for vertex.
 However, we can superimpose these degenerations on the seed graph (through an edges approach) and consider a complete graph (where  we can define a quantum graph). Then the relation  of our graph model with  the standard simplex  appear to be not immediate, because of the vertex valence  2  and  the conformal transformation which in fact reshuffles any ordered relations among  different (ordered) clusters of vertices in one graph (chains of events)\footnote{In a vertex approach, where the graph is isotropic  (edges are equal in measure and colors/flat) but not necessarily homogeneous in algebra, it would be natural to consider the n-vertices  graph as a standard (or unit) (n-1) dimensional -simplex (or $n-1$-simplex), defined   as a $R^{n}$ subset. 
Therefore a seed graph minimal  sequence  can be seen as  a $n$--standard simplex having loops of order (algebra) $\oslash_L$ on a vertex.
 In general we remind that volume under the n-simplex (i.e.n+1 vertices graph) (i.e. between the origin and the simplex in $R^{n+1}$) is
$V_S=1/ (n+1)!$,
%
%
 with $(n+1)$ vertices (and n+1 cells), the simplex has  $n(n+1)/2$ 1--faces (polytope edges), faces number
$n (n^2-1)/6$, with 
 simplex hypervolume $
\oslash_L^{n}.\frac {\sqrt {n+1}}{n!{\sqrt {2^{n}}}}$ and  edge length $\oslash_L$
(Cayley--Menger determinant).
The dihedral angle  is  $\arccos(1/n)$, and the angle that the simplex center forms with its two vertices is $\arccos(-1/n)$,
 graph spectrum is  $n^1(-1)^n$, and
number of cells is $n+1$, with an
hypersurface area
$\sqrt{n 2^{1-n}}(n+1)\oslash_L^{n-1}/(n-1)!$--Figs\il(\ref{Fig:succ}).}. 
The metric graph   should describe events (vertices)  ``bubbling up'' after conformal transformation, where geometro-genesis will be as  an asymptotic re-stitching of the previously  dismantled spacetime structure, producing a metric dynamical structure with the reductions of the symmetries between first and second metric levels. (As pointed out in {QGF} the invariance under permutations of all the vertices
would be lost to the translation group in scale prospective.)
 There is an emerging events relational structure, and associated events and graph definition, arising from the polychromatic loops and from the conformal graph expansion. The conformal expansion  preserves the  graph seed structure ("graph-information"), creating  isomorphic parts of the expanded graph, and  new events depending on probability distribution.
 Thus  the idea of an overlap of metric and dynamical graph model is combined with  the existence of two metric levels and  an emerging metric structure   highlighted by  events emergence from the graph inhomogeneity. The selection  of the  metric level depends on scales and conserved  quantities under some specific transformations.  Further relevant  aspects  of this model  is the graph coloring and probability role, the graph conformal transformation (expansion), the introduction of  vertex algebra  in the metric and graph transformations and  the emerging graph self-similarity, clear  in the graph sequences and chains and graph conformal expansion. More specifically, the events are  macroevents  (homogeneous in colors "loops"  which serve as graph re-parametrization  with a color adapted inner base)  and   as derived events (from the inhomogeneity). These notions lead also to investigate the  graph entropy role, concept of graph immersion and interaction, briefly discussed  here in the metric approach.
 Concerning the general idea framing the model and  in relation to  graph expansion, self-similarity after conformal transformation and transmitted information (graph structure),  we could use the figurative image  of  a ``spacetime-DNA'' for the  original seed graph, which propagates  (through   conformal expansion with constant classical colors probability distribution) to higher graph states,  providing the overlap  of new states of the graph and its multiple  isomorphic parts.
In this ``events-DNA''   analogy, we could more generally thinking possible to write  a geometric  code encoding a  procedure to  extract, duplicate  and, eventually, modify such genetic code of our spacetime, providing a characterization of the  spacetime cells and organism\footnote{Spacetime (or geometric)  organism (also in the  graph pre-geometric phase) as  an organized system,
 characterized by  specific  relations, made up by a set of interdependent connected parts  in functional relations,  preserving and, eventually reintegrating and reproducing  its own form (substantialism).  Although this work does not enter into the specifics of the mechanisms called into action by the   Quantum Darwinism, the general view  of this model  overlaps in some  language proximity; in this respect  a similar analysis would be seen as a
    "Mendellian" investigations,  precursors of modern genetics.} (dealing  somehow  with spacetime knowledge as a spacetime engineering).
Graph seed would have the role to contain such  ``genetic information'' of the spacetime which replicates, propagates and  it is  preserved and,  eventually, may be manipulated. This would be a spacetime code whose information  propagates interacting eventually with other geometric organisms where here
the  idea of duplication is rendered  in the conformal expansion.

\medskip

{ In this work we introduce for the first time  a   graph   for a new spacetime emergent model based on the construction of states for the systems  as modeled by a graph
 Comparing the  model with the similar structures present in   literature we proceed with the construction on main characteristics of the model  discussing possible applications introducing   the concept of "spacetime engineering". We show  that  the doubled metric structure reflects the graph colored structure under conformal expansions and  in some special
 cases  chains   could be related to  strings generalizations.
  As first attempt to present and  describe  the  graph  model, the article focuses on the introduction of a number of new definitions, therefore  the formalism in this   article is  developed in great  details and it is extensively explained in a consistent first part of this manuscript.}

  In this article we   introduce the general idea  underlining the model and we sketch the   graph, whereas further details and developments will be  addressed in\cite{start01}.
 We use a combinatorial  graph  considered  through its polychromatic clusterings, the realizations, the conformal transformation and the different events  levels, reflected then in  a doubled graph metric structure. The graph is metric as we define a doubled  metric structure adapted  to the polychromaticity  of the graph elements, i.e. its vertex and edges.   In a dynamic model  the  graph would be  described by the graph constituents  dynamics,  vertices and edges,  intended  as graph "particles", as \bp-events\footnote{In  the relationalism   of  the graph frame the actual knowledge of  a physical object is based of (two) objects (contact) interaction. We exploit  this frame by considering events and  \bp-events definitions   grounded upon the  graph poly-chromaticity.  Entropy has been  associated in Sec.\il(\ref{Sec:deg-new-w-web}) to   \bp-events emergence  as expression of  graph chains poly-chromaticity.
Homogeneous    (in colors and algebras)  graph chains may be seen as precision (conformally expanded rate)  "clocks", where the existence of a  minimum local algebra  regulates also  the  conformal transformations  role.
These concepts, framed together  in the graph,  may be  simply expressed  in a relativistic scenario,  in a  causal structure terms where  mass  would be  intrinsically related to a time concept  for with no mass (time-like causal relations) it  would imply having no clocks (events)  and, therefore,  no distances, then we would have no metric structure, but  in fact     a conformal structure (light-like causal relations, here related to a topological notion of  graph scale definition)}.
This article is structured in three parts:
 In the first part we introduce some operators and notation, we discuss the  conceptual bases of the graph introduction. Second part builds the model  introducing the  graph
realizations, the  entropy notion and  the analysis of the effects of the conformal  transformation. Third  part introduces the graph metric structures.
In details: in Sec.\il(\ref{Sec:part-w}) we start discussing  the  graph model providing the list of the main graph features,  and  introducing  events as graph vertices. We  give a definition of sums and products of events, evolution, $\dda$, and shift, $\dda^*$, operators defining  graph events chains and sequences respectively. These definitions lead to develop the  concept of macroevents and algebra $\la^a$ of events and  operators algebra $\la^{\hat{a}}$: events compositions and decompositions  follow as example of graph structure and clusterization. We discuss the assumption of minimum algebras and homogenization in algebra of sequences and chains, and particularly  graph conformal transformations and graph loops.
Sequences and chains of colored graphs are thus thoroughly considered in Sec.\il(\ref{Sec:fromz}). Presence of loops and \bp-events emergence for a state graph and after conformal graph expansion are constrained depending on the colors probability distributions. Entropy notions are also explored in Sec.\il(\ref{Sec:deg-new-w-web}).
We then  consider an events matrix $\Qa_{ij}$ preceding definition of the colored metric graph.
Two  metrics levels $\sigma(\cdot,\cdot)$ and $g(\cdot,\cdot)$ are introduced in Sec.\il(\ref{Sec:metri-gr-gra}).  The overview  of the graph metric structures closes in Sec.\il(\ref{Sec:gooUE-dce})    with brief notes on metric graph symmetries and transformations     in relation to vertexes algebra in a first approximation. Concluding remarks follow in Sec.\il(\ref{Sec:mont-uni}). Appendix\il(\ref{Sec:log}) follows on some general considerations on models and vertices construction.


\subsection{Main graph features}\label{Sec:part-w}
 Here  we  list the  main characteristics graph     referring  to \cite{start01} for  further  discussion.
We consider a graph (body) $\Ga$ as the set of $n$ vertices (events) $\ket{a}$ (events notation $\ket{a}$ or $\bra{a}$ is here used  equivalently  for  graph vertexes) and connections (edges) $c_{ij}$ among vertices  $(\ket{a_i},\ket{a_j})$.
 \textbf{The sum }
$\sum_{i=1}^n \ket{a_i}$ is a  {sequence} of $n$ (not ordered and   with zero valence) events $\ket{a_i}$. This is generally associated to  an event $\ket{\mathrm{A}}$   (macroevent as vertex cluster)
 of the order  $n$ with respect to the $\ket{a}$ events,  equal to the sequence {dimension} (cardinality of the set of  events).
A vertex/event can be in general decomposed in a sequence of $n\geq 1$    lower order events. The $\ket{\mathrm{A}}$ decomposition constitutes an   $\ket{A}$  (sub)-structure. A graph $\Ga$ can be composed by vertices which are clusters of other events, providing  therefore the first notion of events clusterization in the graph $\Ga$,  this   leads to the concept of event algebra $\la$ and algebra  $\oslash_\mathrm{A}$ of an event $\ket{\mathrm{A}}$.
  \emph{\textbf{The algebra $\la^n$:}}  The dimension (degree) $Q$ of  an algebra $\la^Q$ is thus   a relative quantity  relating    events  and it corresponds to
the cardinality of the vertexes decomposition:
 the \emph{macroevent} $\ket{\mathrm{A}}$  belongs to the algebra  of the order (or degree) $n$ ($\oslash_A=n$)
 with respect to the algebra of its constituents $\ket{a}$.
In the limiting case  of $\oslash_A=1$,  the  event $\ket{A}$ is in correspondence with, ``decomposed'' in  one event.
   The vertexes of a  sequence can belong   to  equal (homogeneous in) algebras or  different  (inhomogeneous in) algebras,  in this case the algebra degree will be not constant for event decomposition index.
We assume that there is an  irreducible, \emph{ordinary} or \emph{primary}  \emph{algebra} $\la^0$ of the \emph{order}  $0$ considered therefore homogeneous and providing a procedure to define reference  algebras (the minimum  algebra of one vertex).
 \textbf{Evolutive operator $\dda$: }
It is   useful  to introduce the {\emph{evolutive}}  operator   $\dda:\dda\ket{a_i}=\ket{a_{i+1}}$ relating  two vertexes  with the introduction of  a vertex order relation    specified by the index. These definitions have predominantly a conceptual meaning  in this article,  explicating the  creation of clusters  by the color shift and vertices order.
A \emph{chain}, or  graph path,  is the ordered set of elements related by  $\dda$ (directed graph edges, order corresponds to the   $\dda$  action).  Two vertexes of a  chain are {consecutive} (or  adjacent), according to  $\dda$,
if they can be ordered by the action of $\dda$ as  $\ket{a_{i-1}}<\ket{a_i}<\ket{a_{i+1}}$.
The  chain/sequences  concepts can be expressed  by defining the {product operator} $\ddp n $ as follows
$
 \ddp n\ket{a_0}\equiv \prod_{i=1}^n \dda^{i} \ket{a_0}
$
or sums
\bea
\label{Eq:sum-prod}&&
\dda \sum_{i=0}^{n-1} \ket{e_i}= \sum_{i=0}^{n-1} \dda\ket{e_i}=\dda(\ket{e_0}+\ket{e_1}+\ket{e_2}...)\equiv
\\&&\nonumber
 \left(\sum_{i=1}^n\dda^i\right) \ket{e_0}=(\dda+\dda^2+\dda^3+...) \ket{e_0},
\eea
($k$ in $\dda^k$ is the application degree {ordering and relating graph vertices}). Sums and products properties,  in the context of different aspects of the graph construction,  are explored in details\footnote{The combinatorial, polychromatic graph  can be  related to  causal-set approaches is a clear way. The relation order, through the introduction of $\dda$ operator, provides  the polychromatic structure with  a reflexive, antisymmetric (a poset) and transitive order between graph elements,  combined with the  algebra homogenization  and  metric structures  considering  differently  the chromatic  symmetries. The graph is ordered    and characterized by introduction of  the  state conformal expansion and the state realizations contrasting  the arbitrariness  of the   colors order. Here an operator is intended as a quantity performing specific (logical) actions (evidently connecting elements of the same space). It is clear that the action of $\dda$ is  doublefold: 1. to provide a strict (irreflexive) partial order  and thus ensuring the  chain-of-events/chain-of-operators definition and  relations. 2.  Quantity $\dda$  is related to the coloring (values), i.e. it provides a further degree of freedom   and the clustering after chromaticity and identity in loop of monochromatic  vertices. Action of $\dda$ is therefore  related to the definition of events and can be associated to a color transition or not. For this reason in \cite{start01}, we used a colors shift   $\dda^*$ operator, having all composition properties of $\dda$, to distinguish conceptually the two operations. The colors shift produces only the  equality $(\mathrm{\mathbf{Q}}=\mathrm{\mathbf{P}})$ relation  for two events $(\mathrm{\mathbf{P}},\mathrm{\mathbf{Q}})$ (at the base of monochromatic loops definitions and indistinguishable events) and negation $\mathrm{\mathbf{Q}}\neq \mathrm{\mathbf{P}}$.
This structure is complicated by the introduction of  \bp-events, conceptually related to (acceleration notion\cite{start01} and )  no-locality  (intended as vertices  adjacency) affected by the clustering   in monochromatic or polychromatic loops,  seen as  the vertices in a \bp-$\Ga$  graph characterized by algebras, an order relation and two colors. } in  \cite{start01}. Here, however, we  note  
these definitions establish  a vertex/operator correspondence. We can consider an operators chain, writing an events chain as an associated chain
of operators and viceversa.
From  the events  algebra  $\la^{m}$        the "operatorial"  algebra $\la^{\hat{m}}$ follows associated to     $\la^{m}$ where  $\forall \ket{a_i}\in\la^m\; \exists  \dda \ket{a_i}\equiv \ket{a_{i+1}}\in \la^m$, with $\dda\in\la^{\hat{m}}$ and $\ket{a_i}\neq \ket{a_{i+1}}$. The operatorial algebra  $\la^{\hat{0}}$ is associated to the primary events algebra $\la^0$. Similarly to the events, operators can be composed in  elements of superior algebra or also decomposed in inferior algebras.
The shift operator $(\dda^*)$ acts as ``projector'' relating two events in a sequence without an order relation (changing in fact the vertex color in a colored graph). In general, composition rules of $\dda^*$ follow
similarly to $\dda$--\cite{start01}.
A sequence  of events or operators is {\emph{homogeneous}},  if all the elements of the sequence belong  to the same algebra
\footnote{Homogeneity   implies a "linearity"  in the events indices $(\la^{\alpha_1 n+\alpha_0},\la^n)$,  reflected in the algebra transformations. We generally consider $\alpha_0=0$, where  ${m}=\alpha_1 n$ is  the special case of algebra $\la^n$ conformal transformation, events indexes $(a_i,b_j)$ will be related by $j(i)=k_1 i+k_0$, thus $
(j+1)-j=k_1 (i_{j+1}-i_j)$ and
 $j(i+1)-j(j)= k_1$ where $j>i$, $k_1=\alpha_1$ two algebras  degree ratios.}.
Transformations relating    homogeneous chains  will be particularly important as they can leave conserved many properties of  the graph structure.
Chains homogenization  is a  transformation of a   inhomogeneous chain  into an homogenous one by a  proper algebra adaptation to macroevents (colored) structure.
Graph vertexes are generally clustered in macroevents through equal coloring (loops), a colored algebra adapted to this colors structure arises.  An events base is a  set of vertices (inhomogeneous  or  homogeneous algebra  coinciding with $\la^0$  or a conformally  expanded $\la^0$).
 We denote as \tb{bem} an events set (base) with  adapted algebra, the "reference" \tb{ber} ("inner") base has homogeneous algebra,  generally a conformal $\la^0$ "expansion".
For  convenience we consider  \tb{ber} as   bichromatic, we will use and 2-level logic signal (or 4-level, introducing the "complementary-chains", leading more generally   to a $(r+r)$-levels logic signals).

\subsubsection{Graph realizations,  sequences and chains, \bp-events and loops}\label{Sec:fromz}
  We  here  investigate the  properties of the colored graph $\mathrm{G}$,  laying the groundwork for a dynamic graph  \cite{start01}.
To begin with  we  consider one  graph, assuming  an ordinary  $\la^0$ \tb{ber} we discuss  the emergent   inhomogeneous  basis from the graph coloring and the particular case of   conformally related bases, where  indexes are set accordingly to  the { \tb{ber}}  $\ket{e}$.  More precisely, the colored  graph realizations (chains and  sequences) are  characterized by  a \tb{bem} adapted to the colored vertices clusters. Therefore we consider the cardinality $n=\beth\{v_i\}_{i=1}^n$ of the set of  colored vertices in a    basis $a$  of  a graph $\Ga$.
Thus $r\equiv\beth\{a_i\}_{i=1}^r$  is  the cardinality of the set of {distinct}   colors (values). There is  in general $r\leq n$ and $n=\sum_{i=1}^r n_{i}$, where $n_{i}$ is  the {multiplicity} of the  value $a_i$, that  is the cardinality of the set of vertices (in \tb{ber}) for a  color $i$,  evidently there is $n=r$ {iff} $n_{i}=1\, \forall i$.  Accordingly we  {introduce} the  quantities  $p_{i}\equiv n_{i}/n\in]0,1[:\, \sum_{i=1}^r n_{i}/n=1$ (probabilities-colors frequencies), which are, mostly,  independent from $n$. (Constant probability distributions are in \tb{ber} while similarly defined  sequences  degeneration  ratios   $p_I$,   are generally not conserved for  realization transition, in chain or sequence.)
\textbf{The loops:}  Connections among one-color vertices define  a clusterization   through homogenization, and thus  a vertex with a loop $\ell$ ({macroevent}),  defining therefore a new associated algebras of indistinguishable events. 
\textbf{The connections:} Differently-colored vertexes connections have been  associated to
{derived (partial) events}, \bp-events, regulated by the  colors  probability distributions.
We address the  connections   emergence introducing   new event definition  through the clustering after coloring and the vertexes algebras relations.   

\textbf{The realizations: }
A graph  $\Ga$, at fixed values  $\{a_i\}_{i=1}^r$,   $m$ vertices,  and the constants $p_i$, has multiple realizations, i.e.,   it is  characterized by different sequences which are  the sets of macroevents  of vertices with loops. Each sequence is   then characterized by several chains (directed colored graphs), realizing  the same sequence.
In this representation we clearly decompose a complete graph into its paths (also in   paths bundles of disjoint or crossing chains\footnote{A graph  bundle $\mathrm{B}$ is   the  collection of  bounded--i.e. with common fixed boundary vertices/events for any bundle path-- {disjoint}  paths
 (chains) of the  graph state $\Ga_n$--Figs\il(\ref{Fig:succ})-left.
 The disjount paths of the  bundle do not intersect  apart at the two extreme vertexes constituting  the boundaries.
 We can consider a state $\Ga_n$ totally decomposed in  bundles.
At fixed state $\Ga_m$, we can ask if there is at least a couple of   colors (which can be also equal)  such that  $\Ga_m$ can be  totally decomposed in disjoint bundles or one bundle of paths and if this property is transmitted for conformal graph expansion.
 There are some trivial cases:
  for example   $\Sa_{\min}$ where for each couple of colors there are $\beth \Ca_{\min}=(r-2)!$ possible ways to realize the path. The total number of possible  boundaries for such bundle, and  therefore the total number of bundles, is $\beth B_{\min}=r(r-1)/2$ (not considering the symmetric  $c_{ij}=-c_{ji}$).
For a path $\gamma$ to be a complete (total  chain, made of $n$ vertices for a $\Ga_n$ state,  differently from web which is an homogenized chain with colored adapted algebra) $\Ga_n$ replica in $\Ga_m$ (embedded in a chain) it has to be   $m=n\Omega$ where $\Omega\in N$.
Necessary condition for the graph  to be always decomposed in  $\kappa$  disjoint replicas
 is $\Omega=n\kappa$.} in \cite{start01}--see also Figs\il(\ref{Fig:succ})-\emph{left}) with chain and sequences algebra $\oslash_{\Ca}=\oslash_{\Sa}=m$ respectively in  \tb{ber} {(and $\oslash_{\Ca}(m)\in [r,m[$ in adapted \tb{bem})} for a graph  $\Ga_m$  imposing the maximum valence of graph $2$.
 \textbf{The graph state:} A graph $\Ga$ is defined by fixing the values distribution $\{a_i\}_i$ and  the probability distribution $\{p_i\}_i$\footnote{More basis (colorings) can be defined on  a graph. A polychromatic graph can  be clearly decoupled  in $r$ monochromatic  chains, or in an  $(r+r)$-model, with $r$  complementary chains related to each color,  which can be seen as a base coloring, with event value index $\alpha_i\in [1,\oslash_i]$.   We considered in \cite{start01}  a particles/hole  for particles/events $p$ and holes $h = \non{p}$ related to  the complements  for the graph  $\Ga_m$. In this case, the particles total number  in \tb{ber}  (for state and complete realization) is $\beth \{p\}= m$ (independent by the colors number or the probabilities)
and $\beth\{\non{p}\} = m(r-1)$. The ratio $\beth \{\non{p}\}/\beth \{p\}=(r-1)$ does not depend on the  graph state  and  the probability distribution,  not distinguishing different graphs having  equal color numbers and   relating, particularly,  not-isomorphic graphs.
}. The {\textbf{state}}  $\Ga_n$ of a graph $\Ga$ is defined by fixing  the cardinality $n$ of the vertex set.
The graph  state \textbf{realization} is  determined   by its {sequences  $\Sa$  and  chains} $\Ca$.
 However a graph $\Ga$ may be equipped with some constraints governing the clusterization in sequences  or chains, and  in this way reducing the degeneration.
 \textbf{The conformal expansion.}
 Then, if $m\geq n$ is the events number  in the \tb{ber}, the graph  $\Ga$  underwent (transition) from the \emph{state} $\Ga_n$ with  $n$ vertices  to the \emph{state} $\Ga_m$ with  $m$ vertices. Conformal transformation (expanded or contracted graph) between graph states  is related  to  the conformal metric transformations.
To simplify our discussion we   assume the following \textbf{$\mathds{N}$-constraint}:
$m/ n\equiv \ff{ m/ n}$,  where there is $\sum_{i=1}^r m_{i}=m$, we shall assume the conditions  $m_{i}=p_{i}m=n_{i}\ff{ m/ n}$ or $m_{i}/n_{i}=(\sum_{j=1}^r m_{j})/(\sum_{k=1}^r n_{k})=(1/r)\sum_{j=1}^r(m_{j}/n_{j}) =\ff{ m/ n}\; \forall i$, thus  $m_{i}$  is the macroevents number  in the reference basis associated to the color $a_i$.

 For convention here we consider a \emph{seed} graph state as $\Ga_{{n}}$,  with a  minimum $n=\min{m}$.
 In   \cite{start01}  we include more discussion on  graph relational structure intended  as set of realizations after conformal expansion.
 Consider  different states complete realizations, i.e. with  $m$ vertices for a $\Ga_m$ state,  if  $m>n$ then  any realization of  $\Ga_n$  is  embedded in the realizations of  $\Ga_m$, and  $\Ga_n$ has different replicas in $\Ga_m$. Then $\Ga_m$ realizations can be set in one-one  correspondence, in different ways, by a conformal transformation, with the  entire realizations of the  $\Ga_n$-state, in this sense we can say that the set of $m$ realizations contains the $\Ga_n$ realizations (graph self-similarity and scale structures).
On the other hand, new relational structure is generated after expansion--\cite{start01}.
 We characterize a graph by its states and the  degenerations and we show  particularly how  the conformal transformations do not preserve the graph structure in general, but  the sequences of an expanded graph includes all the conformally expanded sequences (the respective algebras are conformally expanded) of the original graph.
 Some realization  properties are not preserved by the application of the group of permutations on $n$ vertices.
Some graph properties   are invariant for  state transition,  whereas other graph features are transformed according to the probability distribution.
On the other  hand, graphs with same $\{p_i\}$ and number $(r,n)$ but different bases have clearly equal structures, therefore we defined these graphs as equivalent (isomorphic) graphs.
We  discuss  here in details the chains and  sequences degeneracy,
much  of this discussion mirrors some aspects of the  combinatorial graphs.
   We  consider the sequences degeneration    and we discuss a criterion for the \emph{eligibility} of sequences and chains as part of  a state degeneration, that is a criterium to establish  those chains and sequences  emerging as  graph state realizations, in adapted base or \tb{ber},  based on the analysis of the necessary and sufficient conditions for the occurrence of equal colors vertices adjacencies. These quantities are then  studied under state transitions, particularly relevant  is the minimal $\Sa_{\min}$ sequence ($r$ vertices for $r$-colors graph  with maximum algebra $\oslash_i$ for each color $i$) and the supremum  $\Sa_{\sup}$ sequences (with $m$ vertices for a graph state $\Ga_m$ with minimum algebra per vertex $\oslash=1$) and associated chains $\Ca_{\min}$ and $\Ca_{\sup}$ respectively, constituting  $\Sa_{\min}$ and $\Sa_{\sup}$ degeneracies.   The existence of these "boundary" realizations, and especially  $\Sa_{\sup}$ and  $\Ca_{\sup}$,   turns to be  a relevant graph feature. We  considered in particular two special cases: the  equiprobable graph where there is  $p_i=1/r$ for any color $i\in\{1,...,r\}$  and secondly  we investigate the effects of presence of a maximum in  the distribution of $p_i$ ratios, and more generally the probability distributions in the graph realizations and conformal expansion.  The exploration of the  two limiting cases  will be crucial in determining the graph loop on vertex and their variation by state, the presence of a maximum $\Sa_{\sup}$ and  $\Ca_{\sup}$, and the \bp-events emergence.
We focus on the loop (macroevent) sets for different graph states and the existence of maximum (algebra $ \max\oslash_\ell$)  loop, discussing the existence of the minimum number, $\min\beth \ell$,  of loops  $\ell$ and their algebra $\oslash_\ell$ and loops emergence after state transitions.  Using these results  we investigate the emergence of  \bp-events and  other aspects of the graphs conformal expansion--Figs\il(\ref{Fig:succ}).
\subsubsection{Inhomogeneities of  the graph  state realizations: degeneracy  in chains, webs and sequences}\label{Sec:deg-new-w-web}
We evaluate the chains and sequences degeneracy   for a   colored graph state  $\Ga_n$  and after conformal transformation. The metric structures and  the spinorial emerging structure would be  eventually adapted to  the colored graph realizations \cite{start01}. There is a generation of  different events and graphs ($\partial \Ga$ whose vertices are \bp-events) associated with the graph  inhomogeneity (differently colored vertices).
 The metric structure rests on the graph states which is made by evolving (conformally related) self-similar blocks.
The change in algebra and colors, in  this framework, is  closely related to  the events definition.
For a fixed state,  the sequences realizations have  different  macroevents (multiplicity $N_i$ for $i$ color) and, the different chains, for a fixed sequence, have different  \bp-events, whose number (for fixed vertex number $n$), and the cardinality $\beth\Ca$ of chains for graph state $\Ga_n$, are  constrained. The graph conformal expansion (a state transition) changes also the maximally decomposed realizations $(\Ca_{\sup},\Sa_{\sup})$  having  maximum number  of differently colored  macroevents in adapted algebra, and the the loops and \bp-events emergence  for the graph  state transition and graph transition from one  realization to another,   constrained by the probability distribution.
In   \tb{ber}, the total number of chains, $\beth \mathbf{C}_n$, of a $\Ga_n$ state  is the multinomial distribution\footnote{
Clearly a graph state $\Ga_n$ of $r$ colors is associated to a polynomial:
$\sum _{k_{1}+k_{2}+\cdots +k_{r}=n}{n \choose k_{1},k_{2},\ldots ,k_{r}}x_{1}^{k_{1}}x_{2}^{k_{2}}\cdots x_{r}^{k_{r}}=(x_{1}+x_{2}+\cdots +x_{r})^{n}$ where
$\sum _{k_{1}+k_{2}+\cdots +k_{r}=n}{n \choose k_{1},k_{2},\ldots ,k_{r}}=r^{n}
$. We  can   represent the  oriented connections,  including loops on vertex (with no-loop substructure),  for a bichromatic  graph  ($r=2$)  for  $(x,y)$ colors  for a  state $m$,  with  $m$ preserved in the connection set but not the colors frequency  (for example in the anyons model developed in \cite{start01}),   as
$\sum_{i=0}^{m-1}\left(x^i y^{m-i}+y^i x^{m-i}\right)$, (sum is understood as a composition and $x^0=1$)}:
\bea\label{Eq:colpr-C}
&&
\beth \mathbf{C}_n= \frac{n!}{\prod_{i=1}^r(p_{a_i} n)!},\quad\mbox{where}
\\&&\nonumber
\forall n\quad
\beth\mathbf{C}_{\min}= {r!}\neq\min\beth\mathbf{C},\quad
\beth{\mathbf{C}_{\sup}}\neq\max\beth\mathbf{C}\quad\mbox{and}
\\\nonumber
&&
\beth\bar{\mathbf{C}}_{\min}\equiv\beth({\mathbf{C}_n}-\mathbf{C}_{\min})\equiv\beth{\mathbf{C}_n}
-\beth\mathbf{C}_{\min}=\beth{\mathbf{C}_n}-r!>0,
\eea
 $\beth \mathbf{C}_n(a)$
 determines in how many ways it is possible to  distribute $n$ distinct events (as \tb{ber} events,  they  are all distinct as ordered)
in
$r$ different boxes\footnote{Algebra inhomogeneities can  be considered  in the decoupled  $r$ monochromatic chains,  coupled with constraints related to the state and conservation of (classical) probability, or can be considered in  an $r+r$ model of monochromatic chains $\Ca_i$ and their complements $\non{\Ca}_i$  (having different state number with
$m(1-p_i)$ vertices). The polychromatic chains (and the graph) decoupling  into $r$ monochromatic  chains (graphs)
is determined  clearly by the  loops partitions,  regulated by the  multinomial  distribution (\ref{Eq:colpr-C}). In the   $(r+r)$ models, the   monochromatic chains and   the complementaries    are not independent  and the latter are subjected to  constraints different from the first. {In first approximation loops are associated to the symmetric part  (in color) of a polychromatic  chain, whereas  \bp-events (loops free), eventually  determining  the spinorial structure,  to the  antisymmetric one. The chains coupling regulates the \bp-events emergence,
  in $\Ca_{\min}$ (having $r!$ degenerations) and in $\Ca_{\sup}$  depending by the probability distribution and different for conformally related states having self-similar structures and  different symmetries reflected in the metric structures.}}, in each of which there can be at most $ \oslash_i\equiv p_{i}n $ different elements (events) for the $i$-th color (therefore  the permutations   {$P^{\oslash_1,\oslash_2,\dots,\oslash_r}_n$  act on the chain vertices}).
In this way, however, we will make  \emph{no} distinction between elements belonging to the same ``box-value'', i.e. in the homogeneous in color  loops.
 (Note that we are neglecting the distinctness of vertices of a monochromatic loops, which are induced by the order relation for application of $\dda$, considering only the distinctness, we discuss this  \cite{start01} in terms of  associated   statistical distribution).
In this counting  the value-boxes are \emph{distinct}, since they have $r$ different (not ordered) labels. The reference, chain   \tb{ber}  events order  the events  in $ r $ classes of $ \{\oslash_i \}_{i = 1}^r $ elements all equal and not ordered. We count macroevents afferent to the same color but different algebras as different vertices because clustered   with different loops.

\textbf{On the webs: }
Cardinality $\beth \mathbf{C}_n$ of the  graph $\Ga$ chains  can be  parameterized in an  colors adapted   homogenized  algebra, where   equal-color vertices having  different algebras are considered equal, leading  to the \textbf{web} definition.
In the webs only the  number     $N_i$ of macroevents, associated with a  color $ i $, is considered ignoring
their respective  loops algebras in the reference base.
 For any sequence of a $\Ga_n$ state, the total macroevents number   is $N\in[r,n]$,
  correspondingly there is  the minimum $\Sa_{\min}(r)$ sequence  ($N=r$) and  relative chains $\Ca_{\min}(r)$  (with $\beth\{\Ca_{\min}(r)\}=r!$) and a  superior sequence  $\Sa_{\sup}$ and chain $\Ca_{\sup}$.  Condition $n=\sup(N)=\max(N)$ is realized only in special circumstances, where  $\Sa_{\max}$ corresponds to  $N=n$ vertices, this property rarely occurs and it is usually lost for  conformal graph expansion depending on the colors  probability distribution--\cite{start01}. We say that all these  $\Sa^{\otimes}_{N}$ sequences  are "included" in the set $\Sa^{\otimes}_{N}\in[\Sa_{\min},\Sa_{\sup}]$, with multiplicity,  $\Sa^{\otimes}$,  for  fixed macroevent multiplicity   but not (algebra) loop  vertex,  defining the  {web} where vertices with different loops (algebras)  are  indistinguishable in \tb{bem}. For  a graph  state is, the cardinality of   the webs
:
\bea&&\nonumber
{\beth} \mathbf{W}_{N, N_i}(a)=\frac {N!}{\prod_{i=1}^rN_i!}-\sum\limits_{s=1}^q\left[\frac{\left(N+s-\sum\limits_{j=1}^
s N_j\right)!}{\prod\limits_{j=s+1}^r N_j!}\right]\quad\mbox{where}\\
&&\label{Eq:B-Gar}N_i=1 \;\forall i\in\{q+1,...,r\}\quad\mbox{and}\quad q\leq r.
\eea
The last term subtracted in ${\beth} \mathbf{W}_{N, N_i}(a)$  of  Eq.\il(\ref{Eq:B-Gar}) is $r!$.
This cardinality  questions how many are the chains of $ N-1 $ connections (\bp-events) for $ N $ macroevents with different multiplicity $ N_i $, then the sequences $\Sa_n^{\otimes}$ (in \tb{bem}), for a  state $\Ga_n$, so that a chain segment does not connect two macroevents afferent to the same value, or  the equal value macroevents  are never adjacent, particularly in the expanded states there could be web chains or \tb{ber} chains, that is, a chain has a multiple multiplicity $(N_i>1)$  where a  (monochromatic) cluster is considered as a vertex, depending on the colors probability distributions and evolving with the conformal transformations \cite{start01}. This is  the  cardinality of the set of chains with adjacent macroevents, subtracted for the algebra  reduction (homogenization), to  sequences with  lower macroevents number,  given by the second term of Eq.\il(\ref{Eq:B-Gar}).
In this counting, the (not distinct) macroevents of multiplicity $N_i\in[1,\oslash_i]$ (instead of $n_i=n p_i$) are considered. Note that $\{N_i\}_i$, in the web, are algebra independent.

Finally, we note that the ordinary $\la^0$ basis definition   provides an "absolute" for the $\beth \Ca$ definition, using  \tb{ber} indexes, the cardinality in  the ordinary base  $\la^0$, is obviously  greater (or equal) than the cardinality where chains are   in different indexes.
It is therefore straightforward to introduce (Shannon) entropy  $S_m$ for the graph state $\Ga_m$ associated to the degeneracy Eq.\il(\ref{Eq:colpr-C})--\cite{start01}, and  similarly we can define an entropy for a monochromatic  sub-chain as follows:
\bea&&\nonumber
\forall i\in\{1,...,r\}\;
S_i= k_i \ln \Omega_i,\quad
 \Omega^c_i=\frac{(m p_i)!}{\prod_{j=1}^{N_i^c}\oslash^i_{J}!},\\&&\nonumber
\Omega^c_i(\min)=1, \quad\Omega^c_i(\max)= (m p_i)!,\;\mbox{(sum on any subchain vertex).}
\\\label{Eq:cfrit-cross}
&&
\mbox{Thus: }
\frac{1}{m}\log \Omega=\frac {1}{m}\left[\log m!-\sum _{i=1}^{r}\log((mp_{i})!)\right],\\&&\nonumber\mbox{where}\quad
\lim _{m\to \infty }\left({\frac {1}{m}}\log \Omega\right)=-\sum _{i=1}^{r}p_{i}\log p_{i},  \eea
{or we have }
\bea\nonumber
S(m)=\ln [(1)_m]-\sum _{J=1}^r \ln [(1)_{\oslash_J}]=\ln[\Gamma(m+1)]-\sum_{J=1}^r \ln[\Gamma (\oslash_J+1)]
\eea
$(a)_n$ is the Pochhammer symbol  and $\Gamma(z)$  is the Euler gamma function.
(The minimal sequence entropy,  not considering loops algebra, could be expressed as
$S_{\min}=\ln (1)_r$.)
For a graph $\Ga$,  there is  minimum entropy $S$  associated to the  graph seed $\Ga_n$ ($n<m$). We  study this quantity after graph conformal expansion.
The entropy (state independent  in the Stirling approximation) as the degeneracy is an algebra and state  dependent concept Figs\il(\ref{Fig:INstap})-- \cite{start01}.

\medskip

\textbf{Adjacencies,  loops and  $\Theta$-criterion}
This  analysis fixes the presence of a totally decomposed sequence, the \bp-events emergence and transformation after conformal expansion,   the minimum  and maximum loop algebra for a $\Ga_m$ state. The $\Ga_n$ realization could have, according to the probability distribution, a necessary adjacency of equal color events in any  chain of $\Sa_{\sup}$.  Therefore, for a  graph state $\Ga_n$,
 there may not be a  macroevents sequence  realizing  the supremum  ($N_{\max}=N_{\sup}=n)$, i.e. a totally decomposed sequence ({in other words we address the issue questioning  the  prevalence of loops or \bp-events after conformal transformation}). The adjacency  would lead to a higher algebra macroevent and consequentially there is a reduction of $ N = n$ vertices according to the number of the adjacencies  and this    sequence must be rejected as coincident with one other sequence of the graph realizations. In other words there is in general   $\max N\leq \sup N$
(while $\exists ! \min(N)=r$ and $\exists ! \Sa_{\min} $. It is  clear that if $\max(N)=\sup(N)=n$, then   $\exists! \Sa_{\sup}$ and,  as for $\Sa_{\min}$, there is no degeneracy  in color  or in algebra, but $\Sa_{\min}$ and its $r!$ chains  is  an intrinsic    (state independent relational structure) graph characteristic.
Then $N=\beth (\eth a)+1$, where $\beth(\eth a)$  is the  \bp-events number  for the  $i$-sequence excluding the boundary of the first and last event of a $n$-events chain of a $\Ga_n$ state the vertex  boundaries in graph  vertex definition is also an issue of the metric graph\footnote{A polychromatic connection (\bp-event) corresponds to a  \bp-$\Ga$  vertex. In contract with a  $\Ga$-vertex, which is always  monochromatic  and with an algebra $\oslash$,  a  \bp-$\Ga$  vertex is dichromatic, with an order (a values $\pm$), and algebras  $(\mathbf{\mathcal{f}}_1(\oslash_1),\mathbf{\mathcal{f}}_2(\oslash_2))$, with $\mathbf{\mathcal{f}}_i(\oslash_i)\equiv a_i+b_i\oslash_i-c_i$   (where  $i\in \{ 1,2 \}$) with values $(a_i,b_i,c_i)=\{ \pm1,0 \} $,  depending on the loop  counting on first or second  (respect to the connection order) vertex .}.).
\begin{figure}
  \includegraphics[width=5.9cm,angle=90]{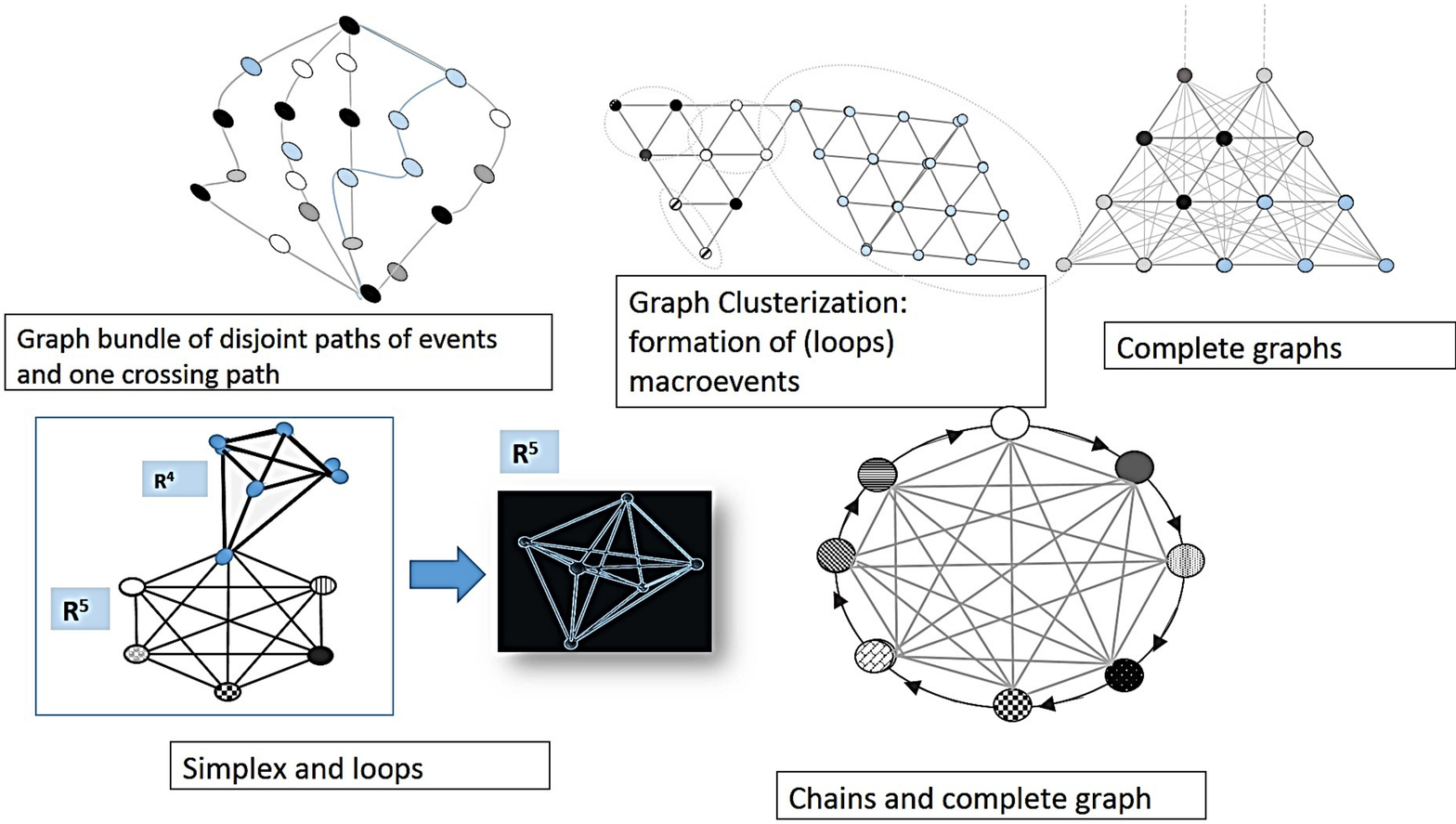}
  \includegraphics[width=5.9cm,angle=90]{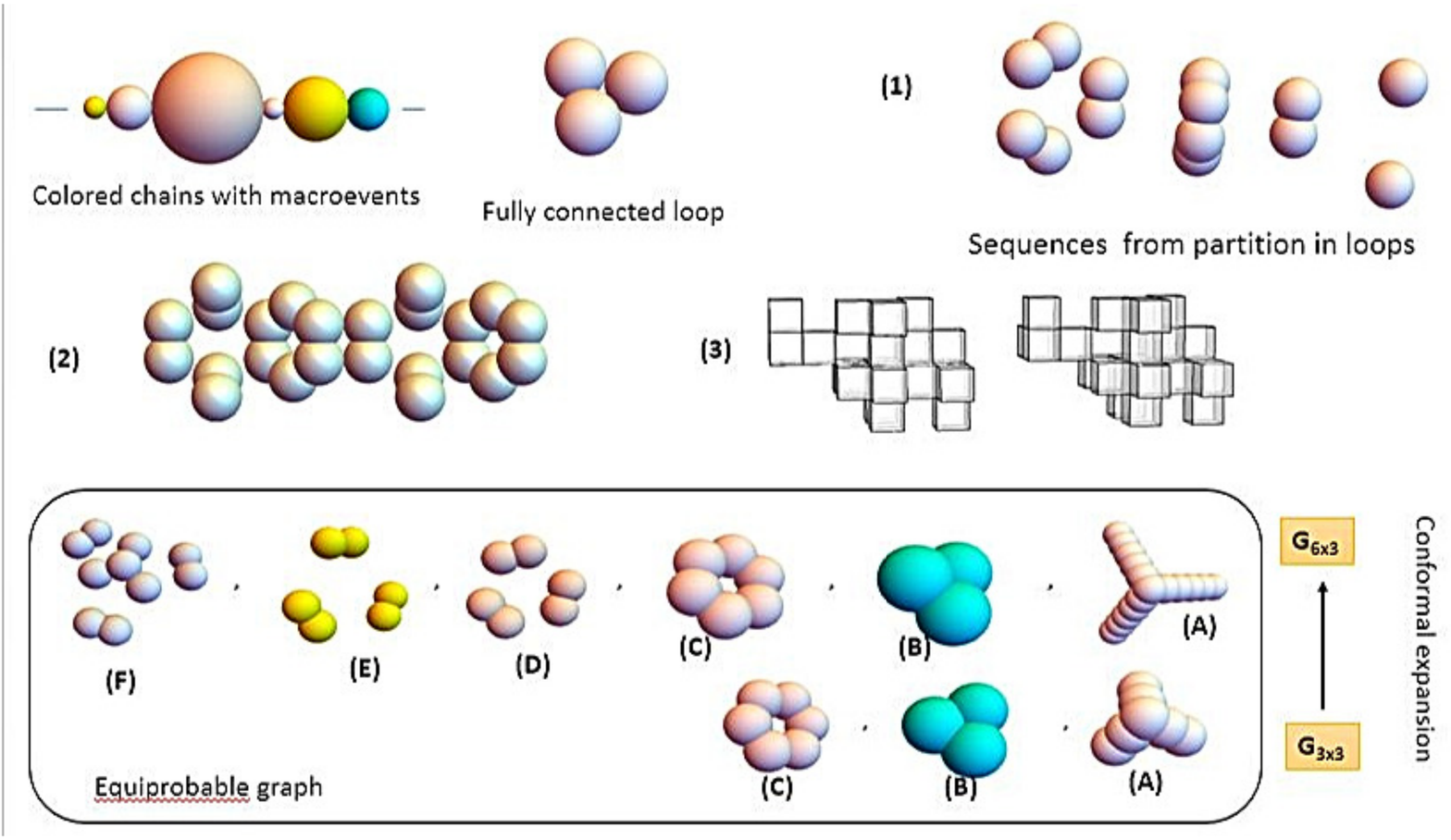}
  \caption{Upper panel: scheme of the graph structure and  different components, chains, bundles and relation with simplex. Bottom panel: \textbf{(1)-(2)-(3)}
  elements. Trinomials $\mod g$.
Multinomial distributions $\mathbf{M}[n_i,n_j,n_k]$, providing the first three positions of the number  $s$
represented as    spheres/boxes location, the similarities in different schemes are evident especially for conformal transformations in the  below box.
\textbf{(1):} $n_i, n_j\in[0,4]$ for $n_k\in[0,16]$, $g=4$,
$s=1$.
\textbf{(2)}-\textbf{(3)} gives the multinomial coefficient where
$n_i=4 - i, n_j=4 - j, n_k=16 - k$ and $ g=4$,
for  $s=2$ \textbf{(3)} and
\textbf{(2)}
  $s=3$.
Below box. Conformal transformation.
Multinomial distribution: $n_i=x-i, n_j=x-j, n_k=x-k, \{i,j,k\} \in[0,x]$.
Below elements:
$x=3$
from  \textbf{A}-to-\textbf{C}
element $s=(1, 2, 3)$, respectively.
Upper elements: $x=6$
from \textbf{A}-to-\textbf{F}
$s=(1,2,3,4,5,6)$, respectively.
}\label{Fig:succ}
\end{figure}

We can write the set of  $\Sa^{\otimes}_{N}$ sequences, ($\Sa_{N}$ degeneration class in  \tb{bem}) as:
{\small{
\bea\label{Eq:S-algenra}
&&\Sa_{N}= \{\Sa^{\otimes}_N\}_{N=r}^{N_{\max}}
=\underbrace{\prod}\limits_{a=1}^r \left(\sum\limits_{N_a=1}^{\oslash_a}  N_a\right) \Hl [1-\Theta], \\&&\nonumber \beth \Sa^{\otimes}= \prod\limits_{a=1}^r\oslash_a-
 \beth\{\Hl [\Theta-1]\}_{a=1}^r,\quad\mbox{where}
\\\nonumber
&& \Theta\equiv\left[\max\{N_a\}_{a=1}^r-\sum\limits_{a=1}^{r-r_{\max}}N_a\right]_{\leq1},\quad \mbox{then}\\&&\nonumber \exists! \Sa_{\min}\equiv(1,..,1)\mathbf{-}[{\oslash_1},...,\oslash_r(n)]_{\oslash},\\&&\nonumber \Sa_{\sup}\equiv(\oslash_1,..,{\oslash_r})\mathbf{-}[(1^{\oslash_1}),...,(1^{\oslash_r})]_{\oslash},
 \eea}}
$a\in[1,r]$  is the color  index, $N_a\in[1,\oslash_a]$ is the number of {macroevents} for a color  $a $, the sum $\sum_{N_a=1}^{\oslash_a}  N_a$ is the decomposition, at the fixed  color, of its algebra (integer $\oslash_a$  partition). In the sum, $r_{\max}$ refers to the color associated to the maximum of the macroevent number $N_{\max}$ in the selected sequence.
$\underbrace{\prod}$ is the  symbol of  composition in sequence,
acting  on the sums by composing elements of the sum  of the partition  with the decomposition (distributive property).
Round brackets,
$(N_1,...,N_r)$ ($r$-entries), denote   the macroevents number   for each color $i\in[0,r]$, and square brackets $[(),...,()]_{\oslash}$,  $[(\oslash_1^1,..,\oslash_1^{N_1})_1,...]_{\oslash}$,  denote the algebra distribution $(\oslash_i^1,..,\oslash_i^{N_i})_i$ between the $N_i$ macroevents  for each equal-value  $i$ macroevent.
 Where  $\Hl[x]$ {is the unit step function (Heaviside step function)}, such that  $\Hl[x]=0$ for $x<0$  and $\Hl[x]=1$ for $x\geq0$.
This equation provides the actually eligible (effective) vertices, eliminating the necessary adjacency, however it does not   discern the difference in algebra between equal color macroevents  in the  sequence, which therefore constitutes a further degeneration in addition to chain degeneracy.
The quantity in $\Theta$ can also be negative and  it serves to eliminate those particular  sequences in  the product, that would not give any macroevent  implying  a necessary equal color vertices adjacency. Hence, the threshold term $\Theta\leq 1$ ($\Theta$-criterion) excludes from the counting the cardinality of excluded sequences, and it    obviously depends on  the graph state.  (This criterium  is investigated in details in \cite{start01} where we study this under conformal transformations. Particularly we explored  equiprobable distributions closely associated  to the conformal homogeneous algebras).
 {The equiprobable systems always have a maximum degeneracy in sequence}.
The degeneracy in  algebra of  supremum sequence is obviously $1$.
We note that this question is tantamount to asking whether, given the particular choice of macroevents   $(N_{1},...,N_{r})$, there is at least one acceptable chain realization (according to the no-adjacency condition) and for conformal transformation.  (If $\non{\exists} \max\{N_a\}_{a=1}^r$ for a sequence,  there is  $N_a(1,1,1..,1)=N_a \Sa_{\min}$ which, in the case of maximal sequence satisfies  the $\Theta$-criterion).
Complete analysis of these cases is in\footnote{In \cite{start01} we particularly  focus on partitions and combinatorial problems  in the construction of graph states and their variation after conformal transformations, particularly  partitions  for the monochromatic  loops the issue of adjacencies and invariants after expansions  Here  however we consider   polynomials for equiprobable distributions, we can describe this problem in terms of an $r$-th degree polynomials in $\ff{n/r}$  variables. There are several expressions with polynomials of this partitions  and combinations problem, here it is natural to consider  for the  equiprobable \emph{associated polynomial}:
$
\mathcal{P}(\Sa_{\odot})=
 \sum _{{k_{1}+\ldots +k_{\ff{n/r}}=r}}{r!\cdot \prod _{{i=1}}^{\ff{n/r}}{\frac  {x_{i}^{{k_{i}}}}{k_{i}!}}},
  $
 summation of all possible $ \ff{n/r} $-plets, whose sum of the elements is equal  to  $r$.
 This follows in fact from  the application of multinomial theorem,
$x_{\alpha}$ stays for the $\ff{n/r}$ variables such that $x_{1}$ corresponds to the number of loops $1$  and $x_{\ff{n/r}}$  corresponds to  $\ff{n/r}$ macroevents. The variable power indicates the color, thus for example  for $x_{4}^2$ the multiplicity of the macroevent associated to the value $a_2$ is $N_2=4$.
We can paraphrase the problem through the interpretation of the multinomial coefficient as  the number of ways in which   $r$ objects can be distributed in  $\ff{r/n}$ boxes, such that $k_i$ objects are in the $i$-box and the permutation of $r$ objects of which are   $k_i$ equals.
 If  $\ff{n/r}\leq r$ then \emph{all} the sequences satisfy $\Theta$-criteria and then  $\beth \Sa^{\otimes}_{\bigodot}=\ff{n/r}^r $. The case  $\ff{n/r}>r$, i.e. the number of polynomial  variables is larger  then the  polynomial degree,
is  considered more deeply in\cite{start01}.
}  \cite{start01}.
A further related  aspect, significant in this analysis, is the loop symmetries (equal color vertices).
Discussion on the loops distribution involved partitions of equal-value graph vertices:
it is in fact  the problem of  partition  $\mathbf{P}(\oslash_a,N_a)$ of an integer $\oslash_a=p_a n$ in a $N_a$  fixed number of slots,
 where $N_a$ is the number of  summands in one of $\oslash_a$ partition.
Algebra analysis   is related  to the identification of sequences loops distribution,  determining  metric symmetries  (evolution) for graph  under conformal expansion and  also the related to  the graph entropy definition--\cite{start01}.
Discussion on loops involves also analysis in terms of specific  polynomials (associated to $\mathbf{D}_{\oslash_\ell}$ dihedral group of order $\oslash_\ell$). For polynomials of  equiprobable distributions, we can describe this problem in terms of an $r$-th degree polynomials in $\ff{n/r}$  variables. There are  however several expressions with polynomials of this partition  and combinatorial  problem.
In the reference algebra $\la^{0}$,
for any  color $i$ there is $\oslash_{\ell_i}\in[1, \oslash_i]$, where $1$ stays for  a  $\la^{0}$ vertex,
 $\oslash \ell$ for the algebra (order) of loop $\ell$, with respect to \tb{ber}.
Thus  there is $\oslash \ell_{J_i}=\oslash_{J_i}$  algebra of the macroevent  $J$ associated to the value $a_i$.
The cardinality  $\beth (\ell_i)$ of the set of loops   with $i$ color,
is the $i$-color loops (macroevents) number.
For a chain, if
$\exists!\; i:\exists!\; \min \beth{\ell_i}=1$, then there is $ \oslash \min{\ell}_{i_{\alpha}}=\Theta_{i_{\alpha}}$, where $\Theta$ is introduced in Eq.\il(\ref{Eq:S-algenra}),
and $i_{\alpha}$ is macroevent associated to the  value $a_i$ to which the minimum loop belongs, and that is also  the \emph{necessary} loop   for that chain.
Therefore, if the minimum order (algebra) of the necessary  loop (according to the $\Theta$-criterion) is  $\min \oslash\ell_i$, then there must be  $\min  \oslash\ell_i$ vertices that admit only (2-valence) connections between them, according to $\Theta$-criteria. The investigation of   the minimum order of loop means,  in the first place, to asses the necessary existence of a vertex with loop, and this case is  realized only by a one single vertex (macroevent multiplicity and loop cardinality equal $1$)  and, as we  discuss \cite{start01},  the minimum necessary loop  has to be unique.
 Graph loops number   and its state-dependence,  the  existence of  $\max\beth\ell$ and the order $\max\oslash\ell$ can be studied considering a decomposition in polygons  according to vertex algebra.
We   faced the problem of determining the maximum number of loops,
$\max\beth\ell$, and the maximum loop order, $\max\oslash\ell$, for a graph and  the variation of these quantities  following a conformal graph expansion in \cite{start01}.
{In particular, for the equiprobable $(\odot)$ case, there is}
$\beth_{\odot} \ell(i)=2^{\frac{m}{r}}-1-\frac{m}{r},\quad\beth_{\odot}\ell(r)=
r(2^{\frac{m}{r}}-1)-m$--Figs\il(\ref{Fig:INstap}).
These relations  explicitly consider    the  change for conformal transformation  through direct dependence on $m$ and the growing of the range $\oslash_{j_i}\in[2,\oslash_i]$,
(note when $m=n$, for the seed state, then $\beth_{\odot} \ell_i(n)=0,\,\beth_{\odot} \ell(n)=0$). 
Therefore we asses in this way  the existence of the minimum number of loops  $\min\beth \ell$  and their order $\min \oslash \ell$ and  $\oslash \min \ell$ and loop emergence.
We particular focus in \cite{start01} on the \emph{minimum order} (algebra) $\oslash \min \ell$  of the \emph{necessary} loop    for a graph state, and  the minimum necessary number of loops, $\min\beth \ell$, and after a graph conformal transformation. The loop algebra is connected with the  metric invariants and the graph inertia $\mu$, moreover these quantities are related to the graph radius definition. These considerations deal also with the graph (self-similarity in sense of chains embedding and graph replicas in higher states) properties after conformal expansion.  Note that vertices in an homogeneous loop considered here are symmetric graph (sub)states.
Considering the  symmetry group related to loop polygons (actually a polygonal with open boundary conditions),
quantity should  transform accordingly to the $\mathbf{D}_{\oslash_\ell}$ associated to loop\footnote{
 The rigid movements of a regular polygon can be thought  as vertices
permutations.
Then, each group is isomorphic to one
subgroup of a group of permutations ($S_n$).
According to Cayley' s theorem if
 $G$ is a finite group of order $n$, then $ G$`is isomorphic to a
subgroup of $S_n$. (i.e. we could say Cayley  theorem  implies that  each group can be considered as a particular group of permutations).
 In general, however, not all the permutations of the $n$ vertices of a regular polygon are induced by symmetries of the polygon.
In general, the symmetries of a regular polygon having $n$ sides are:
the identity $id$,
the rotations $r_i$ (amplitude $2 \pi/n$ (here related to the loop radius))
 around the origin;
$n$ axial symmetries $\{s_1, ..., s_n\}$ with respect to a beam of $n$ straight lines passing through the center the axial symmetries  depending if
if $n$ is odd or even,
$D_n=\{id,r...,r^{n-1},s_1, s_n\}$ is a not abelian--(axial rotations) group of order $2n$ isomorphic to a subgroup of $S_n$ (Cayley theorem).
All the $D_n$ elements   are determined once the rotation $r$ has been assigned and any axial symmetry $s$. Finally, the question if  a (monochromatic)  loop should be considered open or closed, in the sense of graph cluster,  touches some aspects of simplex-graph relations and loop  inner symmetries. Here we associate to any loop a (outer, boundary) valence $2$.} $\ell$,  having proper boundary conditions (closed) for ``reflection'' of indexes $a_{i-\ell}$  (equal color vertex) coincides with the  $a_\ell$ of the  loop with rotation for angle $2\pi/\oslash_{\ell}$ (loop radius)\footnote{
Concerning the  loops and their representations, multisets (or bags) formalism  can be a  valid  homogeneous loops representation,   for sequences but not for chains (i.e. $c_{ijjji}\equiv c_{ij}c_{ji}=c_{II}$)
where the value of multiset coefficients (number of multisets of cardinality $\oslash_\ell$, with elements taken from a finite set of cardinality $\oslash_m$) can be given explicitly as
$ \left(\!\!{\oslash_m \choose \oslash_\ell}\!\!\right)={\oslash_m+\oslash_\ell-1 \choose \oslash_\ell}={\frac {(\oslash_m+\oslash_\ell-1)!}{\oslash_\ell!\,(\oslash_m-1)!}}
$.
}.
Adapting  a  metric structure to the colored graph, we encounter in the problem to  provide a suitable topological structure and  appropriate analysis of graph  isomorphism  under conformal transformation, we discuss these aspects in \cite{start01}.
{For a metric graph  it  would be important to provide a  proper graph ``topology'' notion, in particular concerning a separability definition\footnote{A further point to be addressed is whether  an events chain can be seen as a  Cauchy sequence (in the sense also  of clarifying how and when we can eventually provide  a notion of  Hilbert space or Banach space properties). More generally, given a metric space (X, d), a sequence
$x_1,x_2,x_3,...$
is Cauchy if, for every positive real number e > 0 there is a positive integer N such that for all positive integers m n > N, the distance
$d(x_m, x_n) < e$. (In this model a vertex norm is never null and then the distance between vertices is never null, it can be null instead $\sigma(\cdot,\cdot)$ and $g(\cdot,\cdot)$). Generally, for a  discrete metric, any Cauchy sequence of elements  must be constant beyond some fixed point, and converges to the eventually repeating terms. The notion of convergence of sum of vertexes  has to be considered through an appropriate topological notion adapted to this graph framework. A further related problem consists in the research of graph ``fixed-points``:given graph property $ f_{\Ga} (m)$,  we can ask if it is satisfied  $f_{\Ga} (m)=f_{\Ga} (m_0)=f_{\Ga} (\kappa m_0)$.
We face this (persistence) problem also in \cite{start01}).}.
Since  the graph  admits isomorphic parts and  propagates (in the sense of conformal expansion) transmitting  seeds-copies and  isomorphic parts then  a question to be addressed  regards the events graph  isomorphism\footnote{More precisely  here intend a bijective application $f$ from the vertices of $\Ga$ to the vertices of $\widetilde{\Ga}$ that preserves the ``relational structure'', in the sense that there is an edge  from the vertex $a_{\bullet}$ to the vertex $a_{\circ}$ if and only if there is a similar connection from the vertex $f(a_{\bullet})$ at the vertex $f(a_{\circ})$ in $\widetilde{\Ga}$. This notion has to be considered into a dynamical graph framework for transformations preserving relational structures. The relational structure is  transmitted in the graph conformal expansions to larger states while a new  relational structure appears, expansion of the graph structure as discussed in \cite{start01}.
We could use a conformal expansion model   with    constant graph vertex  number  after expansions,    and equal to the  seed   initial state graph  $\Ga_n$, the expansion would be on charge of the connections number. Conformal  expansion connects  vertexes of the seed  graph  $\Ga_n$ to the twin  graph $\Ga_{n}$ for the first step of expansion. At the $\mathrm{(m-1)}$-step there is the $\Ga_{n}$ $\mathrm{(m-1)}$-copy    (completely) connected to the
$\mathrm{m}$-$\Ga_{n}$  copy (all the $\Ga_{n}$ vertices,  which  have always zero valence in  $\Ga_{n}$, are connected with all the vertices of the next-and according to the model to the former- copy $\Ga_{n}$). The connection between the two copies in the 1-step process inherits  the  connections of the former  states $\Ga_{n}$ with the $\mathrm{(m-2)}$ copies, according to the specific transmission model fixed by some  "boundary data". (Since connections are all "flat" (equal) this model closes  the simplex).
Particularly useful in the equiprobable graphs representations, this model can be used also  in  the conformal expansion with  expanded  vertex number, in the study of the isomorphism of complete realizations replica in larger states. In general, there is then,
$N^{out}_v(m)=N_v^{out}(0)\prod_{\chi=m}^1(r-r^o(\chi))(r-r^i(\chi))$  and   $N^{in}_v(m)=(r-r^i(m)) (r-r^o(0))\prod_{\chi=m}^1(r-r^o(\chi))(r-r^i(\chi))$ where
$N_v^{out}(0)=r-r^o(0)$,  for the number of connections per vertex outgoing (ingoing) $N^{out}_v(m)$ ($N^{in}_v(m)$) from the  $\mathrm{m}$-step ($\mathrm{(m-1)}$-step)  graph $\Ga_r$ copy  to the  $(m+1)$-step ($m$-step) copy $\Ga_r$   in a conformal expansion  process from state   $\Ga({m_0})$ to $\Ga(m)$ (obviously here $(m, m_0)$ are the expansion step from the seed $(m_0)$ to the step ($m$) the union $\bigcup_{i=m_0}^{m}\Ga(i)$ can be directly related to the $\Ga_m$ from the conformal expansion in the variable vertex number mode). Then $N_v^{out}(\kappa)$ provides the total number of outgoing connections per vertex from the state of (colored direct) graph  $\Ga_\kappa$ (with "memory" in the sense of transmitting structure/connections as explained below) from former state $\Ga_{m_0}$ to  expanded  $\Ga_\kappa$  ($\kappa>m_0$) inheriting or not previous connections (of former state of the expansion)  depending on the values attributed to the boundary data $(r^i(x), r^o(x))$, $x$ is a graph state variable
and stands for the number of steps in the conformal expansion, while  $r$ number of graph vertices assumed to be colored. There can be  then $r^o(\kappa)=\{0,1\}$ and  $r^i(\kappa)=\{0,1\}$  for each state (particularly $\kappa=m_0$)and  depending on the chosen model;  each connection is differently considered according to be a loop or antisymmetric, conditions  on $(r^o(\kappa),r^i(\kappa))$ deals also with this aspect,  $r^i(0)=\{0,1\}$ considers the possible previous history to a zero state $m_{0}$,more specifically we intend a graph symmetric -connection, the absence of this condition $(1)$ would give rise to a not completely  connected graph copies and can be seen as a model for signal transmission from a source-vertex to others of a  simplex, it is also clear that the outgoing-sign of the "directed" conformal  expansion can be also read as a back-transmission, a "reply"-direction, to the simplex vertexes which are always "trapped"  in a (constant) $\mathbf{R}^{r-1}$    dimension (or superior) spaces thus the condition of not-reflexive connection.
 The explicit  choice of $(r^i(x), r^o(x))$    makes  possible to change the structure transmission model for expansion in accordance with each process state.     To obtain the total connections number (during a transition) it is necessary to multiply $N_v^i$ or $N_v^o$   for  the number of vertices per state which is always $r$.
Clearly any graph  chain  of $m$  vertices can be a well chosen path in the graph sequences $\sum_{i=0}^m\Ga_r(i)$, the sum is clearly intended here in the sense of clusterization introduced for graph vertexes.}.  Two graphs are isomorphic if one can be transformed into the other simply by renaming its vertices, this  has  be considered  in the metric graph context\footnote{{However, the problem of isomorphism of subgraphs (appearing here in the scale invariance  and conformal transformation) is $NP$-complete, although it is evidently in $NP$, it is suspected that the graph isomorphism  problem  is neither $P$ nor $NP$-complete. Note that a graph is said to be strongly connected or disconnected if every vertex is reachable from every other vertex. The strongly connected components or disconnected components of an arbitrary directed graph form a partition into subgraphs that are themselves strongly connected. It is possible to test the strong connectivity of a graph, or to find its strongly connected components, in linear time. Note that this problem of connectivity is indeed strongly related to  locality problem through valence notion. We address this issue in relation to equidistant (connected with equal edges) vertices  in \cite{start01}.} }.
The loops study   faces  a different aspect of the conformal transformations and on the $\mathds{N}$-constraint and on the choice of color base\footnote{ Then, a proper characterization of the graph state transition has to be included for the case  $m=\infty$.
Loops distribution,  $P(\ket{\oslash_0})$,  has always cardinality greater then $\oslash_0$;
thus being $\oslash=\infty$, the  (countable) cardinality is cardinal.
On the other hand, the number of ways in which $n$ events of $r$ colors can be combined,
considering  also a loop  on one value with  $\oslash_i=n$ (not satisfying the conservation of $p_i$ at $\Ga_n$, having role in the conformally expanded graph),  is  $\beth \mathrm{G}=r^n$ (homogeneous  loops).
If the events $n\in \mathds{N}$ are infinite there is $\beth  \mathrm{G}=\aleph_1=r^{\aleph_0}$.
If  colors  $r\in \mathds{N}$ (but infinite) then  $\beth  \mathrm{G}=\aleph_1$ holding also if
  $r$ is infinite and not numberable  and 
if  $n$ is finite or infinite but numerable.
On the other hand if $n$  (\tb{ber} events) is infinite \emph{not} numerable then  $\beth  \mathrm{G}= \aleph_2$.} $r$ (finite and discrete). (A further related problem is in what extent the events chains in conformal expansion could be considered as Markovian chains).
{Large conformal expansions would represent the asymptotical states, while the  inhomogeneity changes  with  realization at equal state. We could say that the larger is the inhomogeneity (in color or algebra) and more articulated and complex   are  graph  (and  metric)  levels, more articulated is the emerging spinorial structure depending on the colors probability distribution.
 Chains transitions are related to change in the loop-\bp-events  distribution, where  loops presence is balanced by the \bp-events in the chains.}

\medskip

\textbf{Connections, \bp-events and inhomogeneous vertex aggregates}

 Inhomogeneous in color connections, \bp-events, and  more generally not-identical aggregates of any order, arising  from permutation (and rotation) groups representations, would play a relevant role in the metric and dynamical frame considered in this approach, with the analysis of  different representations  of graph chain realizations\footnote{
We could introduce a ``graded structure`` in terms of  "annihilation" and "creation" operators (related to  $\dda$ evolution), written as    as graded anticommutative  product  ({{Superalgebra a $Z2$- algebra graded.}}). Considering also algebra inhomogeneity:
 $[a_{p\sigma}^{\dag},a_{q\kappa}^{\dag}]_{\lambda}=a_{p\sigma}^{\dag}a_{q\kappa}^{\dag}-\lambda a_{q\kappa}^{\dag}a_{p\sigma}^{\dag}=0$
where
  $\lambda=\Theta^{(pq)(\sigma\kappa)}\equiv\Theta^{(pq)}\Theta^{(\sigma\kappa)} -2\bar{\delta}^{pq}\bar{\delta}^{\sigma\kappa}$,  in the connections of the order $2$, \bp-events, where   macroevents are considered adjacent,  $(\sigma,\kappa)$ denote the algebras of two events, $(p,q)$ are colors indexes, $\bar{\delta}^{xy}=\delta^{xy}-1$, $\Theta^{xy}=\bar{\delta}^{xy}+\delta^{xy}=e^{\imath (1+\delta^{xy})\pi}=(-1)^{\bar{\delta}^{xy}}$. The ``truth table''  emerges as $A\lor B$, $\mathrm{OR}$ logic gate, with the
the  logical disjunction as an operation on the  two logical values $\pm1$ (a connection can be represented by the function $\psi(a_2,a_1)=(-1)^{1-\delta(a_1, a_2)}\psi(a_1,a_2)$, where $a$ is a base index, $a_i$ value.)
}.
In this  frame it can be simpler to  explicit the related Clifford algebra:  using such representations  for antisymmetric connections, events-aggregates, according to the permutations,  obey
Clifford statistics, thus vectors as  plexors (for permutation) and spinors (for rotation) would appear, where spinors
can be seen in some way as  plexors  of  aggregates--\cite{start01}.

 We conclude with some considerations on the emergence of polychromatic  connections, whereas we deepen these issues in  \cite{start01}.  Let $\beth c(m)$ be   the cardinality of the set of  all possible  edges in a (fully connected) graph of vertices $m$, including the symmetric connections, and $\beth{\mathbf{C_i}}(m)$ the number of connections per vertex of the graph  (valence in a fully connected graph).
 We note that for  $\bar{m}(\ell)$, number of effective vertices for a general  graph state $\Ga_m$,  the following relations {always} holds:   $\beth c(m)=\bar{m}(\ell) \beth\bar{\mathbf{C_i}}(m)$ and $\beth\bar{\mathbf{C_i}}(m)=\bar{m}(\ell)-1$--Figs\il(\ref{Fig:INstap}). We need however to establish  {the exact  form of  $\bar{m}(\ell)$ and the transformation for  a conformal expansion,    from  an initial state  $m_0$  and  reduction in  the effective  number of vertices by considering  loops $\ell$.
The \emph{effective number}  $\bar{m}$  of vertices for sequence  $
 \beth \bar{m}=m+1-\min \oslash\ell(m)$  where $\min \oslash\ell(m)=\Theta(m) H[\Theta(m)-2]$,
 (these relations, including the $+1$ term  hold properly considering the algebra of a vertex without loop equal to $\oslash=0$ or vice versa $\oslash=1$, in fact we consider these possibilities differently in this work, this choice depends on how the loop boundary is considered)--\cite{start01}.
These quantities are regulated by the
  $\Theta$-criteria and the  $\Ga$ probability distribution.
It  is thus important    to asses   \bp-events transformation with conformal graph expansion. As a general result,  the
number of \bp-events is minimum in the minimal sequence $\Sa_{\min}$ (a state invariant-conformally expanded in algebra), then there is
\bea\label{Eq:sa-y.assa}
&&
\min \beth  \eth c= r-1\, \forall \, m\quad \mbox{for}\quad \Sa_{\min}, \quad\mbox{and}\\&&\nonumber
\forall m\quad \exists  \Sa_{m}:\; \partial_m \max \beth  \eth c(m)>0,
 \eea
 and the following two cases are possible
\bea
&&\nonumber
\mathbf{(1)}\quad
\max \beth  \eth c(m)=2m(1-p_{\max})<m-1,\\&&\nonumber \partial_m \max \beth  \eth c(m)=2(1-p_{\max})<1\quad\mbox{if}\quad\exists !p_{\max}\geq\frac{1}{2}+\frac{1}{m},
\\
&&\mbox{otherwise}\\
&&\nonumber\mathbf{(2)}\quad
\max \beth  c(m)=m-1\quad\mbox{and}\quad\partial_m \max \beth  c(m)=1
\eea
(total  order of  two connections)--Figs\il(\ref{Fig:INstap}).  The \bp-events number (and density  $\beth c(m)/m$ for  $\Ga_m$)  increases with a graph conformal expansion.
 The extremes in cardinality is an graph property depending on   $r$ and  $\{p_i\}$ (the maximum probability $p_{\max}$).
Regular permutations build the degeneracy  class of the  minimal sequence $\Sa_{\min}$,
this  is not the case for other  sequences.
The supremum occurs  in maximally decomposed sequence, $\Sa_{\max}$,  and  the  \bp-events number is bounded in $\beth \eth c\in[r,\bar{N}-1]\subseteq[r,m-1[$, being  $\bar{N}$ in the effective number of vertices in a sequence (homogenized algebra). In general the degeneracy in chain is \emph{not} invariant under  permutation, accordingly to the $\Theta$-criteria.
 We   could use {$\Theta$} as an index of the peak of the colors probability distribution and  the colors homogeneity.
The situation depends on the {gap} in the probability distribution with a maximum (or the presence of more maxima), for a gap increasing, the number of \bp-events decreases. Loops  number on vertex can vary at  $\max \beth  c(m)=$constant--Figs\il(\ref{Fig:INstap}).
The higher is  the  probability distribution peak (maximum probability gap) and the greater is {$\Theta$} and lower is the  inhomogeneity and chain length (number of events in value frame)  in value, that will reflect on the metrics $\sigma(\cdot,\cdot)$ and $g(\cdot,\cdot)$. 
The lower is the maximum and  the greater is the number of decomposed (lower event algebra) chains with the minimum case of equiprobable distribution where the quantities transform conformally.
The results found here can clearly be generalized for  $ N $ macroevents, addressing this aspect of the graph realization for  sequence in index of macroevent.
However,  given $N$ {macroevents},   the  {total} number of  \bp-events, including the symmetric chains\footnote{Chains  are (colored) oriented realizations.  $\Ca^+$ with an a-priori ordering (orientation) always admits an inverse $\Ca^-$, equal to  $\Ca^+$ but with an opposite orientation of  vertices  ordering.
$(\mathbf{C}^{(\pm)}_n(a)= \mathbf{C}^{+}_n(a)\cup\mathbf{C}^{-}_n(a))$  is  a (not-homogeneous) loop on the
first (last) vertex of $\mathbf{C}^{+}$ ($\mathbf{C}^{-}$), this is  a  "not-abelian" composition, i.e. $(\mathbf{C}^{(\pm)}_n(a)\neq(\mathbf{C}^{(\mp)}_n(a)= \mathbf{C}^{-}_n(a)\cup\mathbf{C}^{+}_n(a))$ that is a loop  on the first (last) vertex of the
$\mathbf{C}^{-}$ ($\mathbf{C}^{+}$).
In a graph, \emph{cycles} or totally symmetric  chains, $\Ca^+=\Ca^-$, are possible (an homogeneous loop  of algebra  $\oslash_i$ is  a  cycle of length $\oslash_i$ endowed with several symmetry properties.
Necessary condition for a chain to be  a cycle  is  the periodicity (according to color, algebra) with a $ 1 $ step.
We say that the (open) chain is totally symmetric if-(necessary condition) there is at most one element, a part the boundaries with multiplicity (macroevent number in adapted frame) $1$, if this is unique it is called the chain center $a_{\circledast}$, or  symmetry pole.
With the exclusion of $a_{\circledast}$, which has multiplicity 1, there is   $\min N_i=2$ for any vertices of the chain, particularly for the chain boundaries.
 Then  $\forall a_i \exists ! a_j: a_j<a_i<a_j$, \emph{microreversibility} or local symmetry with center $a_i$.
According to this scheme if there is a chain then there is also a reverse, except if it is totally symmetric where $\Ca^+=\Ca^-$.
Note homogeneous loops are always reversible. We discussed extensively the influence of chain ordering in the graph metric symmetries, an interesting question is  how, considering a probability distribution,  the number of symmetric and reversible chains, or the micro-reversibility increases for superior states under graph conformal expansion.}} is
\bea
&&\nonumber
 \beth c_{\eth}(\bar{N},N_i)= \bar{N} (\bar{N}-1)- \sum_{i=1}^rN_i(N_i-1)=\bar{N}^2-  \sum_{i=1}^r {N}_i^2
\\&&\nonumber\mbox{and}\quad
\sum_{i=1}^rN_i=\bar{N}=N,\quad \beth \ell_2(\bar{N})=\sum_{i=1}^rN_i(N_i-1),
 \\\label{Eq:65-nn-mm}&&
 \beth \ell(n)=n\left(n\sum_{i=1}^rp_i^2-1\right),\\&&\nonumber \beth c_{\eth}(n)=
\bar{n}^2\left[1-\sum_{i=1}^r p_i^2\right]=\bar{n}^2 \sum_{i=1}^rp_i(1-p_i)\\
&&\nonumber\mbox{and}\quad  \bar{n}=n+1-\min\oslash\ell(n),
 \eea
and particularly we find
\bea
&&
\mbox{for}\quad\Sa_{\min}\quad\beth c^{\min}_{\eth}({N},N_i)= {r} ({r}-1)\quad \mbox{and for}\\&&\nonumber\Sa_{\odot}\quad\beth c^{\odot}_{\eth}(n)=
{n}^2\left[1- 1/r\right],\quad
 \beth \ell_{\odot}(n)=n\left(\frac{n}{r}-1\right),
 \\\label{Eq:2-case-Report}
 &&\mbox{for }\quad  r=2 \quad
 \beth c_{\eth}({N},N_i)=2 N_1 N_2\\&&\nonumber\mbox{and}\quad\beth c_{\eth}(n)=
2 {n}^2 p_1 p_2,
 \eea
(These relations  were related to graph  inertia $\mu$ and radius $R$\cite{start01}). We considered  the  \emph{effective} macroevents ("normalized" by the algebra homogeization considering  the equal color events adjacency).
Then $\beth \ell$ is the  {total} number of possible loops (of the order 2); we take  into account all the possible connections, for the  sequence,  ($N (N-1)$) and  \emph{all} loops are subtracted. Thus, quantity
 $\beth\ell(N)=\sum_{i=1}^rN_i(N_i-1) $ and, respectively, $\beth\ell(n)=n\left(n\sum_{i=1}^rp_i^2-1\right)$ are the supremum  $\max{\ell}(n)$ and $\max{\ell}(N)$, for that sequence. In this counting, however, we have considered  all multiple events related to the same  color as necessarily separated, while as we have seen  not all combinations allow the separations but some adjacencies emerge, then a second level of  rationalization or reduction in effective vertices is needed. %
Connections $\sum_{i=1}^rN_i(N_i-1)$ are the totality of those connecting   equal values (loop), while $N (N-1)$ is the  totality of connections,
 and  there is  $\Theta$
$-\bar{N_i}(\bar{N_i}-1)=-(N_i-1)(N_i-2) H[1-\Theta_i]$.
 That is,
we excluded loops derived from not-necessary connections between  same vertices, however we did not ruled out
   the symmetric $(\pm)$ connections or   connections between identical vertices providing  a distribution of macroevents with multiplicity greater than $1$. %
(Importantly  this implies that these procedures can be applied  also for  inhomogeneous  algebra respect to $\la^0$).
The introduction of a colors algebra, changes also  the metric structure associated to the original not colored \tb{ber} graph, and consequently we can characterize a graph according to the different (dynamic)   metrics $\sigma(\cdot,\cdot)$ and  $g(\cdot,\cdot)$  defined  on the different graph paths (chains).
Below we  introduce  an events matrix, preceding
definition of the graph first level metric based on the vertices composition and decomposition and the \tb{ber}-\tb{bem} relation, through this formalism metrics transformations can be also discussed.

  \textbf{Matrix $\Qa$ of the events}
Before addressing  the discussion on the metric graph we introduce a matrix $\Qa$, this will serve to the presentation of some conceptual features of the graph that have findings in the graph adapted metric structures.
For the   colored graph, the $n\times m$   matrix $\Qa$  
  has   $m$ columns  of    (\tb{ber}) events  in chains,  $\widehat{n}$  being the degree of application of  $\dda$,   $m$ refers to an  adapted index (rows are made of  sequences elements,  
  for a multi-bases graph there is a multiple matrix  $\Qa_{ijk....}$).
Events matrix $\Qa_{ij}$  corresponds  to  an operator matrix   $\hat{\Qa}_{ij}$ of  elements $\dda^i\times\dda^j$, according to the vertex-operator correspondence:
%
\bea\label{Eq:yo-te-ho}
&&
\dda^k \Qa_{ij}\equiv(\dda^k\times 1) \Qa_{ij}\equiv\Qa_{i+k\;j}, \\&&\nonumber 
\Qa_{ji}\equiv{\Qa_{ij}}^*\equiv{(\dda^i\times \dda^j)}^*\Qa_{00}\equiv\\&&\nonumber{(\dda^j\times \dda^i)}\Qa_{00}\equiv(\dda^j\times 1)(1\times \dda^i)\Qa_{00},
 \eea
 %
 There is
$\hat{c}/\hat{r}\in[{0},{1}]$ for the  ratio between column  (shift) and row (evolution) operator  \emph{algebras}, related to  $\dda^{\hat{r}}\times\dda^{\hat{c}}$ on the initial vertex $\Qa_{ij}$ whose indexes are   not constrained (a colored graph vertex  has an algebra greater or equal then the \tb{ber})
\footnote{According to the  constraints on the operators indexes, two matrixes definitions could be given: \textbf{(1)} A matrix \emph{excluding} not-homogenous loops: $0<\hat{r}\geq \hat{c}\geq 0$. (For $\hat{c}=0$    there is an  homogeneous loop). There is then: $\mathbf{(\odot)}$ $0< r_2-r_1=\hat{r}\geq c_2-c_1=\hat{c}\geq0$, where $r_i,c_i$ are indices following the first or second application (according to  chain order) and  $\mathbf{(\otimes)}$: $c_2\geq c_1$ and $r_2-c_2\geq r_1-c_1$  for any starting element.   \textbf{(2)}A matrix \emph{including}  not-homogeneous loops (in some ways with  "memory property"):  $0<r_2-r_1=\hat{r}\geq \hat{c}$ and $\hat{c}\neq c_2-c_1$. There is  $\mathbf{(i)}c_2-c_1=0$, an homogeneous loop; $\mathbf{(ii)}c_2-c_1\lessgtr0$, then if  $c_2\in\lhd\{c_i\}_{r_2}=(c_2\in\{c_i\}_{r_2})$ ($c_2$ belongs to the "past" of indices $c$) it is a loop ($\hat{c}=1)$, or   if $c_2\in\rhd\{c_i\}_{r_2}=(c_2\non{\in}\{c_i\}_{r_2}$ ($c_2$ belongs to the "future" of indices $c$) then $c_2=\hat{c}+\max{\lhd\{c_i\}_{r_2}}$. Algebra relation $\hat{\sigma}(\hat{\tau})$ could be generic according to constraints. 
}.
 With an initial  event $\Qa_{ii}$  on the ``diagonal'',   the upper-triangle  part of the matrix is ``inaccessible `` (paths confinement), and  evolution of $\Qa_{ii}$ element,  and all the graph chains are confined. 
%
%
Chain  $Tr(\Qa)\equiv\gamma_{\delta}$, made of elements on the main diagonal, corresponds to   a maximally decomposed chain $\Ca_{\max}$, and any chain  $\gamma_g$ could be recovered as deformation of  $\gamma_\delta$, locating  the $\gamma_g$  upper boundary in the sense of matrix confinement
\bea&&\label{Eq:tracc}
\gamma_{\delta}=Tr(\Qa)=
\delta_{ij} (\dda^i\times \dda^j)\Qa_{00}\equiv\delta_{ij}\Qa^{ij}\\&&\nonumber \gamma_g=g_{\tilde{i}\tilde{j}} (\dda^{\tilde{i}}\times \dda^{\tilde{j}})\Qa_{00}
\equiv(\delta_{ij}\mathcal{G}^i_{\tilde{i}}\mathcal{G}^j_{\tilde{j}})\Qa^{\tilde{i}\tilde{j}},
\eea
 (a further term can be added to the $\delta$-term to consider \bp-events, here  $\dda^i\times\dda^j:\Qa_{00}\mapsto \Qa_{ij}$),  
 $\delta_{ij}$ is the  Kronecker delta (sum on repeated index is intended composition in chain), while  $g_{ij} $ is a $\delta_{ij}$  deformation and $\mathcal{G}^j_{\tilde{j}}$ are deformation matrices  constrained by the assumption on the algebra degrees $\hat{r}/\hat{c}$ mirrored   in the metric graph approach. 
\begin{figure*}
  \includegraphics[width=5.5cm]{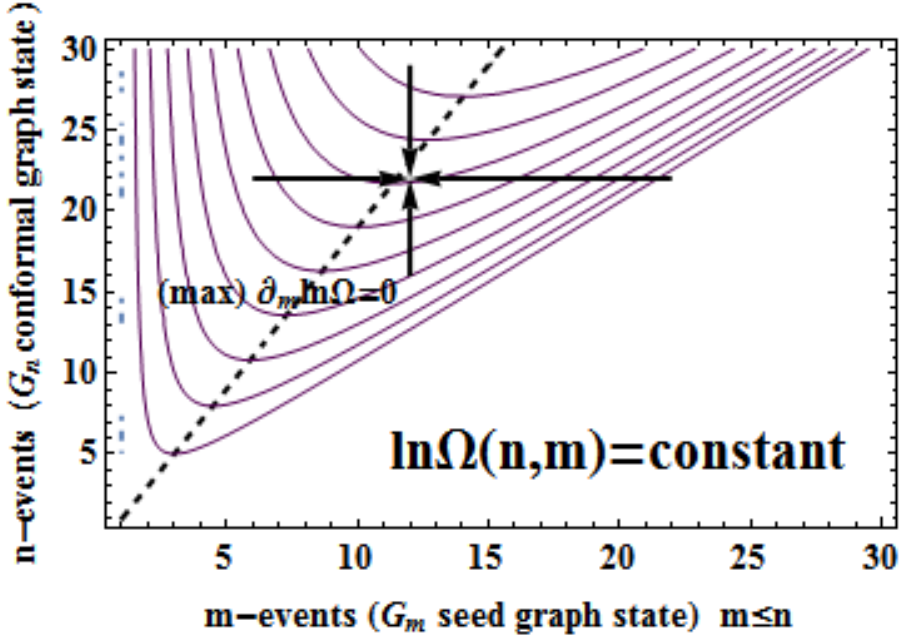}
  \includegraphics[width=5.5cm]{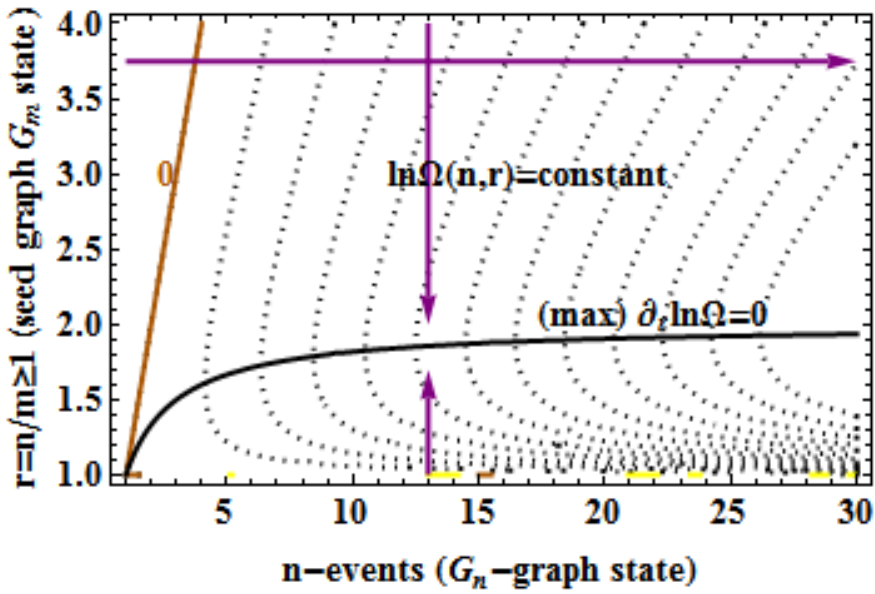}
  \includegraphics[width=5.5cm]{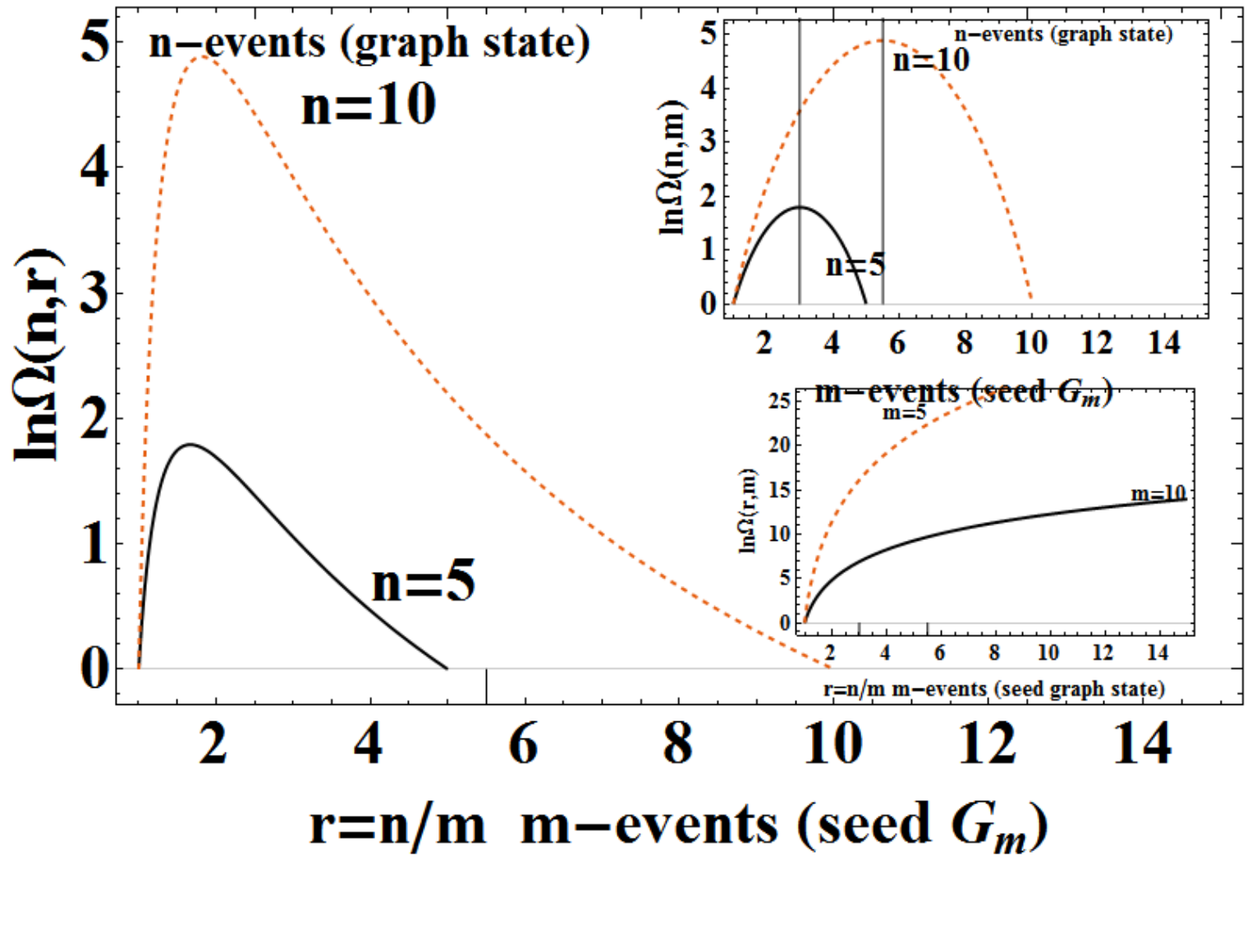}\\
   \includegraphics[width=5.5cm]{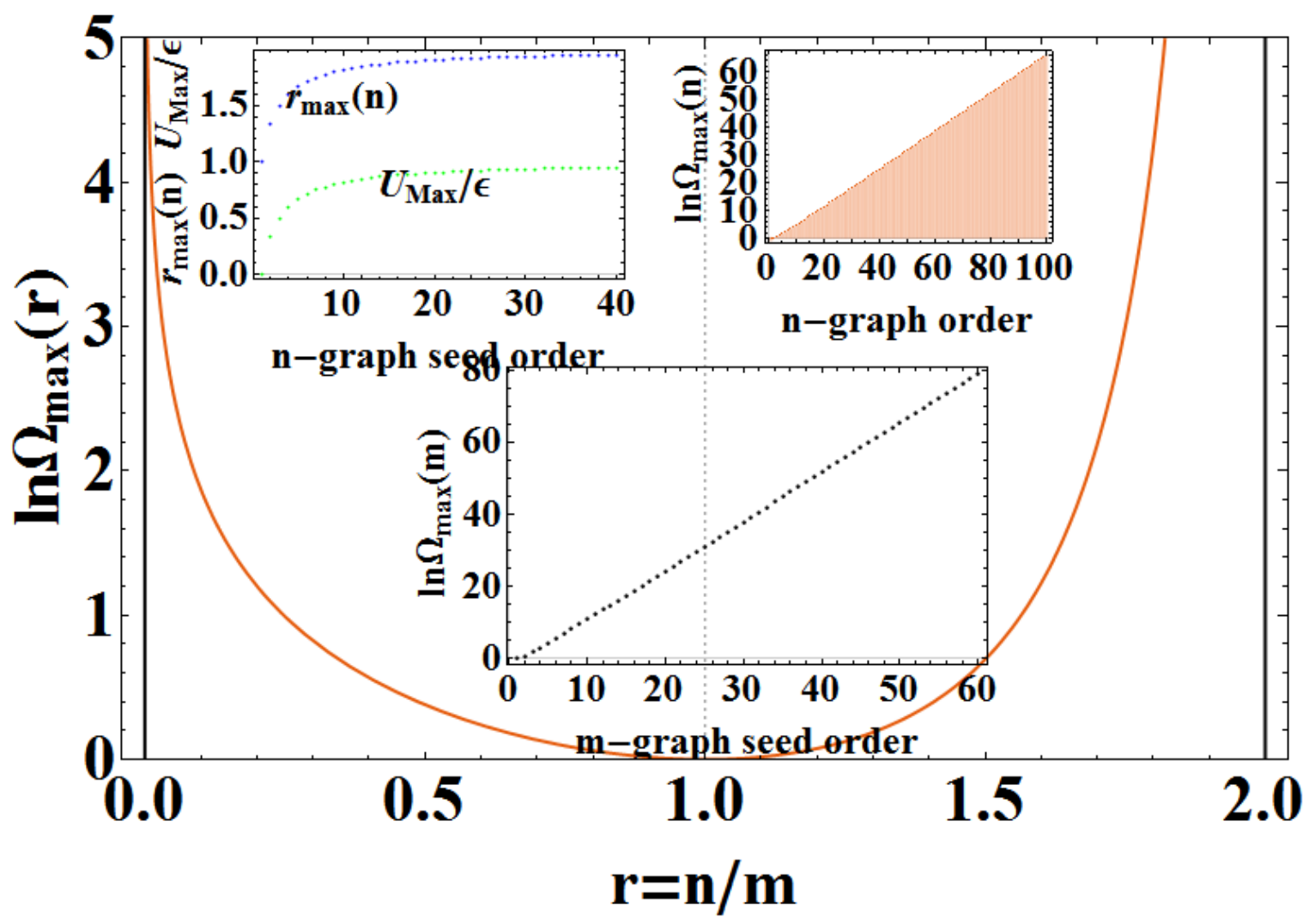}
   \includegraphics[width=5.5cm]{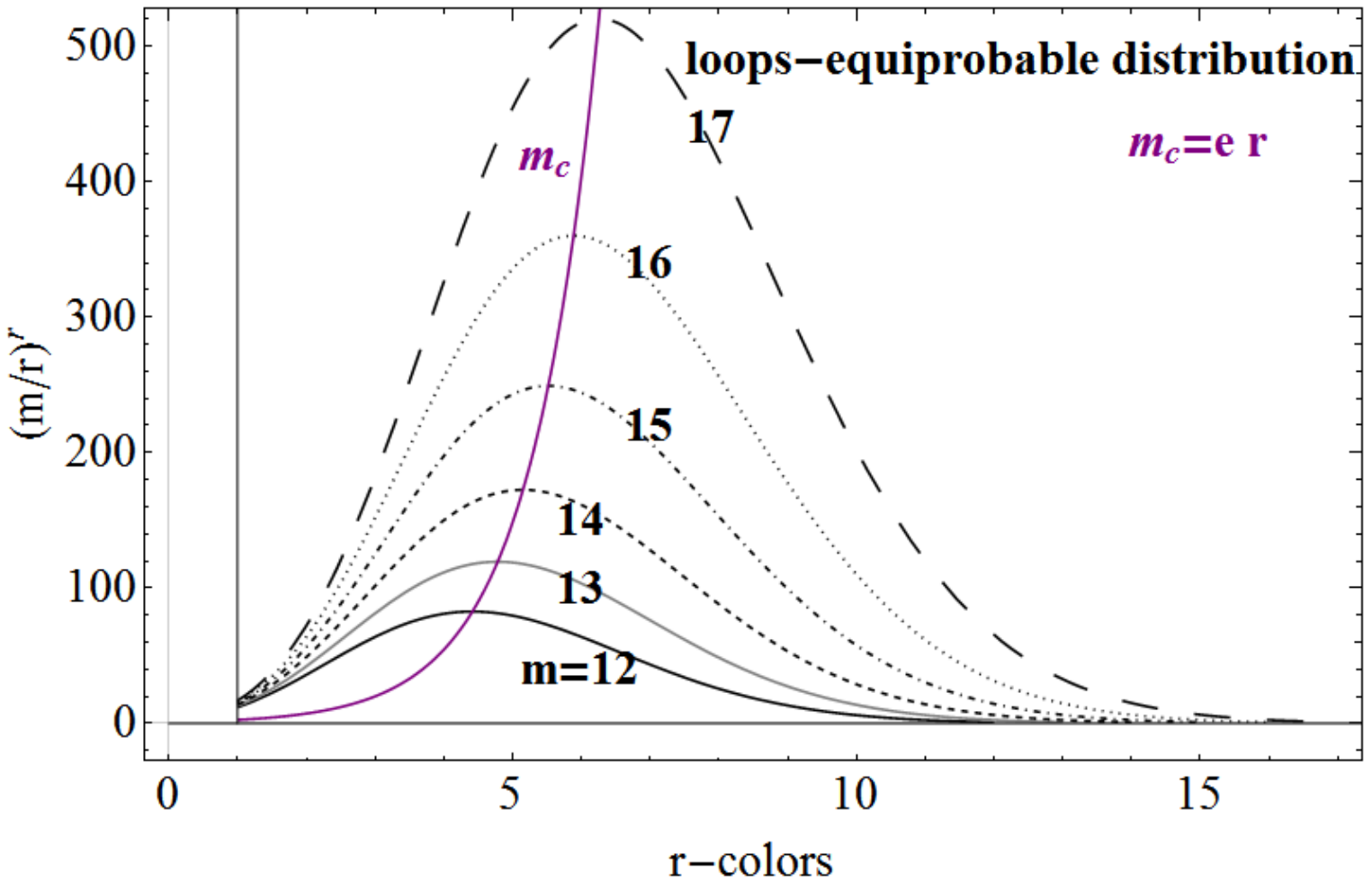}
    \includegraphics[width=5.5cm]{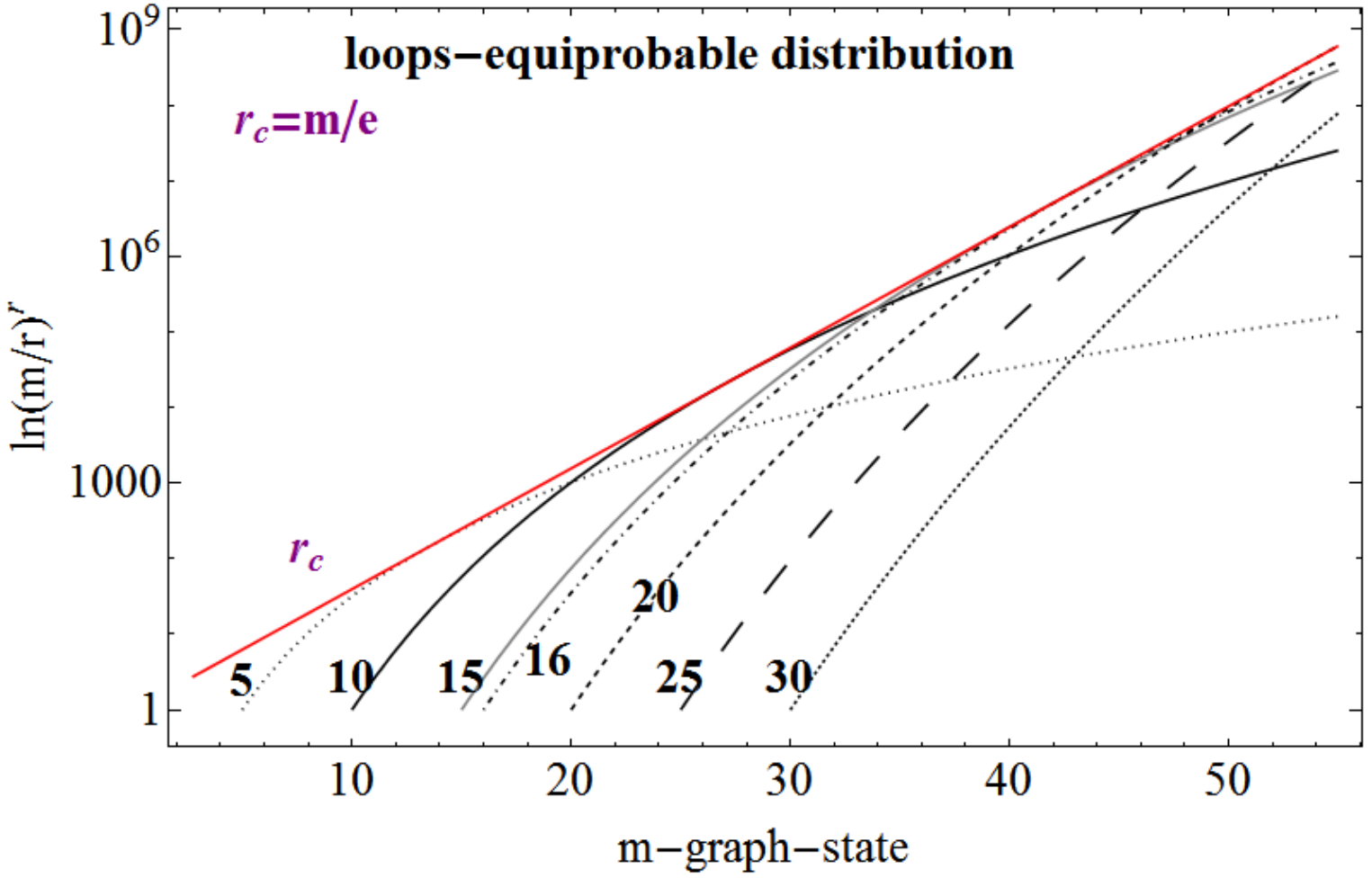}\\
   \includegraphics[width=5.5cm]{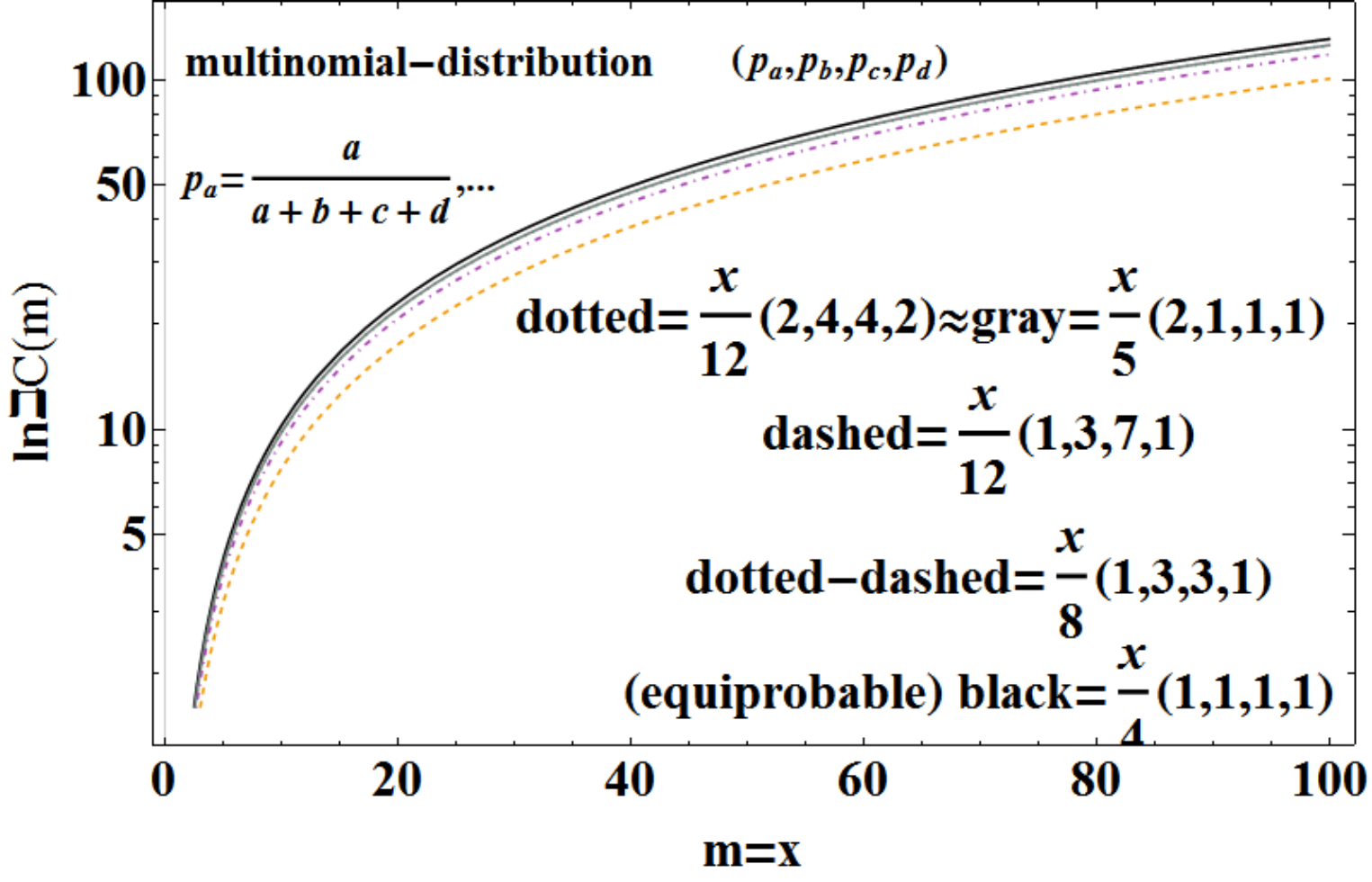}
  \includegraphics[width=4.5cm,angle=90]{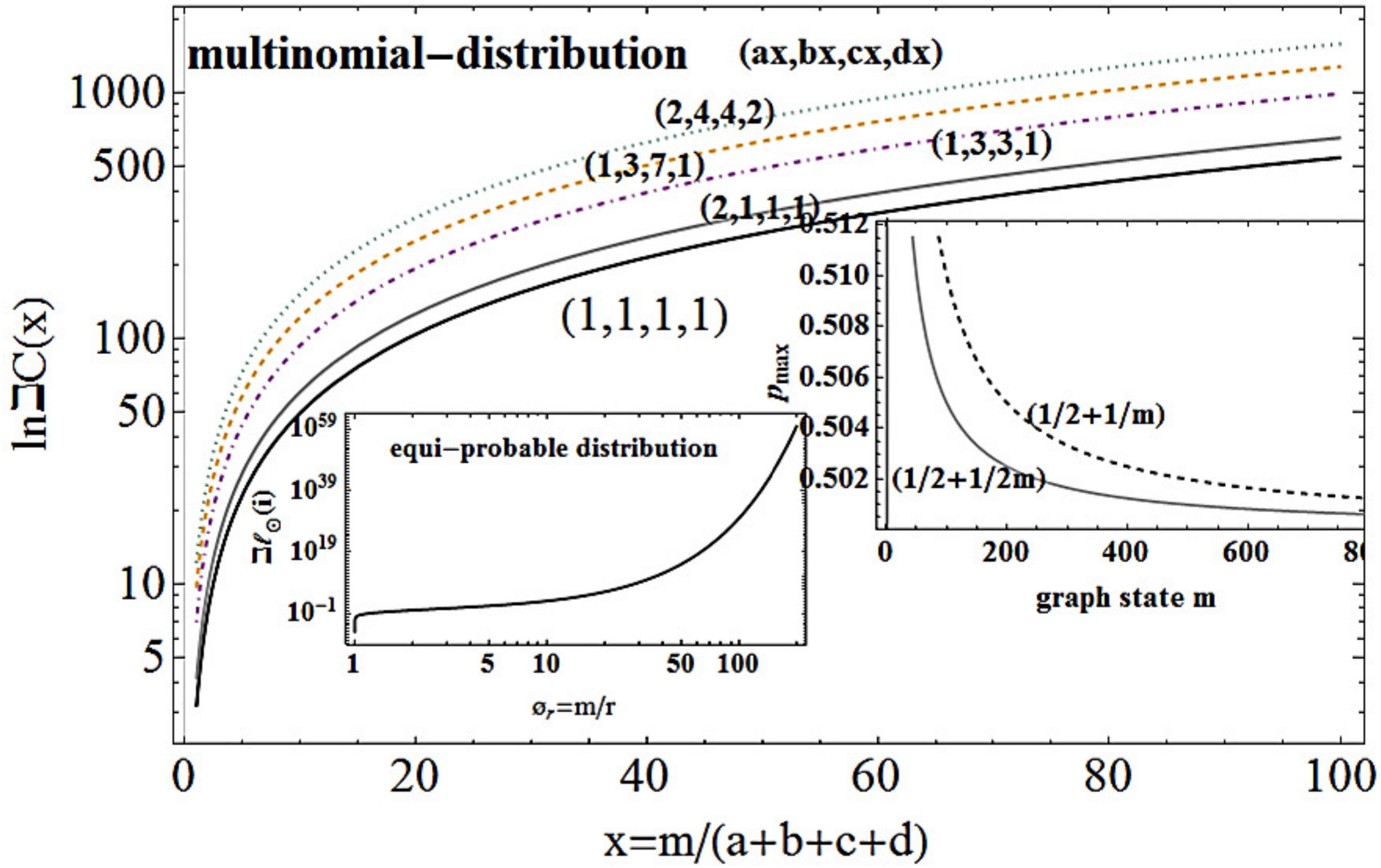}
   \includegraphics[width=5.5cm]{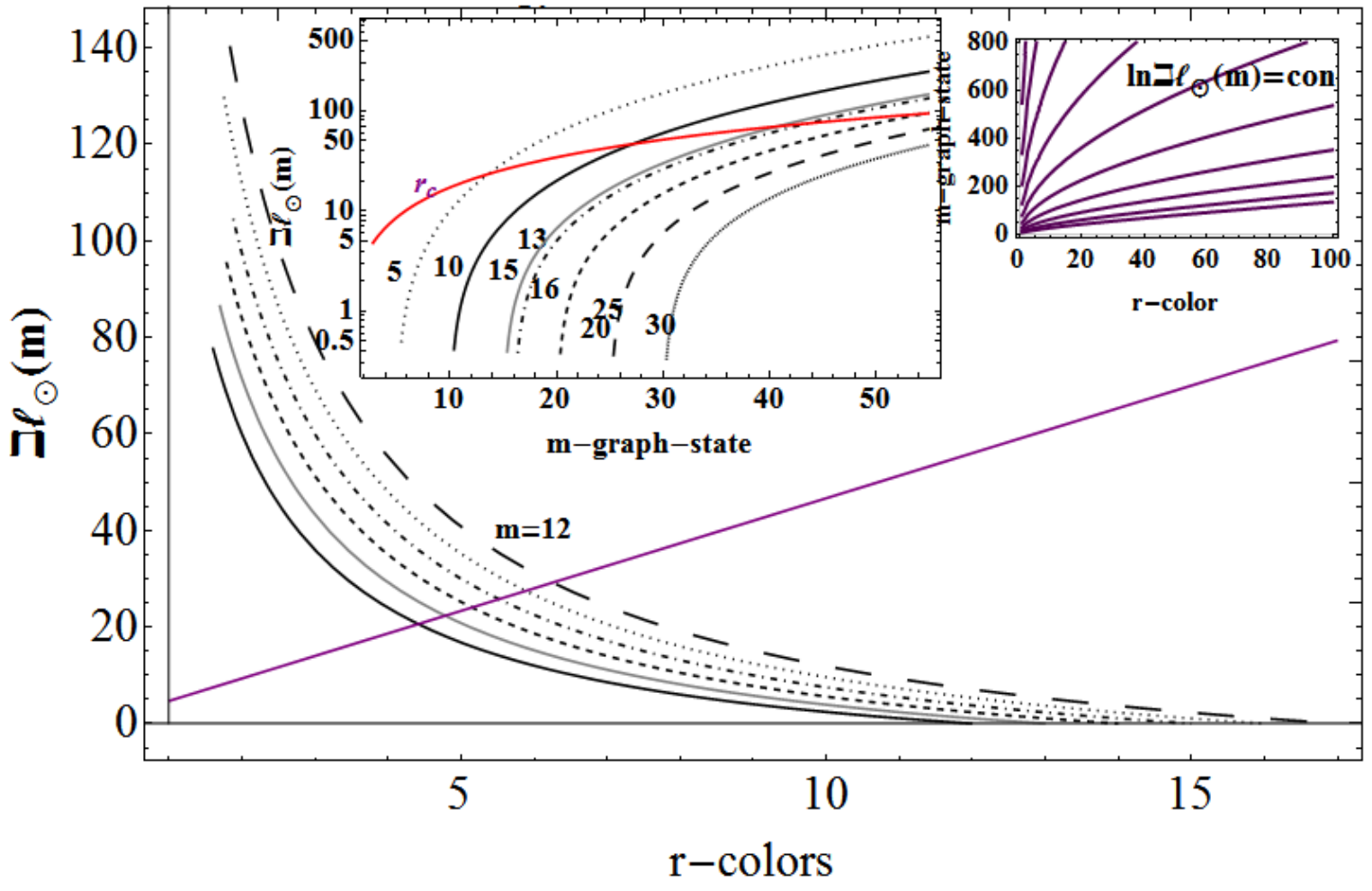}
  \caption{Cardinality $\beth\Qa$ of the set of  $\Qa$-quantities and algebra (order) $\oslash \Qa$  of the element $\Qa$, where $\ell$ is for the  graph loop, $\Ca$ to chains,  $\odot$ indicates  equiprobable distributions, $r$ is the colors number (where not otherwise specified), $(m,n)$ refer to graph states $(\Ga_m,\Ga_n)$, $\Omega(\Qa_1,\Qa_2)$ refers to the distributions considered in Sec.\il(\ref{Sec:fromz}) and particulary Sec.\il(\ref{Sec:deg-new-w-web})  re-parameterized in terms of quantities $(\Qa_1,\Qa_2)$. We highlighted particularly  the presence of maxima and minima of  entropy and cardinality, the peculiarity of equiprobable distributions and, in the case of not-equiprobable graph,  the cases with a single or multiple maximum of probability, the dependence on colors number  $r$ and the change for conformal transformation. The curves relate different graphs at equal states creating classes of different graph with similar properties.
$U/\epsilon$ refers to a system $\ell$ of $\oslash_\ell$
indistinguishable, not-interacting (no \bp-events), particles, let $\epsilon(n)$ denotes the
energy of a particle (a quantity related to the
algebra $(\1 \otimes \1)$ in \tb{ber}). Since particles do not interact,
the total energy  ($U$) is the sum of  single-particle energy.
This  explores the  system of $m$ oscillators describing, in the limit of infinite $m$ a   Klein-Gordon scalar field. The fundamental difference in our case  here is that we do not consider the possibility
that a macroevent could be associated to $0$ event.
The Stirling approximation can be used within the hypothesis $m \gg 1$ and $k = n- m \gg 1$,
implying $r \gg 1$ (large conformal expansion). We note that $S\approx \ln \Omega$ in the Stirling approximation is symmetric for
change of $k$ and $m$ ($U/\epsilon \approx k/m$). Last panel: shows   $\beth\ell_{\odot}(m)$ function of $r$
--\cite{start01}. }\label{Fig:INstap}
\end{figure*}
\section{The metric graph $\Ga_n$}\label{Sec:metri-gr-gra}
In this section we   introduce  the metric graph,  discussing   aspects of the polychromatic graph symmetries, while we will deepen  this feature of the graph model in \cite{start01}. After introducing the double metric structure, we then conclude   in Sec.\il(\ref{Sec:gooUE-dce}) by exploring  the color symmetries in the polychromatic metric graphs.

  A doubled metric structure  is overlapped to the colored graph $\Ga$ realizations with   two chain-adapted  {metric} definitions, $\sigma(\cdot,\cdot)$ and $g(\cdot,\cdot)$, to be defined on $\Ga_n$ chains, and  therefore differently reflecting     chains symmetries and chains  transformations\footnote{Multilevel structures associated with the  higher levels  graphs should be also considered.  More generally the concept of structure has become increasingly important over the last few decades, becoming also one of the fundamental notions of modern mathematics, and particularly  structure in a relational theory is clearly a predominant concept.
In the spacetime (Lorentzian) maniflold the ({one-level}) metric structure overlaps in some extent to the causal structure,   describing the causal relation between points in the manifold, rendered and reflected also  by the metric definition. As described here,
metric structure is the (polychromatic) graph structure reflected and readjusted, in the sense of the role of symmetries, in ${g}$-metric. In $g(\cdot,\cdot)$ the  invariance for colors inversion results in  a gluing of realizations in chains which are  related by certain predetermined transformations.
This approach can be considered therefore a  form of  structuralism where  objects "intrinsic" properties are defined by their (external) relations.
Structure, intended as  information organization, in the case  of the  combinatorial graph we are considering here   involves the way in which a graph  in  partitioned into interrelated components.
An event structure represents in a broader sense, the set of objects  included in $\Ga$ and their relationships, understood as  reciprocal relations of $\Ga_m$ constituent elements, where  different levels and different clusters emerge.
This   structure includes a hierarchy  of relations (conformal levels- monochromatic clusters, or the  polychromatic \bp-events)  where the   higher-level structure  contains  multiple copies of the  lower-level structures (the replicas),
 then in this sense we can speak of events (scale or conformal) structure of the spacetime.
 How this structure relates to the it-from \emph{qubit}-models, is a matter  of how the clusters and distances  are considered. {Cayley graph proposed here could be used for a    complexity
geometry approach, using
bits, qubits, and logical gates. A metric on quantum states  has been provided as inner-product distance where  states orthogonal are at maximum distance. Length may be framed in gates concept.
For  $n$ bit the maximum entropy is $ ln 2^n$
(as for qubit).
For $n$ qubits, a complete basis for the tangent space of SU(2n) can be constructed by $2^{2n }- 1 $ generalized Pauli matrices. 
}}.  A metric transformation connects chains of the same graph state or graph different states,  reflecting the isomorphism and replicas of  all the $\Ga_m$ states of a seed and  conformal expanded graph \cite{start01}.  Algebra homogenization  is also translated into metric transformation.  Metric  $\sigma(\cdot,\cdot)$ depends on and discerns the colors (and eventually   algebras) symmetries  of the polychromatic chains, as defined in Sec.\il(\ref{Sec:fromz})  mirrored in the {metric} {structure} defining metric symmetries. Metric  $g(\cdot,\cdot)$ is invariant for a set of chain transformations,  we define this  metric-level as emergent from $\sigma(\cdot,\cdot)$ level, describing  classes of  graphs realizations  related by specific assumed transformations, in Sec.\il(\ref{Sec:gooUE-dce}) we discuss  this aspect  considering a  simple  frame.
 The $\sigma(\cdot,\cdot)$-metric  and $g(\cdot,\cdot)$-metric can be written in different forms as:
 \bea\label{Eq:spe-r-vec}&&\sigma(\cdot,\cdot)-\textbf{metric}\quad\sigma=h_{AB}\omega^{AB}_x; \quad \sigma=h_{AB}\omega^{AB}_{xy},
  \\&&\nonumber \clubsuit\; \sigma=h_{AB}\omega^{A}_{x}\omega^{B}_{y},
\\\label{Eq:spe-r-vec1}
&&g(\cdot,\cdot)-\textbf{metric}\quad
  g=h_{ABCD}\omega^{ABCD}_{xy}, \quad
 g=h_{ABCD}\omega^{AB}_x\omega^{CD}_y;
    \\&& \nonumber\quad \qquad \qquad \qquad \qquad g=h_{ABCD}\omega^{ABCD}_{xyzt}, \quad \spadesuit\; g=h_{ABCD}\omega^{AB}_{xy}\omega^{CD}_{zt},\\&&\nonumber g=h_{ABCD}\omega^{A}_{x}\omega^{B}_{y}\omega^{C}_{z}\omega^{D}_{t},
  \eea
coincident in special cases  depending on the  adaptation to graph realizations.
  Here we do not include the \bp-events explicitly and
we  concentrate, in the following, mainly on the  $\clubsuit$ and $\spadesuit$  forms. An  "induced metric" level has be also defined
   \bea\label{Eq:fles-milt-par}&&\sigma(\cdot,\cdot)-\textbf{metric}\quad\sigma^{AB}=(h)^x\omega^{AB}_x; \\&&\nonumber\sigma^{AB}=(h)^{xy}\omega^{AB}_{xy},
  \quad \sigma^{AB}=(h)^{xy}\omega^{A}_{x}\omega^{B}_{y},
\\\label{Eq:fles-milt-par1}
&&
  g(\cdot,\cdot)-\textbf{metric}\quad g^{ABCD}=(h)^{xy}\omega^{ABCD}_{xy}, \\&&\nonumber  g^{ABCD}=(h)^{xy}\omega^{AB}_x\omega^{CD}_y;
 \\\nonumber
 &&
\qquad  \qquad\qquad g^{ABCD}=(h)^{xyzt}\omega^{ABCD}_{xyzt},\quad g^{ABCD}=(h)^{xyzt}\omega^{AB}_{xy}\omega^{CD}_{zt},\\&&\nonumber
 g^{ABCD}=(h)^{xyzt}\omega^{A}_{x}\omega^{B}_{y}\omega^{C}_{z}\omega^{D}_{t},
  \eea
(here we do not focus on the induced metrics  but it is clear that an  induced metric structure exists to which a graph can be  associated. In some way, also $g(\cdot,\cdot)$ and $\sigma(\cdot,\cdot)$ metrics  in Eqs\il(\ref{Eq:spe-r-vec}) and
(\ref{Eq:spe-r-vec1})  are composed by properly chosen  induced $(h)$-metrics).
Quantities  ${h^{\daleth}}_{\digamma}$, with  $h^{\daleth}\in \{h,(h)\}$ are  general matrixes  we detail below, $\digamma$ are for indices $\{A, B, C, D, ...\}$ of {\tb{bem}} or colors in   polychromatic  or doubled monochromatic approaches to graph   chain realizations. Indices $\{x, y, z, t, ...\}$  are "inner indices"  ({\tb{ber}}, defined in an "inner space-graph"). These quantities will be discussed in more detail in Sec. (\ref{Sec:gooUE-dce}) with an example which, adopting a fictitious exemplification,  will focus especially  on symmetries.

  \textbf{Quantities  $\omega^A$ and  $\omega^{AB}$}

  Quantity   $\omega^{AB}$   refers to  distance notion  for a couple of colored graph vertexes  (we drop the inner indexes for  convenience while sum on repeated indices  and contractions  are clear from the context). Element $\omega^A$  in the first level graph  $\sigma(\cdot,\cdot)$ considering  mainly loops,  refers to a vertex in colored bases indexes $A$.
We have implicitly considered an homogenized algebra, whereas in a monochromatic loop  internal symmetries act (group of polynomial $D$) which  we will ignore (undistinguishable vertices).
Equivalently, because of the vertex/events definitions we intend $\omega^{A}$ as measure related to  colored indexed vertex,
 or  event index  and  base index,  related to  polychromatic or couples of monochromatic chains with complements, and in first approximation providing the vertex algebra in a given base.
  Decomposition $\omega^{AB}=\omega^A\omega^B$  follows from assumption on the   $\omega^{AB}$ evaluation  and the action of matrix $h^{AB}$.
  Similarly for  $\omega^{ABCD}$, this general  quantity refers to a couple of vertexes (a chain  connection in  some colored algebra). Actually metric $g(\cdot,\cdot)$ couples polychromatic chains related by certain transformations. Decomposition $\omega^{ABCD}=\omega^{AB}\omega^{CD}$ binds together specifical couples, eventually we bind through this assumption $g(\cdot,\cdot)$ to $\sigma$--metric,

 \textbf{On the matrices  $h$  and $(h)$}

Here we consider matrices $h$  and $(h)$ in Eqs\il(\ref{Eq:spe-r-vec},
\ref{Eq:spe-r-vec1}) and (\ref{Eq:fles-milt-par},\ref{Eq:fles-milt-par1}) in $\sigma$- and $g$-metric.  Matrix  $h_{AB}$ can be  been considered  be
   symmetric  or antisymmetric,  we can  include an  antisymmetric part clearly related  to the \bp-events or connections. Using a first approximation,  in Sec.\il(\ref{Sec:gooUE-dce}) we consider "Lorentzian" metric $h_{AB}=\eta_{AB}=\mathrm{diag}\{1,-1\}$. A  second  main assumption concerns the form of  $h^{ABCD}$.  We already partially   addressed  this issue discussing  quantities $\omega^{AB}$.
Considering  the symmetries in $g(\cdot,\cdot)$ metric structure, we  consider the   following two possibilities:
\bea\label{Eq:qua-miel-por}&&
 \mbox{\textbf{$\eta$-form}}:\quad
\mathbf{{g}}=
h_{ABCD}\omega^{ABCD};
\\&&\nonumber \mbox{\textbf{$\sigma\Qa\sigma$-form}}:\quad
{g=\Qa_{\alpha\beta...}\sigma^{\alpha}\sigma^{\beta}...};\\&&\nonumber\maltese\;
g=\sigma \Qa\sigma=h_{AB}\Qa h_{CD}\omega^{AB}\omega^{CD}.
\eea
In  \textbf{$\eta$-form}, $h_{ABCD}$ is a general matrix, in the  \textbf{$\sigma\Qa\sigma$-form}, metric $g(\cdot,\cdot)$  is related to $\sigma(\cdot,\cdot)$ metric, emerging  from a set  $\{\sigma^{\alpha}\}_{\alpha}$  of $\sigma(\cdot,\cdot)$  metrics  through the combination of metrics on which generic transformations $\Qa$ act, adapted to chains of the same  graph state $\Ga_n$  related by permutations.
   These chains  are not distinguished in  $g(\cdot,\cdot)$ metric structure.
   (Symbolic expressions of $\sigma^{\alpha}$  in the second term of  (\ref{Eq:qua-miel-por}) indicate in a generic way these metricized chains.)
   In general $\Qa$, defined on a color base, corresponds to  an
inversion of   the  graph dichromatic connection orientation,   acting here  directly on the matrix $h_{AB}$ indexes (in this case $\Qa$ acts on the  right terms) and
 we mainly consider chain couples and their colors inversion  using the  $\maltese$ form.

\textbf{The inner  indices}

General  inner indices  $(x,y)$ can be  as  $$\{(x,y)\}=\{(a,b),(a_A, b_B),(a_A,a_B)\}.$$  Within these different choices, from Eq.\il(\ref{Eq:spe-r-vec}) and
(\ref{Eq:spe-r-vec1}) we obtain
\bea\label{Eq:cons-parti-zza-gust}
&&\mathbf{\sigma}\mbox{-\textbf{metric}}:\quad \sigma=h_{AB} \omega_{a_Ab_B}^{AB}\equiv\sigma_{a_0b_0a_1b_1};
\\\nonumber
&&\mathbf{\sigma}\mbox{-\textbf{metric}}-\mbox{\textbf{chain-form:}}\quad \sigma=h_{AB} \omega_{a_Aa_B}^{AB}\equiv \sigma_{a_0a_1}=\sigma_{a_0b_0a_1b_1}\delta^{ab}
\\\nonumber
&&\mathbf{\sigma}\mbox{-\textbf{metric}}-\mbox{\textbf{string-form:}}\quad
\bar{\sigma}_{ab}\equiv h_{AB}\omega^{AB}_{ab};\\&&\nonumber\mbox{and if } \quad  \diamondsuit\eta_{AB}=h_{AB},\quad \mbox{then } \quad\bar{\sigma}_{ab}={\sigma}_{a_0b_0}=\sigma_{a_0b_0a_1b_1}\delta^{01}
\eea
(where $(A,B)$ have values in $(0,1)$).  It is worth noting that the particularizations  presented in Eq.\il(\ref{Eq:cons-parti-zza-gust}) refer to assumptions on inner indices, and on the particular inner space. These different cases turn    equivalent   only  under  special assumptions.
 The  indexes  $(a_A,b_B)$  and $(a_A,a_B)$  carry color base indexes, the inner and color spaces are consequently attached to each other.

 $g-$\textbf{metric}

 Options $\{(x,y)\}=\{(a,b),(a_A, b_B),(a_A,a_B)\}$, present  in  $\sigma$-metric   are reflected and   inherited by  second level  $g-$metric, according to Eqs\il(\ref{Eq:spe-r-vec}),
(\ref{Eq:spe-r-vec1})
and Eq.\il (\ref{Eq:qua-miel-por}). We note that there is a clear  formal analogy  between $\sigma(\cdot,\cdot)$ metric of $g(\cdot,\cdot)$ metric in \textbf{$\eta$-form}  of Eqs\il(\ref{Eq:spe-r-vec}),
(\ref{Eq:spe-r-vec1}). Because of this, we can   understand, in this discussion,  properties described for  $\sigma(\cdot,\cdot)$ metric  holding, in form, also  for  $g(\cdot,\cdot)$ in {\textbf{$\eta$-form}}.
In general however, we can specify the indices introducing notation ${g}_{\upsilon \upsilon'}$,  where $\upsilon\equiv (a_0b_0a_1b_1)$, notation $(')$ for the second pair of indices would  fit the \textbf{$\sigma\Qa\sigma$-form} with the $\maltese$ form of Eq.\il(\ref{Eq:qua-miel-por}), where $\upsilon$ and  $\upsilon'$  are related to a  $\Qa$  transformation acting on the capital, (color) indices ($\Qa$  acts on $h_{AB}$ metric). Therefore with  ${g}_{\upsilon \upsilon'}$ we  mean this  special \textbf{$\sigma\Qa\sigma$-form} with the $\maltese$ case as the generic case where the indices are  $\{a_A,b_B,c_C,d_D\}$--i.e $\upsilon'=(c_C,d_D)$.
We consider different cases: a "chain-form" $g_{a_A a'_B}={g}_{a_0a_1a'_0a'_1}={g}_{\upsilon \upsilon'}\delta^a_b\delta^{a'}_{b'}$ corresponding to indexes $(a_A,a_B)$  from  Eq.\il(\ref{Eq:cons-parti-zza-gust}).
With particular assumptions on $h_{AB}$, we obtain,
${g}_{a_0a_1}={g}_{\upsilon \upsilon'}(\delta^a_b\delta^{a'}_{b'})(\delta^0_1)$, reducing  to the string form $g_{abcd}=h_{ABCD}\omega^{AB}_{ab}\omega^{CD}_{cd}$  and to $g_{abcd}=h_{AB}\Qa h_{CD}\omega^{AB}_{ab}\omega^{CD}_{cd}$, depending on transformations acting  \emph{only} on the color capital  indices. We  note that all the variants  can coincide, depending on assumptions on the  inner space of small indexes.

 \textbf{Inner indexes   and the inner  space definition}

 We can introduce a general quantity
$\chi_A^{\varrho}=(a_A,b_A)$ or $\chi^{\varrho}=(\mathbf{a},\mathbf{b})$, where
  $\mathbf{a}=a_A \hat{e}^A$ and  similarly for $\mathbf{b}$ ($\hat{e}^A$ refers to the  inner space-color space attribution). Then, the index choice (\ref{Eq:cons-parti-zza-gust}) can be reduced   through an appropriate choice of inner  indexes, relating $\mathbf{a}$ to $\mathbf{b}$ indexes, and transformations on the  colors ones,  through assumptions on  ${h^{\daleth}}_{\digamma}$ matrices and  $\Qa$ transformations. Mainly,  we  consider adapted bases in $\mathrm{(1+1)}$-models, where $(a,b)\in \{\sigma,\tau\}$ and  the  capital color indexes vary in $(0,1)$.
  Schematically, we set the indices in accordance with
$(x,y)\equiv\chi=(a,b)$, string form, where  in general  $\chi\in \Sa_a\times\Sa_b$
 ($\Sa_x$ being  index $x$ definition  values space), and
  $(x,y)=\chi=(a_A,b_B)$ for the general  case, where  $a_A\in \Sa_{a_A}$  and $b_B\in \Sa_{b_B}$, and finally in the   chain-form    with $(x,y)=a_\upsilon$ where $\upsilon=(A,B)$.
In Eqs\il(\ref{Eq:cons-parti-zza-gust}) the inner indexes could be independent from the colors indexes, as in the  string-choice,  or  also  related to the color indexes. (These  different cases  overlap and  coincide in special cases. In fact the indices $a_0,b_0,a_1,b_1$ are not fixed, in other words   the indexes  $a_A$ can be as  $a_0\in \Sa_{a_0}$ which may  coincide with one of the others, depending on the specific model and inner space).

  \textbf{Inner indices   in $\omega$ quantities}

The inner indexes  appear   in  $\omega^A$, $\omega^{AB}$ and  $\omega^{ABCD}$ quantities  in the different forms adopted in  Eqs\il(\ref{Eq:spe-r-vec}) and
(\ref{Eq:spe-r-vec1}).
We explicit the role of the  small, inner indices in quantities $\omega^A_x$ and  $\omega^{AB}_{xy}$, by  introducing new quantities  associated to $\omega$s, and associating the inner indexes to the action of a derivative new  operator   (conceptually related to $\dda$ operator), acting on these new quantities and  having different interpretations according if acting  on the quantity  related to $\omega^{A}$  (for a vertex) or differently to  $\omega^{AB}$ (for a connection).  More specifically,  there are
 $\omega^{AB}_{x}$ and  $\omega^{AB}_{xy}$ as in  Eqs\il(\ref{Eq:spe-r-vec}) and
(\ref{Eq:spe-r-vec1}), which coincide depending on  derivative definition.
 Introducing a quantity $X^A$, with an abuse of notation, we could write,   $\partial_a X^A\equiv X_{,a}^A\equiv \omega^A_x\equiv\omega^{A}_{,x}$, and analogously
 $\partial_{xy}X^{AB}\equiv \partial_{x}X^{A} \partial_{y}X^{B}\equiv\omega^{AB}_{xy}$ (construction of $\partial_{x}X^{A} \partial_{y}X^{B}\equiv\omega^{AB}_{xy}$ implies the specification of the action of the derivative  in $\omega^{AB}_{xy}$). These definitions are inherited in  $\omega^{ABCD}$ from   $\sigma$-metric as in Eq.\il (\ref{Eq:qua-miel-por}). Quantities $\omega$ should reflect the colored vertex order relation    on the colors indexes (therefore related to assumptions on $h$ matrix)   and on  the inner indices implying assumption on the inner space.  We mainly consider $\omega^{AB}_{xy}$ with the assumption that the derivative acts by taking differently into account  the indices order, acting on the first or second vertex of the chain polychromatic connection. This implies that this definition reflects  the chromatic order relations of the graph chain connection   and thus  considers the chromatic symmetries of the chains. This construction clearly refers to the situation where we take into account the inner indexes reflecting the \bp-events structures. (In this section we use an approximation where we neglect the \bp-events and  the metrics reduce to  algebras relations. For example, concluding this section, we discuss a simple doubled metric model  using  the $\Qa$-transformations  as  color order inversion to overcome these issues, evaluating  loops, the  monochromatic, symmetrical connections. 

 However, to explicitly consider the  higher  terms derived from the  polychromatic vertices, as raised from  first level \bp-event and the
  (colors) symmetries describing \bp-events,   we can define a quantity $\partial_x\mapsto(\partial_x+ \blacktriangle)$, considering connections  as new vertices, with the introduction of $\blacktriangle$ term we specify below).
%
 Therefore, considering  $\sigma(\cdot,\cdot)$ and $g(\cdot,\cdot)$, for the $\sigma(\cdot,\cdot)$ metric we could write  $\sigma(\cdot,\cdot)=
 h_{AB}\omega_{x}^{AB}$ or   $\sigma(\cdot,\cdot)=
h_{AB} X_{,x}^A X_{,y}^B\equiv h_{AB}\omega_{xy}^{AB}$. For the $g(\cdot,\cdot)$ metric structure, considering the different forms,  we can write
\bea
&& g_{a_A b_B}\equiv h_{ABCD} \omega^{AB}_{a_Ab_B}  \omega^{CD}_{c_C d_D} \equiv h_{ABCD}{\omega_{a_A}^A}{\omega_{b_B}^B}
{\omega_{c_C}^C}{\omega_{d_D}^D}\\&&\nonumber\equiv{g}_{a_0a_1b_0b_1}=
\hat{h}_{a_Ab_Bc_Cd_D} \omega^{AB} \omega^{CD};
\\\nonumber
&& g=h_{AC} \omega^{AA}_{a_A b_A}
 \omega^{CC}_{c_C d_C}= g_{a_A b_B}\left(\delta^{AB}\delta^{CD}\right);
  \\&&\nonumber g=h_{AC} \omega^{AA}_{a_A}
 \omega^{CC}_{c_C}= {g}_{a_0a_1b_0b_1}\left(\delta^{AB}\delta^{CD}\right)\left(\delta^{ab}\delta^{cd}\right)
 \\\label{Eq:av-fat-perc}
 &&\\
 &&\nonumber\mbox{\textbf{string-form-}} \bar{g}=h_{AC} \omega^{A}_{a}
 \omega^{C}_{c}={g}_{a_0a_1b_0b_1}\left(\delta^{AB}\delta^{CD}\right)
 \left(\delta^{ab}\delta^{cd}\right)\delta^{01}
 \\\nonumber
 &&\mbox{\textbf{chain-form-}} {g}_{a_0a_1b_0b_1}\left(\delta^{AB}\delta^{CD}\right)
 \left(\delta^{ab}\delta^{cd}\right)\delta^{ac}=
\mathbf{g}=h_{AC} \omega^{A}_{a_A}
 \omega^{C}_{a_C}\\&&\label{Eq:cgain-milt-on}(\mbox{where}\quad
 \bar{g}\delta^{ac}=\mathbf{g}\delta^{01})
\eea
(we used the metric  operator  $\hat{h}_{xy}\equiv \eta_{AB}\partial_{x}\otimes\partial_{y},$  and grouped  the colors indexes  $ \omega^{AA}\rightarrow \omega^{A}$,
Several assumptions  on color indexes in fact are equivalent to assumptions in the symmetrical part of the matrix and quantities  $\omega$.
 Analogously,  for the induced,  $g$-metric, considering Eqs\il(\ref{Eq:fles-milt-par1}) , for example we find
$g^{ABCD}=(h)^{{xyzt}}\partial_xX^A\partial_yX^B\partial_zX^C\partial_tX^D=(h)^{{xyzt}}
\omega_{xy}^{AB}\omega_{zt}^{CD}$ where, eventually,  $(h)_{xyzt}=(h)_{xy}\Qa (h)_{zt}$, using the  $\Qa$-form  for an induced metric.
%

\textbf{Notes on $\sigma(\cdot,\cdot)$ and $g(\cdot,\cdot)$metrics}

Graphs  chains metric notion, as in Eqs\il(\ref{Eq:av-fat-perc}) and Eqs\il(\ref{Eq:cons-parti-zza-gust}), shows similarities in form as strings generalizations. Nevertheless $\sigma(\cdot,\cdot)$ and $g(\cdot,\cdot)$ metrics result to be conceptually very different.
Metric  $\mathbf{\sigma}(\cdot,\cdot)$ can be considered a natural graph metric adapted to the graph coloring for vertices  defined on a graph ordered realization.
Metric $\mathbf{\sigma}(\cdot,\cdot)$, acting directly on vertices,  compares different events, mixing the event and basis indexes. On the other hand, ${g}$-metric   can be constructed using   $\sigma$-metrics--see Eqs\il(\ref{Eq:qua-miel-por}). Thus   $\sigma(\cdot,\cdot)$  metric  as in Eq.\il(\ref{Eq:spe-r-vec}) constructs  vertex-"distances" with elements with different color indexes. %
%
More specifically,
consider   $X^A=\{x, y\}$ with  $\omega^{00}\equiv \dd x\equiv x\otimes x$ for a base $x$, and  notation $\otimes $  denoting composition in metric, then  metric $g(\cdot,\cdot)$, in form Eq.\il(\ref{Eq:qua-miel-por})-$\maltese$,  could include term  $\dd x \dd y$ 
 in contrast with   $\sigma(\cdot,\cdot)$ where elements    $\{{x}\otimes{y},\dd x\equiv x \otimes x,\dd y\equiv y \otimes y\}$ can be considered. The term $x\otimes y$ is not defined in  $g(\cdot,\cdot)$ of last Eq.\il(\ref{Eq:qua-miel-por})-$\maltese$,  defined by connections and not  vertices. Note that    if   $h$  is a diagonal matrix, then  ${\sigma}(\cdot,\cdot)$  has no  $x\otimes y$ term.
 Metric $g(\cdot,\cdot)$, compares distances in events of different algebras and  event basis relating ${g}$-metric transformations to   events transformations.
%
%
Metrics $\sigma(\cdot,\cdot)$  and $g(\cdot,\cdot)$  differ also in the role of symmetries  for directed colored  connections.

\textbf{Transformations and symmetries}

 We can consider transformations  on inner indexes and transformations on colored indexes. These different  transformations can be  related in some cases.
 \cite{start01}.
We can  also  introduce  quantities $\mathcal{L}_\sigma$ and  $\mathcal{L}_g$,  for $\sigma$ and $ g$ metrics, respectively, from contraction of
Eqs\il(\ref{Eq:spe-r-vec})-(\ref{Eq:spe-r-vec1}) with a matrix $h_{\natural}$  or $(h)_{\flat}$, where
$\natural\in\{A,B,C,...\}$ are colored indices, and  $\flat=\{x,y,z,...\}$ are inner indices, assuming different forms depending on the model chosen  for the inner indices  and  on the matrices symmetries. (Here we  can take advantage of  the  formal analogy   between $\sigma(\cdot,\cdot)$ metric of Eqs\il(\ref{Eq:spe-r-vec}),
(\ref{Eq:spe-r-vec1}),  and $g(\cdot,\cdot)$ metric in \textbf{$\eta$-form}. We can adopt a general matrix for the contraction as in $\mathcal{L}\equiv g^{ABCD}h_{ABCD}\equiv g^{ABCD}h_{AB}\Qa h_{CD}$, and similarly for  other  cases as $
\mathcal{L}_{g}=h_{AB}\omega_{ab}^{AA}\omega_{cd}^{BB}(h)^{ab}(h)^{cd}$, and transformations on these indexes can be represented as graph  vertex transformations  as in $\mathcal{L}=h_{AB}\Lambda^A_C\Lambda^{B}_D X_{,c}^CX^D_{,d}\lambda_{a}^c\lambda_{b}^d (h)^{ab}$, mixing inner and colored indexes.}

More generally we could in fact construct     a  "Nambu--Goto"-like action of the form
 $
 S_{NG}\propto\int (\sqrt{-\det (h)_{xy}} ) \dd \Sigma$, or a  "Polyakov-like" action  form $S_P\propto \int \left(\sqrt{-\det (h)_{xy}} \mathcal{L}-
(p-1)\right) \dd\Sigma\), as  \(\left(\mathcal{L}\equiv (h)^{{xy}} \eta_{AB}\omega_{,{x}}^{A} \omega_{,{y}}^B\right)
$
 (where $p$ is a term due to the dimensionality (1+p) of "worldsheet area"--the inner space--in the "covariant Polyakov" action which we will here  not explore and  we  can consider $p=1$. However here we do not  specify the  "worldsheet invariant area"  $\sqrt{-\det (h)_{xy}}$ and the proportional factor  \footnote{Note that string  loops   in space  are due to closed bosonic string, considering  {closed string} theory compactified on a circle of radius R:
$X \approx X + 2\pi R$
implying  momentum is quantized (string can wrap the circle:
$X(\sigma + 2\pi) = X(\sigma) + 2\pi R w, w \in Z$).
 This (topological) notion of open or closed string must be translated into the definition of loop (closed chain) and connection in graph representations. A further issue  consists in the  conditions imposed for example for open strings
(Lorentz invariance-  leading to
Neumann or Dirichlet conditions).
For closed string  we can parameterize the  equation of motion ($X^A$, $h_{ab}$):
using $\xi^{\pm}=(\pm)\sigma\pm\tau$ obtaining $\partial_+\partial_- X^A=0$, $\partial_a (\sqrt{-\det (h)}(h)^{ab}\partial_b X_{A})=0$  with a solution (with "chirality")
$ X^A=X^A_L(\sigma+\tau)+ X^A_R(\sigma-\tau)$.
}.

Concluding we note that
 graph $\Ga_n$ metric structures is adapted to the each directed path $\Ca_m$, which can be considered {closed or open},  and grouped chains bundles of Figs\il(\ref{Fig:succ}), and then considered after conformal transformations.
The  $\Ga_n$ cluster substructure plays an important role  in $\sigma$-metric, in  the "inner" structure of  $X^{AB}=X^AX^B$, being  loop or connection where color  inhomogeneity has to be considered through \bp-events definitions, therefore as transformation on vertex index.
{(Vertices of the first level graph are  monochromatic, in different  levels  there is an articulated structure due to the  multiple chromaticity.)}
Aspects of this graph model, as the   loops- and connections sets after  paths and states transitions, and  the  generations of different \bp-events levels (graphs),  approach some conceptual and formal aspects of supersymmetric models--\cite{start01}.
We can consider additional (anticommuting) terms in the $\omega$ quantities  related to homogeneous loops  (with an internal symmetry linked to $\mathbf{D}_{\oslash}$ group)  and  the  antisymmetric connections.  Conveniently, specifying the $\blacktriangle$ object introduced above, we could use $\partial_a X^A\mapsto\left(\delta^A_C\partial_a+\Theta_{aC}^A\right)X^C\equiv\Gamma_{aC}^{\hat{w}\phantom\ A}X^C
$ 
 enclosing  coupling terms, and $\hat{w}$ indicates the  \bp-events level
(any higher levels $\widehat{(w+1)}$ are related to $\hat{w}$ one), related to the vertices of a sequence of derived graphs\footnote{Thus, there is
 $\Gamma_{aC}^{\hat{w}\phantom\ A}\Gamma_{bD}^{\hat{k}\phantom\ B}X^CX^D=\partial_aX^A\partial_bX^B+
\Theta_{aC}^{\hat{w}\phantom\ A}\Theta_{bD}^{\hat{k}\phantom\ B}X^CX^D +[\partial_aX^A.\Theta_{bD}^{\hat{k}\phantom\ B}X^D+\Theta_{aC}^{\hat{w}\phantom\ A}X^C
\partial_bX^B]=\sum_{w=0}^{w_{\max}(m)}\hat{w}\partial_{a_w}X^{A_w}
\partial_{b_w}X^{B_w}+[\emph{coupling}]$.
Note there are no terms $\partial_{a_w}X^{A_w}
\partial_{b_{w'}}X^{B_{w'}}$ in the sum ($w'\neq w$).  In  strings theory models   we could use light-like gauge for the inner indices, then
$\mathcal{L}=\eta_{MN}\partial_- X^{m}\partial_+X^N+\{
(d_{1\alpha})\partial_+\theta^{1\alpha}+
(d_{2\alpha})\partial_-\theta^{2\alpha}\}$   containing additional  terms. Eventually   using 
$\Pi^M_{\alpha}=\partial_a X^M+\Theta_a^M$ with  $\mathcal{L}\approx  h^{ab}\eta_{MN}\Pi^M_{a}\Pi_{b}^N + L_{WZ}$, where $L_{WZ}$ a typical  Wess-Zumino-Novikov-Witten like term (with $\Theta_a^M=-i\bar{\theta}^A\Gamma^M\partial_a\theta^A$).}.
We stress here that   in supersymmetric models there are as many (physical) fermions as  (physical) bosons (in  versions  for each scalar $X$ there is a
 Majorana spinor--for example each of these fermions a
two-component Majorana, {real spinors}).  Considering the antisymmetric terms we could consider then   $S=\mathcal{L} +(h)^{ab}\partial_a X^M \partial_b X^N \Phi_{MN}$
(in string-like formulation  $\Phi_{MN}$ would be  formally an  antisymmetric spin 2 tensor
similarly to a Neveu--Schwarz
"B"-field).

  \subsection{Exploring symmetries in graph metric structures}\label{Sec:gooUE-dce}
Metric structures $\sigma(\cdot,\cdot)$ and $g(\cdot,\cdot)$  reflect differently the chromatic (and in some cases algebraic) symmetries  of the graph chain realizations discussed  in  Sec.\il(\ref{Sec:fromz}).  Color  symmetries  interest colored  graph vertexes   and polychromatic clusters,  and they are  mirrored in the graph metrics  playing, eventually, a role in the  dynamical graph.
 Metric structure $g(\cdot,\cdot)$   does not distinguish chains related by   a set of  $\Qa$ transformations as in Eqs\il(\ref{Eq:qua-miel-por}). In the  example explored in this section,    we consider $\Qa$   as chromatic inversion in the ordered graph.
On the other hand, it is clear that generalized permutations reflect   on the metric as the conformal graph expansion  (following  conformal metric transformations\footnote{The issue of  a metric structure definition for the   different scales  events graph  crosses the problem of   a topological space definition
 (which  also underlines  a differential structure).
To some extents we can say that each metric space is a  topological space, and  we could by far generalize saying that  every metric space is a Hausdorff  (T2) space, implying a strong focus  on  the  separation axioms, here  somehow  addressed   introducing  colored vertices   clusters and  clusters bundles. (Broadening the discussion, in this event definition (more close to Parmenide's frame)   leading to somehow degenerate metrics , the identity of indiscernibles is not guaranteed in an immediate way).
However,  metric graph is well studied concept also on latex-graph, and various notions of metric in the quantum setups are considered}).
To discuss some  aspects on  algebra and colors chains symmetries  in metric structures
we adopt here a simplified model for the graph metrics considering only the loops. In this way we also point out some  relations between the  $\sigma(\cdot,\cdot)$ and $g(\cdot,\cdot)$ metric levels. To simplify our discussion we  have chosen  the  chain-form introduced in Eq.\il(\ref{Eq:cons-parti-zza-gust}) and Eq.\il(\ref{Eq:cgain-milt-on}), considering $h_{AB}\equiv \eta_{AB}=\mathrm{diag}\{1,-1\}$.
In first approximation the metric could  be expressed into  colored vertex algebras relations, by introducing    a graph vertices metric     and  vertex norm, with   a (natural) metric element  
 adapted to the graph coloring and   related with the  vertex algebra $\oslash$--\cite{start01}.  A  minimal (not-null) norm  is assumed associated to  the   ordinary events $\ket{e_0}\in \la^{0}$ (and  "minimal"  distance   $\dd e^{01}$ between two adjacent  vertices  of the  minimum $\la^0$ \tb{ber}).  A   \tb{ber} can be of local  (related to a embedded  sub-graph concept)  $\la^0$ minimum algebra 
\footnote{A colored $\Ga$ graph  has a  structure composed by loops and \bp-events, from these a new \bp-$\Ga$ level is generated with  a related  derived  metric structure. As a first approximation and to fix the ideas, the actual distance notion   for    connection  is not specified. In complexity models   distance definition is differently provided and widely debated. This  depends  from  the emerging spinorial structure. In this sense $(\sigma(\cdot,\cdot),g(\cdot,\cdot))$
 are to be  considered  approximations of  more complex metrics including
greater levels  \bp-$\Ga$.}--\cite{start01}.   Conveniently, in this simplified framework    the sequences and chains   homogeneity conditions  can be   reduced  to a metric condition, through an analysis of algebra relations where   metrics  provide  a definition  of  and a criterion to establish the  algebra homogeneity. (A chain can be inhomogeneous   respect  to  the \tb{ber} assumed  homogenous, in this case the metric  could be "evolved" considering  action of evolution $\dda$ on its elements\cite{start01}).  
Exploiting,  with $\Qa$ applications, the colors symmetries for the  directed connections, we consider on  similar levels in the metrics,      connections $(c_{ij},c_{ji}=-c_{ij})$ and loops $(c_{ii}, c_{jj})$ (color index $i,j$). 
 We  discuss these aspects in  details for the combinatorial graph in \cite{start01} with a graph $\Ga$ characterized by
   a base of  $r$ colors, having therefore a total number of $r^2$ distinct kinds of connections ($r$ loops plus $r(r-1)$ connections)  and one \tb{ber}.
In these approximations  we discuss the   $\sigma(\cdot,\cdot)$ and $g(\cdot,\cdot)$  transformations. Using  explicitly the notion of algebra  $\oslash_{\ell}$ of a  loop $\ell$ on a graph $\Ga$ vertex, we compare graphs algebras and  structures.
We consider here a  $1+1$ dimensional framework 
  a \tb{ber}-\tb{bem} couple, including  an  emerging  ordinary algebra associated to an emerging base: by considering  metrics transformations  a new   \tb{ber} is  related to a derived colored base.
  Two main approaches can be  explored, depending  to the decomposition base,  the algebra relations and the polychromatic order.
We start by considering $\sigma(\cdot,\cdot)$ in chain form with $h_{AB}=\eta_{AB}$ and using the simplified  notation $\omega^{xx}\equiv\dd x$  for the  base   $x$.
We explored,  for the  directed graphs,    the  quantities
$\dd x_{mn}(\mathcal{o}_\vartheta)=\pm\oslash_x^{\epsilon_{\vartheta}}\dd t_{mn}(\mathcal{o}_T)$  for the elements  of $\xi\equiv\mathbf{\natural}(\mathbf{\sharp}\dd x_{mn},\dd t_{mn})$ with ($(\mathbf{\natural},\mathbf{\sharp})=\pm$),   where $  \epsilon_{\vartheta}=\pm1$  and   $(
\vartheta=\{x,t\})$, for $(\mathbf{\theta},-\theta)$  respectively,  being $\mathbf{\theta}\equiv (c_{mn},\pm c_{mn})
$, notation $\oslash_L$ is for  the ordinary algebra $\la^0$. We mainly adopt inner bases   with $\vartheta=t$ and  $\epsilon_{\vartheta}=-1$.  A  relevant quantity within this   choice of  signs is $\Gamma\equiv \epsilon_1(\oslash_x^{-a}-\epsilon_2{\oslash_L^{-a}})$, where $(\epsilon_{1},\epsilon_2)=\pm1$ and $a=1(2)$ for  $\sigma(\cdot,\cdot) (g(\cdot,\cdot))$,
expressing  the graph inhomogeneity  and  symmetries  of $\sigma(\cdot,\cdot)$ and $\mathbf{g}(\cdot,\cdot)$  metrics--\cite{start01}. In \cite{start01} we also consider $\epsilon_{\vartheta}=+1$   adapted for the description of the  graph realizations properties
related to the  colors probability distribution.

\textbf{{From $\sigma$-metric to $g$-metric}}

A $g$-metric  transformation  can be reduced to  $\sigma$-metrics transformation and these to a part  of the graph (algebra) conformal transformations (and generalized permutations on vertexes).
By using   Eq.\il(\ref{Eq:qua-miel-por})-$\maltese$ as decompositions in $\sigma$-metric, a graph with a general  colored $\la^x$ algebra is compared with a $\la^0$-graph which can be considered an  embedding (cell) graph, since  the graph  with $\oslash_x$ algebra can be seen  an immersed graph or an  overlapped coloring, of the original, embedding,  graph.
To enlighten this situation and the   colored graphs symmetries    we can write  the $\sigma(\cdot,\cdot)$ metrics as follows:
$\sigma_-^{\upuparrows}(x):
 \sigma_-^{\upuparrows}(L)=\sigma_-^{\upuparrows}\circ y (L)\equiv\Gamma^{\uparrow}_-$, where
arrows indicate the connections $c_{mn}$ sign (fixing quantity $\xi$), subscript $(\pm)$ refers to the connections sign relation for the  directed graph $X^A\equiv(x,y)$ colors bases.     Then metric $\sigma_-^{\upuparrows}(x)$, for  a $x$-base graph  with  algebra $\oslash_x>\oslash_L$, can be written as  the combination  of  a  $x$-base metric  $\sigma_-^{\upuparrows}(L)$, with  ordinary algebra  $(L)$, and   $\sigma_-^{\upuparrows}\circ y(L)$ in for a $\la^0$  graph in a fictions  derived $y$-base.
Transformations $\Qa$ as colors inversion  $P$ and  $T$ for the color base $(x,t)$ respectively relate  $g(\cdot,\cdot)$ and $\sigma(\cdot,\cdot)$  graph metrics  and, consequently, graphs endowed with ordinary $\la^0$-algebra base to a general $\la^x$-algebra one. Metrics $\sigma(\cdot,\cdot)$ and   $g(\cdot,\cdot)$  transform differently for $\Qa$-applications,  involving  differently colored connections symmetries  for the   directed-antisymmetric graph connections.
General $\la^x$ algebra metric can be related to  minimal $\la^0$ metrics composition describing  $\sigma$-to-$g$  transformations for a $\la^0$ graph  and for a $\la^x$ graph as  follows
{\small
\bea\label{Eq:iav-lost}&&
(T-1) [\sigma_-^{\upuparrows}-\sigma_-^{\upuparrows}\circ y](L)=0,\\&&\nonumber
\sigma_-^{\upuparrows}(x)=\frac{(T+1) [\sigma_-^{\upuparrows}-\sigma_-^{\upuparrows}\circ y](L)}{2}=
\frac{(1-P)\sigma_-^{\upuparrows}(L)}{2}-\frac{\dd t}{\oslash_X},\\&&\nonumber \frac{(T-P-2)\sigma_-^{\upuparrows}(L)}{4}
=\frac{\dd t}{\oslash_L}.
\\\nonumber
&&
g_{-}^{\upuparrows}(L)=\sigma_+^{\upuparrows}(L)T\sigma_+^{\upuparrows}(L)=\sigma_-^{\upuparrows}(L)
T\sigma_-^{\upuparrows}(L),\\&&\nonumber
\sigma_-^{\upuparrows}(L)T^{2n}P^{2k+1}\sigma_-^{\upuparrows}(L)=-g_{-}^{\upuparrows}(L)=0,\\&&\nonumber
\sigma_-^{\upuparrows}(L)T^{2n+1}P^{2k}\sigma_-^{\upuparrows}(L)=g_{-}^{\upuparrows}(L)=0,\;
\eea}
($k, n\in \mathds{N}$, and action of  $\Qa$ follows  associative and distributive property on the elements on which it applies).
Metrics $\sigma(\cdot,\cdot)$ and $g(\cdot,\cdot)$     transformations
by the chromatic inversions   $\Qa\in\{T,P\}$,   differ  for a graph with   $\la^x$ algebra  or  the minimal $\la^0$ algebra. On this ground, we introduce  concepts of local ordinary algebras in metric frame, immersion graph  emergence.  The $g$-metric emerges   from $\mathbf{\sigma}$-metric as in Eq.\il(\ref{Eq:qua-miel-por}) with  $\Qa$  applications, differently for $g\circ x(L)$-metrics or general algebra.
The chain inhomogeneity (i.e. $\Gamma$s introduction, also defined     for graph conformal transformations)  affects both   symmetries and  algebras $\la^x$ and $\la^0$ relations. No inhomogeneity can be introduced for a metric related to  $\la^0$ algebra (graph). Therefore the $\la^0$  $\sigma$-metrics, related  by $\Qa$-applications,  are  equivalent 
i.e.  they describe  the  same graph structure (connection $c_{ij}=-c_{ji}$, \bp-events distribution...). This property characterizes the metric structures adapted to  the colored graphs with colored $\la^0$ algebra (maximally decomposed $\Ca_{\max}$ chains).
Metric $g(\cdot,\cdot)$  and $\la^x$-graphs  do not capture  many of  the graph structure properties, even for $\la^0$ maximally decomposed sequence  level,  which are instead considered in    $\sigma$-decompositions \cite{start01}.
The properties  appear   in the  $\sigma(L)$-decompositions as "degeneracies" (a set of $\sigma$-metrics correspond to a colored  adapted $g$-metric).  The inhomogeneity $\Gamma$-terms  introduction  destroys part of these symmetries. Similarly, however, it should be noted that the graph  inhomogeneity, generated   by the colors,
is reflected  by combining  the   $x$ and (colored adapted) $y$ bases, with  an $\la^0$ graph.
 This fact, together with  the   $\Gamma$ definitions  leads, by enucleating
the ``ordinary algebra part'' of the $\sigma(\cdot,\cdot)$ or $g(\cdot,\cdot)$-metrics, to define  a ``local''  $\la^0$--\cite{start01}. Eventually $\sigma\circ y(L)$ becomes  the metric  with  an ordinary algebra ($\oslash_L$) in the colored adapted  $y$-base, implementing    the concept of local algebra of a colored graph
 ($\sigma\circ y (L)=0$ and respectively $g\circ y (L)$=0).
 The inhomogeneity  term   becomes characteristic of the graph coloration   or  the graph path. 
 We consider  the   ordinary $\la^0(y)$ algebra,  decomposed in   $\la^0(x)$ as the local algebra of the embedded graph in the embedding $x$ basis graph. 
Metric  $\sigma\circ y (L)=0$  in the adapted base, indicates that if the graph is assumed as embedding, any other "immersed" graph  has to have algebra to be compared with the local  graph algebra which is  actually the local  minimum algebra.
In \cite{start01} we explore also a particle/graph relation where to each embedding graph   we can associate   a particle/event with an  algebra  $\oslash_x=\tilde{\oslash}_L$
(model scales emergence).
(The  $g$-metric  graph with ordinary local base with deformed $\tilde{\oslash_L}$ could be represented as a particle with an inertia immersed\footnote{{For a $g$-metric a definition of loop area can be as $
 A=\pi R^2=\pi\tilde{\oslash_L}^2=
 \pi{\oslash_L}^2{\omega_{xx}}/{\omega_{tt}}$, where
$R$ is a loop radius definition and $
 \mu^2=\sigma T\sigma/dt^2= -({\omega_L^2-\oslash_x^2})/(\omega_L^2\oslash_x^2)\in[0, \oslash_L^{-2}]$  can be defined  as  the square of loop inertia,  $\omega_{xx}$ is a  conformal factor of decomposition of a color base in the reference one. We provided different definitions of  maximum distances in the graph: for two  vertices afferent to   equal color probability $p$
  there is $
 m(1-2p)$ or, depending on boundary  vertices  in the loops,
 $ m(1-p)+1$ (these quantities are evaluated on a lattice graph mode).
 These definitions, which can be interpreted in the graph/particle correspondence, can be re-scaled for $m$, as quantities  independent from $m$ for large conformal expansions.}} in a $g\circ y(L)$  graph with ordinary algebra
within   therefore the
particle-graph
equivalence, we can   introduce an  $S$ (action) quantity and inertia). The $g(\cdot,\cdot)$-metric decomposition   reflects in the \bp-events  structure,
where emerging element $\dd y$ represents  a measure of graph homogeneity together with $\Gamma$s.

\textbf{Metric structures, transformations and symmetries}
We end  this section with further notes on metric graph transformations,  already partially addressed   concluding the Sec.\il(\ref{Sec:metri-gr-gra}) with the introduction of the quantities $\mathcal{L}_{\sigma}$  and $\mathcal{L}_g$.
Then, considering   metric  $\sigma(\cdot,\cdot)$ and  $g(\cdot,\cdot)$ transformations,
we can look  for   matrices,  $\Lambda$ and $\lambda$,  in colors and inner bases respectively,    considering  three possible cases:
(1)  $
\dd \widetilde{\sigma}=\eta^{A'B'}\Lambda_{A'}^A \Lambda^B_{B'}\lambda_{a'}^a \lambda^b_{b'}x_A^{a'} x_B^{b'}=\dd \sigma$;    (2)  $
\dd \widetilde{\sigma}=\eta^{A'B'}\Lambda_{A'}^A \Lambda^B_{B'}x_A x_B=\dd \sigma$, or (3) $
\dd \widetilde{\sigma}=\eta^{AB}\lambda_{a'}^a\lambda^b_{b'}x_A^{a'} x_B^{b'}=\dd \sigma$  and  similarly  for  $g(\cdot,\cdot)-$metric (where it can also be $\Lambda=\Lambda(\lambda)$, it is clear that an analogue problem can be addressed for the induced metric in Eqs\il(\ref{Eq:qua-miel-por}) and on the other hand, we  explore the conditions for the three cases coincide).   
%
 Metric transformations  are connected with the spherical wave transformations   (the
conformal group includes the  subgroups of Lorentz    and  the  Poincare group and
 the Laguerre group, while Cayley's group theorem ensures a connection with   the symmetric group).
 The associated  symmetries can be  explored by considering the  permutation group action (and Cayley group)--\cite{start01}.
We can write metric transformations from   the graphs colored structure, using the notation $\omega^{AA}=\dd x$  for the color base   $x$ and   $de_{ij}\equiv e_i\otimes e_k$,   for an ordinary bases  with inner indexes $(i,j,...)$,  ($\eta=\mathrm{diag}\{1,-1\}$), we used also notation  $\otimes$  for the base composition, thus
$\sigma\equiv\sigma^{ij}\dd x_{ij}\equiv\eta^{AB}\mathbf{\mathcal{G}}_{AB}^{ij}\dd e_{ij}$.  Matrix $\sigma^{ik}$  arises as $\eta_{ab}$ deformation through the ${x}$-basis   events decomposition  in {one} {only} \tb{ber}  ${e_i}$, with  coefficients e $\mathcal{G}_{AB}^{ij}\equiv [(x_A)^i(x_B)^j] $ as "soldiering objects", thus reducing   the bases indexes into event indexes (the relation between base-vertex indices does not automatically translate  into an  ``up-to-down'' index correspondence, as it is not meant as a  change between the  adapted bases, a \tb{bem} into a \tb{ber}).
%
Thus, for the $g$-metric, we  find
$g_{-}^{\upuparrows}(x)=g^{sl mn}_{(ij)(ij)}\dd e_{sl}\dd e_{mn}=\eta_{AB}T\eta_{CD}(x_{i}^{A})^{m}(x_{j}^{B})^{n} (x_{i}^{C})^{s}(x_{j}^{D})^{k}\dd e_{mn}\dd e_{sk}
\equiv(\sigma_{(ij)}^{sl}T\sigma_{(ij)}^{mn})
\dd e_{sl}\dd e_{mn}=
-\left(\oslash_L^{-2}-\oslash_x^{-2}\right)\dd t^2\equiv-\left(t^{ij} \dd e_{ij}\right)^2\equiv\dd s_{-}^{\upuparrows}\dd s_{+}^{\upuparrows}\equiv- \left(s^
{ij}\dd e_{ij}\right)^2 $,  or $\dd s^2=  \eta_{ab}{\mathcal{G}^{aa}_{mn} \mathcal{G}^{BB}_{lk}\dd e^{mn} \dd e^{lk}}\equiv g_{mnlk} \dd e^{mn}  \dd e^{lk}$. 
%
 Doubled notation in round brackets $(ij)(ij)$  clarifies the origin of the double terms with inner indices.
 We used  the arrow-sign convention in Eq.\il(\ref{Eq:iav-lost}), introducing  the emergent base $s$ (correspondent "light-cone frame") decomposed in \tb{ber}.
%
Metric  deformation reduced to graph vertexes  deformations. (However,  metric   deformations are  also  written  metric operator deformations,  since the  graph  vertexes-evolutive  operator (graph edges) correspondence considered in Sec.\il(\ref{Sec:part-w})
\footnote{
The correspondence between  $\Qa_{ij}$ matrix  and  metric $ \sigma(\cdot,\cdot)$  and $g(\cdot,\cdot)$ implies particularly that   $\Qa_{ij}$  constraints (derived from  a minimal algebra assumption $\la^0$) are inherited at metric levels   as metric constraints.
 Recalling that  $\Qa_{ij}$ elements are events with  indexes referring to  chain \tb{ber}  events and  \tb{bem},
  we adopt the following compact representation:
%
$\Qa^{A}\equiv(\Qa_{\bullet \star},\Qa_{\bullet \star}^*),\;
\dd \sigma=[\1 \otimes (1\times \dda)] \Qa_{\bullet \star}-[\1 \otimes (1\times \dda)^*] \Qa_{\bullet \star}^*=\Qa^{A}\hat{\eta}_{AB}\Qa^{B},\;
\hat{\eta}_{AB}=\mathrm{diag} \{\otimes (1\times \dda),-\otimes (1\times \dda)^*\},
$
$ (\dda^n\times\1)\otimes  (\1\times\1) \Qa_{00}=(\dda^n\otimes \1)\times(\1\otimes\1) \Qa_{00}=\dd \Qa_{(n0)(00)},\; (\dda^n\times\1)^*\otimes  (\1\times\1) \Qa_{00}=[(\dda^n\otimes \1)\times(\1\otimes\1)]^* \Qa_{00}=
\dd \Qa_{(00)(n0)}=\dd \Qa_{(n0)(00)}^*
$
 metric $\eta_{ab}$ (actually $\sigma_z=\sigma_3$  Pauli matrix in two dimensions corresponding to  a \tb{bem}-\tb{ber} bases) replaces
$\delta_{ij}$ in Eq.\il(\ref{Eq:tracc}),   and    $\otimes$  the composition $\times$   implementing thus a path ``metricization''.
Note that  we exchanged  symbols  $\times$ and $\otimes$, we used in this section, indifferently $\dda\otimes \1(=)\1\otimes \dda$,  we will  include signs associated to differently colored edges (algebraic valence) and   related to the  symmetries-role  in the colored metric  graph,  and   \bp-events in the  $(\sigma(\cdot,\cdot),g(\cdot,\cdot))$ metrics.  Metric transformations are related to  $\Qa_{ij}$  transformations   $\Qa_{hk} S^h_i S^k_j=\tilde{\Qa}_{ij}$.
Constraints imposed on the matrix $\Qa_{in}$ in Eq.\il(\ref{Eq:tracc})  translate into metric constraints .}.
\section{Concluding remarks}\label{Sec:mont-uni}
We considered   a   polychromatic multi-scales graph model  with     conformal graph expansion defining  the  graph  states.  Growing  of events (vertices) for   conformal expansion leads   to  the self-similar graph states notion, defining   graph  scales and  events  clusterings  (structure). Different  events are defined depending on the  polychromatic  clusterings of vertices. Parts of the graph can be coupled and used differently in an approach with not-classical color probability.
 A loop  can be  seen  as a structured subgraph with the limiting case of a  minimal $\la^0$ algebra.
We explored the effects of graph conformal expansion,  with   bubbles of clustered events growing and transmitting information (graph structure), and the persistence of the graph seed
after conformal transformation and the emergence of a  new relational (polychromatic) graph  structure.
The new events relational structure   emerges  from the
conformal graph expansion  governed by the  fixed (classical) colors probability distribution on graph.  Lower graph states (scales)   admit fluctuations, understood as variation of algebra distribution (algebra and color inhomogeneity  in the  adjacent chain vertexes).
The vertices are propagated (in sense of conformal expansion) with
emergence of new events and  symmetries, where the  conformal
expansion  preserves the graph seed structure (information), creates isomorphic parts of the expanded graph and  new events governed by the  probability distribution. The presence of a maximum in  the probability distribution  is important  in determining the graph vertex loop,  the presence of a maximum of sequences $\Sa_{\max}$ and chain $\Ca_{\max}$ , and the   \bp-events emergence  and effects of conformal transformation on graph structure.
 We  also analyzed  the special case of an equiprobable graph.
 The (manifold decomposed in a) graph has therefore different
scale structures\footnote{The framework we consider here is mainly  a relational frame, where a   topology definition is a particular  relevant issue.
Metric space, as  a topological space, requires  a notion of separability, grounding also a differential structure.  Thus, a topological notion, in our context, is grounded on considerations on symmetric  and reflexivity relation  given  through the chromatic clustering of vertices/events in the graph, with conformal transformations and  definition of different realizations,  grouping  together all the ordered connections, preserving the graph self-similarity  and generation of new structure.   It this scenario in fact we remain with a  conformal geometry and not properly a metric geometry. A further aspect of this frame is the existence of an ordinary algebra  constraining the   graph vertex relations (with a somehow  degenerate metric notion).}. 
Conformal transformation is connected to minimum not-zero event algebra as minimum limit to the (spacetime texture and) event definitions.  The graph is conformally expanding with new events, replicas of its structures and new structures.
At certain scales,  the (manifold-)graph  can be   treated as
a collection of interacting graphs, the most simple interaction to consider could be  the case of graph $\Ga_i$
immersed in a graph
environment
being  then a sub-structure of the
super-graph.  Geometro-genesis could come  as an asymptotic re-stitching of the previously dismantled
(logic signal) spacetime structure, producing a metric dynamical structure:  the crystallization phase, typical of "dynamic approaches" as QFT,  expects a restitching  of the graph constituents to a smooth  (in some way rigid)  background spacetime.

This  polychromatic graph model  is characterized by  the idea of
an  overlapping of metric
and (dynamical) colored graph, the use of two metric levels and  symmetries role, including  multiple events definitions, macroevents related to loops, and the
definition of multiple levels of events at every scale consequent to linking the events emergence to color exchange which implies also the elaboration of more structured levels of events where metric structures can be defined according to these. Color change in the sense of  {inhomogeneity} is the essential process defining the events.
(A color and algebra change, with respect to \tb{ber}, is associated to an event emergence, "if there is no change there is no clock", conformal transformations are adapted to this situation). {Sec.\il(\ref{Sec:metri-gr-gra}) and particularly  Sec.\il(\ref{Sec:gooUE-dce}) explores some features of the graph in an illustrative approach,  considering several  approximations. }

 We  investigated the  persistence
of isomorphic parts of the graph under conformal expansions, detailed in \cite{start01}, where  we also introduce new  (existential quantifications) operator. We  include here a brief outline of these aspects, which are deepened  in \cite{start01}, emphasizing some  conceptual features of the graph  and anticipating further developments of this model. "Negation"  operator, related to the complementary chains, increases the graphs state possibilities and therefore the associated metric structures.  The inclusion of new existential operators increases in graphs states possibility in the logic signal representations of chains\footnote{The model is based on a  strong  "logic" frame   typically  of it-from-bit approaches.   (The   introduction of $\exists$ and $\non{\exists}$  operators,  in the logic signal formulations,  refers to the  view of   gravity as a manifestation of the "existence"  in the sense of \cite{WH,STARTRECK}.
The formalization of this  general  assertion  (existence role) is connected to events definition (related, more precisely,  to antisymmetric relation),
then gravity must emerge as a phenomenon  related to graph events.) Graph composite states,   considering combinations of vertices clusters, change the concept of probabilities used here (eventually to explore from pure to mixed states in the quantum approach). Generally,  we can consider  chain is a collection of (homogenized) connections $c_{ij}$, and we   use   mainly a 2-level logic signal representation with a  set of  (color independent) events of reference chain,  as doublets $\{(0,1), (1,0)\}$.  We introduce therefore    $\exists$ and $\non{\exists}$,  and \emph{negation} $\non{\Qa}$-operators on a quantity $\Qa$ (in  a  graph  diagram  constructed in \cite{start01}  there are  doubled regions, $\mathcal{I}$ and $\mathcal{Re})$, where vertices can be defined introducing a further possible definition).
 Introduction of $(\non{\Qa},\exists,\non{\exists})$   allows  to expand and enhance  the events interaction possibilities by combining
  different  possible realizations.
In general, for two  events   \srt{$\mathbf{Q \neq P}$},  there are  four  states
{\scriptsize{\textbf{$(\mathbf{\exists Q, \exists \non{P}, \non{\exists} \non{P},\non{\exists} \non{Q}})$}}}. These are used from a transition from \srt{${\mathbf{P}}$ }-to-\srt{ ${\mathbf{P}}$ } (loop), and  from-\srt{${\mathbf{P}}$}-to-\srt{${\mathbf{Q}}$}.
(Events quadruplets and quadruplet transitions can be formed from  doublets as follows:  {(i)}
\srt{$\mathbf{(Q)=(\exists Q,\non{\exists} \non{Q})\equiv (\non{P})=(\exists \non{Q},\exists{P})}$},
and similarly for  \srt{$\mathbf{(P)  (\non{Q}})$};
{(ii)}Secondly there is   \srt{ $\mathbf{\exists Q=(\exists Q, \exists \non{P})}$},  or non-existence  \srt{$\mathbf{\non{\exists} P=(\non{\exists} {P}, \non{\exists} {\non{Q}}})$}.
{(iii)} Finally, we can  consider "values-couples" as  \srt{$\mathbf{(\exists Q, \non{\exists}\non{Q})}$}, and \srt{ $\mathbf{(\exists P, \non{\exists}\non{P})}$}.
  Within the  correspondences  $[\non{\exists}\non{\ket{a}}\rightleftharpoons\exists \ket{a}]$  and
 $[{\exists}\non{\ket{a}}\rightleftharpoons\non{\exists} \ket{a}]$,
with  the operator $\mathbf{/}:\;\;/\cdot(\exists\ket{a})=\non{\exists}\non{\ket{a}}$, we obtain   $2\dda[({\dda^{2n}\exists \ket{a}})({2n})^{-1}]=(1+/)\dda \exists \ket{a}$. (We remind that, according to Eq.\il(\ref{Eq:sum-prod}) the application on vertex of $\dda^\kappa$ correspond to $\kappa+1$ vertices).
The introduction of these operators are linked to the generation of the complementary chains for $r + r$ monochromatic chains  model. The introduction of quadruplets also enlarges the order relations provided by the evolution operator.
The following relations can be then considered:
$
\dda^n \exists \ket{a}=\sum\limits_{k=0}^n \binom{n}{k}\non{\exists}^k \non{\ket{a}}^n$ with $ k\geq1$ and   $ (1+/) \dda^n \exists \ket{a}=\sum\limits_{k=0}^n \binom{n}{k}(1+/)\non{\exists}^k \non{\ket{a}}^{n-k}=\sum\limits_{k=0}^n \binom{n}{k}\left[\non{\exists}^k \non{\ket{a}}^{n-k}+\non{\exists}^{k+1} \non{\ket{a}}^{n-k+1}\right]
$
and we can introduce
$
\theta^{2n}\equiv\left[\dda^{2n}-({2n})^{-1}\right]$,  with
$  \theta^{2n}\ket{0}=
\dda \left[\dda^{2n}-({2n+1})^{-1}\right]\ket{0}=0 $ and
$\dda \theta^{2n}\ket{a}=-(2n (2n+1))^{-1}\dda \ket{a}$.
(in some ways we can say that in these relations $\dda^n$    acts so to ``retain memory'').
Last relation is on  chains with periodic boundary conditions on value $a$, step $2n$ and $\dda \ket{a}=\non{a}$,
thus we  write $[\dda, \theta^{2n}+(2n(2n+1))^{-1}]=0$ and $
[\theta^{2n},\dda]=+(2n(2n+1))^{-1}{\dda}$. With  the new operators,  framed in the polychromatic or monochromatic $r+r$ chains,
   we would have a replica or negative copy respectively of the states of the polychromatic/monochrome graph (in the  DNA-graph-code  figure they would really be as specular graph filaments). Introduction of  $(\Re,\Im)$ planes accounts for states persistence, considering  $\non{ \Ga}$ or  $ \non{\Sa }$ elements.}--\cite{start01}.
  In this transcription of the graph realizations where, for example, a bichromatic graph chain is a   4-levels (from 2-levels) logic signal  the introduction of a pair of  (existential) operators takes into account the fact that the graph, considering its chain structure, is not  complete and, therefore reduces the graph structure to a "chains overlapping" (there is an algebraic valence $(\pm1,0)$) considering a  persistence of each realization in the graph state, increasing therefore the logical levels.
  The graph frame opens up a serie of possibilities to be discerned  concerning the role of the two metric levels, the "lower-level" $\sigma$-metric  or  the "higher" level $\mathbf{g}$-metric depending also on the emergence of different  symmetries, the spinorial structure (eventually defined for  the polychromatic structure) and the choices on the index composition in the explicit  chains metrics approach addressed in Sec.\il(\ref{Sec:metri-gr-gra}). For a general  metric graph there is currently no indication on a metric level prevalence choice. A further aspect to be explored  is the  \bp-$\Ga$ role in the metric approach and the coupling terms   $\Theta^{A}_{aB}$, which we  visualized here in a  possible supersymmetric model and more generally at  the core of the spinorial structure  analysis  through  vertexes clusterization (which is also related to the  from multi-bit to multi-quibit frame transition).
It  should be pointed out then  a quibit-related spacetime texture might be associated to the topological structure following the notion of   spacetime  and quantum complexity and  particularly Black Hole (\textbf{BH}) complexity\footnote{Graph  entropy  has been  related to graph energy, in the graph--particle equivalence introduced briefly in  Sec.\il(\ref{Sec:gooUE-dce}) for the embedded graph in an embedding graph with different local bases and  maximum (and minimum) entropy. Entropy is upper bounded  by the area in the sense that    entropy is related to geometry and, on the other hand, energy is related to geometry, there is a connection between entropy and gravity.  In fact,  concerning, eventually, the  emergence of an Hilbert space we should consider that
  the maximum entropy reachable in a region of a space  appears to be in fact  the \textbf{BH}  entropy
which we may correspond to the  Hilbert space  dimensionality , i.e. the entropy $\approx e^S$
where $S$ is the  \textbf{BH}  entropy,
which  is therefore  importantly  finite in
every  manifold region.}.
One crucial  aspect of  approaches  considering  collection of qubits  as fundamental languages consists in the determination of the   quibit (and multibit) organizations and representation\footnote{Gravity  has been variously    related  to a  ("statistical")  complexity grown where, in these approaches,  complexity definition stands as one key issue. Complexity geometry, for example \cite{Nielsen},  defines the  computational
complexity within differential geometry formalism with a notion of complexity distance, which can be considered
in some ways more refined then metric on quantum states provided by the  inner-product distance (a large
inner product meaning  close states, with a  supremum distance in the orthogonal state case).
So called  "gate complexity" corresponds to  the number of primitive gates in the smallest path (quantum
circuit) that leads to  the two states transformation with non null
tolerance  $\epsilon$ to reach the target
(while the length of the shortest path between state gives the computational complexity).
One key issue however of this frame is the rapidity or, more generally,
 how in the
computational complexity
the time-evolution operator evolves, and this is relevant in \textbf{BH} complexity too.
 A \textbf{BH} could be modeled by $m$ qubits with a \textbf{BH} complexity of unitary evolution operator linearly growing (a conjecture) with
time exponential in $m$.
Therefore  $2m$ qubits maximally entangled state can be   written as a product of $2m$ Bell states.
In general   the considered unitary group   is $ SU(2^m)$ for $m$ qubits  with a complete basis  provided by
$ 2^{2m} - 1 $  generalized Pauli matrices (thus for 1 qubit
the base is  the 3 Pauli
matrices plus the  identity), and  they  possibly correspond to elements of the graph.
Notably, as mentioned above the  quantum complexity for a system of $m$ qubits is similar to the
entropy of a classical system with $2^m$
degrees of freedom.
 The systems
is a set of $m$ classical bits,  $m$ binary digits string (element of space state here a chain graph). 
}. 
 The second crucial  aspect would be  thee dimensionality of the associated Hilbert space (for  a space of states of $n$ qubits, the relative Hilbert space has dimensionality $2^n$,
 number $n$ would be associated to
 order the \textbf{BH} entropy analogously to the classical  entropy  for a system defined by $2^n$ degrees of freedom).
 Concerning the relation with spin network, in the approach considered here we would have, for example, a
$ SU(2)^n$ for a bichromatic (r=2) graph of state $\Ga_n$  (not related to valence).
More generally addressing the colored graph and particularly the conformal transformations, it is clear then that one could think to  express states as representations of $SU(r)^n$ (or   a reduced $SU(r)^{\check{m}}$ with $\check{m}<m$).
 The basis states are the $2^n$ chains  of $n$-events  and an overlapping of these.
 In  here we could eventually  interpret the graph realizations in terms of quantum bits (qubits) or more precisely $r$-levels $n$-multibits (to be multiqubit) for a complete sequence of graph state $\Ga_n$ with $r$ colors
  \footnote{
A physical qubit is   a quantum system with dimension
two, then  a classical bit--i.e. two distinguishable states or a bichromatic graph with $r=2$-- can be
embedded into a qubit with corresponding
classical  probabilities appearing  along the diagonal of
the density matrix, of course  off-diagonal elements in the
density matrix  shall remain, here rendered through combination/organization of different states through conformal  transformation).
The
  dimensionality of the Hilbert space for a system of  $n$ qubits grows as $2^{n}$ or a reduced $2^{\check{m}}$}.


{In this article  the  graph  and   the general ideas underling  the  model are  discussed, while  in the future  investigation using  the  logic signal formalism  we study the  dynamic graph  addressing  the  role of higher level events by exploring  the   spinorial texture emerging from the graph poly-chromaticity, considering again  the graph-particles formalism of Sec.\il(\ref{Sec:gooUE-dce})}

In conclusion we can distinguish in this model the  following three main  aspects: 1. the multiple events definitions,  defined as macro-events or loops (monochromatic clusterings of graph vertices) or polychromatic clustering  and \bp-events (events are intrinsically related to concept of color change)   conceptually identical  to event-loops at any scale. 2. A second characterizing factor is the  admission that  gravity interests  the existence,  (substantialism and universalism) in the sense \cite{WH,STARTRECK},  hence  a  geometric  (relational) description of gravity. 3. The conformal expansion and the isomorphism of graph parts, defining  the  events scales structures,  constitute a   third  relevant element\footnote{Spacetime manifold is dissolved into more fundamental elements and structures, the  graph vertices (events) and their relations (polychromatic connections), which are set conceptually in our frame on the same level of vertices/events by introducing  concept of   \bp-events. This setup addresses  therefore also the issue of  event  definition. Different  events  definitions, emerging geometry, gravity  and notion of "existence"--\cite{WH,STARTRECK}--are    in the same graph frame. {Minimal distance is translated in logic terms,  $\Ga$  inertia is related to existence of local algebra.} The gravity "universality",  in the  general relativistic transcription as a    geometric spacetime,  is formally understood as   logic entity  made of events and a property relative to existence--\cite{WH,STARTRECK}, where an event (in this view as an "existence atom") could  be a point in  a    spacetime transcription.  In the  graph model, a  loop could be seen  a conformal  expanded  point (reported to special relativistic frame we could think to  light-like distances as closest picture  of graph conformal event/vertex  expansion). 
Conformal transformation in this sense  turns to be  a gravity intrinsic property.
   This aspect is at base of   the scale introduction and re-scaling   after a conformal  expansion,  graph self-similarity and generation of new structure. Event is defined  as objects  interaction,   gravity is then doubly related to events and their relations. Observations are interpreted as a set of  (related) events,  determined by the  observer (base), transcripted as logic signal in a  graph, where the order relation is defined  by its  poly-chromaticity.  At certain scales we break the  (classic)  logic signal order, considering the combinatorial graph to construct  the realizations. The   clusterization is  then determined by the dynamics (in this context, considering also notion of graph embedding as a sort of  spacetime cell, as in $\mu$matter there is not  distinction between matter and geometry concepts).
On the other hand, entropy is naturally related to geometry and here,  through graph poly-chromaticity, entropy     connects more events levels.  The gap in colors  probability distributions regulates the inhomogenity and the persistence  loops (as in Sec.\il(\ref{Sec:fromz}) and Sec.\il(\ref{Sec:deg-new-w-web}) in the chains  $\Ca$, the maximum probability  color  regulates the  monochromatic clusters, while at the other extreme case the  equiprobable cases admit a local, in the sense of graph scale, inhomogenity in colors.
  In the equiprobable  graph at certain scales there are fluctuations of clusters  with respect to the homogeneous minimum or maximum  chains.
 }.
   As in different  approaches, the  prevalence of event/existence concept  leads the distance between matter and spacetime  to   disappear  in a graph model in the sense of $\mu$matter of QFT conceiving an embedded graphs represented with the figurative idea of  spacetime cell and DNA   for  the graph and graph structures, (a collateral consequence of relationalism and  the logic  approaches of these frames, yet decoration, graph coloring, is not dissociable from such  ``spacetime-DNA'' containing  the instruction and  the information--manifold substantialism)--\cite{WH}.
  Graph state or graph realizations have, at any state, embedded
replicas as graph $\Ga_n$
``seed-genetic-code'' of the events structure reproducing at any state transitions. Following a
similar idea then, a graph  $\Ga_m$
contains, at inferior scale,
as its substructure a disjoint collection of different $\Ga_n$ graph
seeds, as $\Ga_n$ clusters.
We might  use   the figurative image of a "geometric code" as a kind of ``spacetime-DNA'',  to express the graph and  its properties (within this  picture,
 we might say that this investigation is more a  proposal description of the ``spacetime-phenotype'' rather than the  ``spacetime  genotype'' of such geometric code).
A``space-time  genetics'' would come here   from the   observations on the hereditary characteristics. In this model, considering this analogy, we  may  deal  similarly with  inheritance (replicas of the original seed graph)  and graph (geometric code)  modifications. The possibility of describing, and eventually manipulating, such spacetime  geometric  code would appear to be an intriguing applicative prospect, as a ``spacetime engineering'' starting as   theory of transmission and coding of information.

\appendix

\section{Some notes on   logic nature  of  spacetime   notions in   graphs and relativity}\label{Sec:log}
\begin{figure}
  \includegraphics[width=9.6cm]{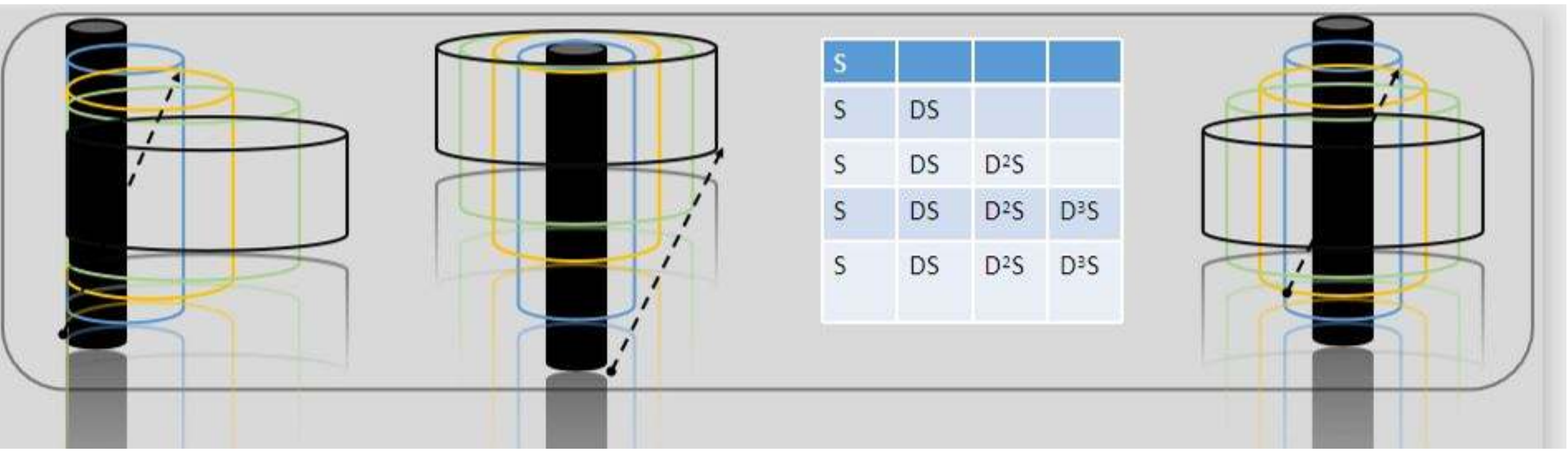}
  \includegraphics[width=7cm]{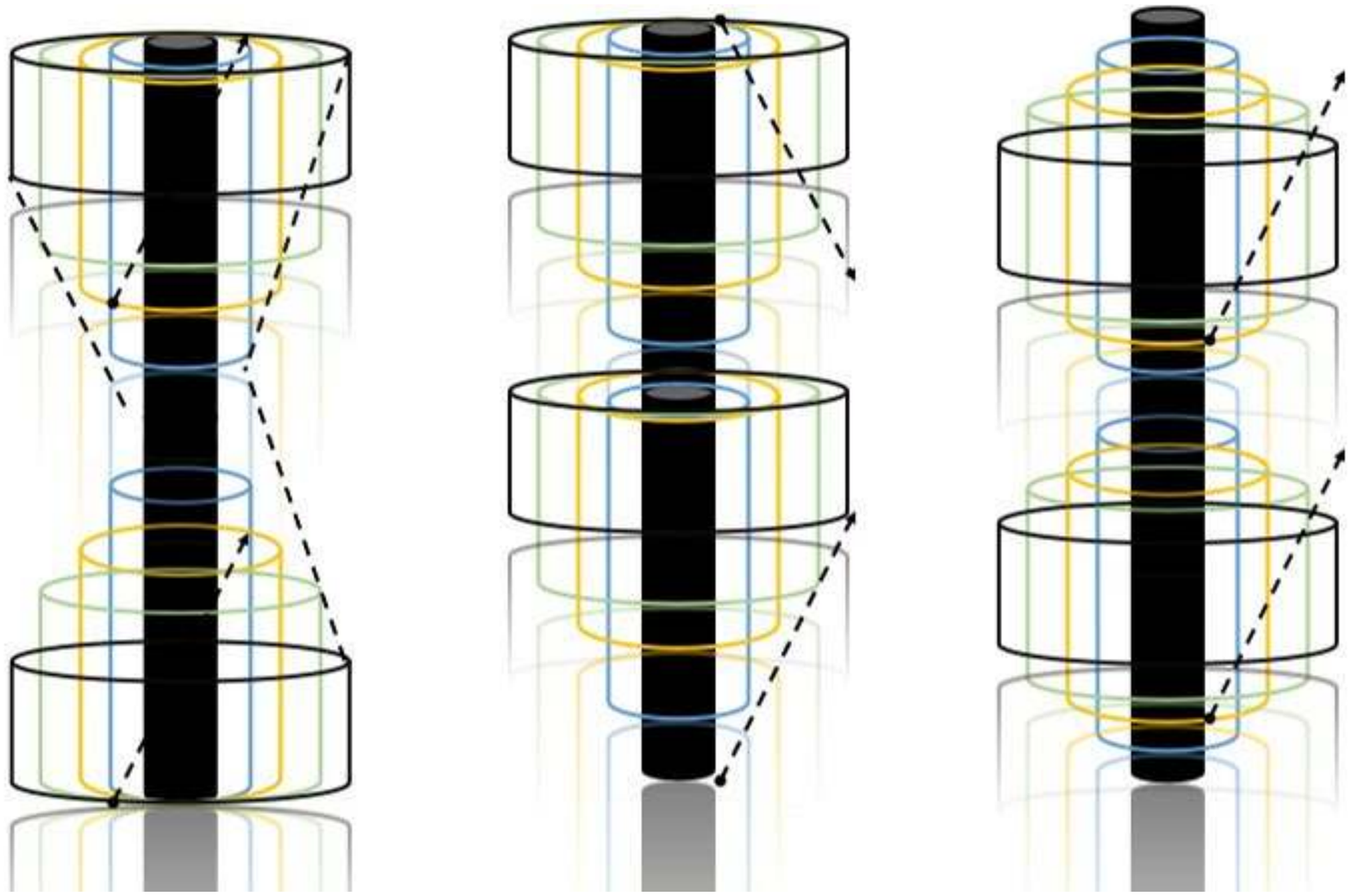}
  \caption{Schemes for the representation of the states persistence in conformal expanded graph states, each cylinder is a different   part (vertex) of a graph. The event matrix $\Qa_{ij}$ is represented schematically in the left panel with the role of evolution operators representing the cylinders. The different configurations represent different  chains configurations and therefore different symmetries, the irreversibility and persistence of parts of the graph is evidenced.}\label{Fig:2cap}
\end{figure}

%
The  polychromatic   graph of events is  framed as
 background  independent,      spacetime emergent  model close  to   it-from bit ideas, where the geometrization of gravity is passed in a logicization of it. We  can then   retrace a      logic  nature   for general concepts of  time and space  notion   in the graph, which are  closed here by the   colors order relation, the events decomposition and composition       in   sum  $\Sigma$ and product $\Pi$ in Sec.\il(\ref{Sec:part-w}), related to   shift $\dda^*$ evolution $\dda$ operators respectively,  and correspondingly clustered    in sequences  $\Sa$ and chains $\Ca$  of  Sec.\il(\ref{Sec:fromz}) .
Minimal (non-zero) algebra  $\la^0$  (formalized by the introduction of logical quantifiers) grounds the graph  spacetime texture (related  to  relativistic light-like distances,  rephrasing   graph conformal transformations), determines the    vertex (cluster) algebra  and loop radius  and constrains the particle/graph inertia and the logic signal  wavelength.
 As discussed in Sec.\il(\ref{Sec:part-w}) a body  (event-graph),  associated to a logic signal,    is  decomposed  (decoded) in the polychromatic graph, and viceversa a set of events (bodies) are  composed in one object  (graph realizations as in sums, sequences $\Sa$)  at any scale, and decomposed in the polychromatic graphs-clusters.
 Realizations  and states constructions are regulated by colors probabilities. In the   equiprobable case,  local  (colors-algebras) fluctuations, in the sense of  Sec.(\ref{Sec:deg-new-w-web}), could be observed. The presence of a  maximum of the color probability is  a further relevant   case  (the color associated to the maximum  of probability  is  considered as a  color loop "attractor", governing  the color permanence and heritage in states and their realizations).
Thus  generalized permutations and  algebra homogenization,  investigated in Sec.(\ref{Sec:deg-new-w-web})  are rendered  as  metric transformations in  the approximation of Sec.(\ref{Sec:metri-gr-gra}) and are frequency and phase  transformations of the associated  logic signals.
 In  \cite{start01} we recovered the (homogeneous) Lorentz group as algebras transformations, in terms of   $\oslash^{\epsilon}$ with $\epsilon=\{\pm1,\pm 2\}$  for a   $\sigma$ metric   ($\epsilon=\pm1$) or $g$-metric ($\epsilon=\pm2$), function of the logic signal wavelength. Thus,  within the particle/graph relation  the  corresponding  equations of motion relate   metric graph bases.
 In this sense  the correspondent relativistic frame is a theory of (minimum) information, where  as in Eq.\il(\ref{Eq:colpr-C}) and (\ref{Eq:B-Gar})   (graph-state) entropy
is indeed related to $ \oslash_L$.
  The equivalence in Eq.\il(\ref{Eq:sum-prod}),    \emph{$\Pi\Sigma$--equivalence}, provides insight on the notion of body and evolution. To enlighten conceptually this fact in  \cite{start01} we used  an  events   composition/clusterization    operator,  acting on  a set $\{a_i\}$   of events  $\hat{f}: \{a_i\} \mapsto\Sigma_ia_i=a$   composing in sum and in product  the  event ${a}$ and decomposition $ \hat{g}:\; {b}\mapsto\{b_j\} = \Sigma_jb_j$   with compositions   $\hat{H}=\hat{f}\hat{g}$ and inverse $H^{-1}=\hat{g}\hat{f}$.  Action of $\hat{f}$ and $\hat{g}$, in the logic signal  formalism can be interpreted as  wavelength shift (read, in the metric approach,  as a{ "graph Doppler-effect"}).
In this context the meaning of the  $\Pi\Sigma$--equivalence, interesting   the graph the structure in monochromatic or polychromatic loop, may reveal of a far reach significance.
 Evolution of body (the chain $\Ca$)  corresponds to  the object  set (the sequence $\Sa$)  defined by the  body evolution.
 Properties  of sequences or body chain  evolution (elements of dynamic graph) are similarly related. The $\Pi\Sigma$--equivalence   refers    to the chain/sequence relations, thus $\dda$ and $\dda^*$ action  that can be followed along the column and rows of  $\Qa_{ij}$ of Eq.\il(\ref{Eq:yo-te-ho},\ref{Eq:tracc})\footnote{We could define  $g_{-}\equiv  \partial^2_t f-\partial^2_r f\equiv \Box f=\sigma_{-}\sigma_{+}$  referring to Sec.\il(\ref{Sec:gooUE-dce}), where however we considered  a set of functions  associated to the graph,  modified for the  inhomogeneity terms $(\Gamma, \dd y)$ and decoupling also  the embedding and embedded graphs.
 (Colors symmetries, and  particularly
   color reversibility considered   in the metric definition for zero level graph, do not appear to be a symmetry   for higher order events). In this example we introduced a  function $f(t,r)$, where $r$   colors space index, $t$ a \tb{ber}  index, according to the convention used in Sec.\il(\ref{Sec:metri-gr-gra}), we can write
$ \partial_t f=\pm\partial_r f$ ,  quantities  $\sigma_{\mp}\equiv  \partial_t f\mp\partial_r f$ refers to the first level graph $\sigma(\cdot,\cdot )$,  in general a no-zero  quantity, ($\partial_t$, $\partial_r$  associated to evolution-shift.) sign  $\pm$ refers to color symmetries.
}.
We can think, two bodies/graphs $\Ga_i=\{s_i^{a}\}_a$ and $\Ga_j=\{s_j^{a}\}_a$
 are composed into in $ \Ga_{(ij)}=\Ga_i\cup \Ga_j=\{s_{(ij)}^a\}_a$, and  decomposing an associated logic signal.   We consider the evolution  of a body in the sense of $\dda$ application
$\Ga\equiv \sum_{n=0}^\kappa\dda\Ga_n=\sum_{n=0}^\kappa\dda^{n+1}\Ga_0$, applying  Eq.\il(\ref{Eq:sum-prod}) (note the existence of operators decomposition mirroring the events decompositions as discussed in Sec.\il(\ref{Sec:part-w})). The evolution leads to graph states with graphs (sphere) radius regulated by  $\oslash_L$  which regulates and establishes  the irreversibility of the process (in the sense of chains symmetries), the conformal transformations, the persistence of structure (inheritance through conformal transformation-self-similarity), the generation of new structure (through conformal expansion, in this sense primordial idea of space-time generation).
There is
$\dda^nS_0\subset \dda^{n-1}\Sa_0$
at any scale $(n)$ discussed in  Sec.\il(\ref{Sec:fromz}) where we also discussed the (chain-structure) irreversibility   regulated by $\oslash_L$.
These properties are represented in  the matrix in Eq.\il(\ref{Eq:yo-te-ho}) and  Figs\il(\ref{Fig:2cap}).

The monochromatic loop of order two  is  totally reversible in the context of the persistence of states in the ordered chain. Graph structure is regulated by entropy in Eq.\il(\ref{Eq:colpr-C}),
   energy and inertia as \cite{start01}), the larger  $m$, the larger the inhomogeneity colors-(constrained  by existence of  $\Ca_{\sup}$) the greater the inertia and the graph  characteristic  irreversibility.
 The decomposition constituents  $s_{ij}$  graph of $\Ga_{(ij)}$, depend on the graph state   and mostly on the probability distribution coming as a "color-attractor" in the union-graph realization.
 In the graphs union concept, metric transformations read in the particle/graph equivalence  reveals  useful to discern, as  (equivalently relativistic longitudinal and
 transverse) Doppler effect,  the change in frequency (and wavelength) of the associate logic   signal
\footnote{A mechanical analogy for the persistence of states and thus the copy of states, the realization of sequences and chains we could refer to the phenachistoscope  creating the illusion of movement, based  on the physiological persistence of the vision in the human eye,
together with the chronostasis  an  illusion  where  following the introduction of a new event  to the brain can appear to be extended in time. We also mention the so called  phi-phenomenon for illusion of  apparent motion  observed if two nearby spatial  events (with a constraints on frequency).
In this approach, however, we must note that there is no causal relation in the copies.}
Considering  Eq.\il(\ref{Eq:sum-prod}) an object $S$ is decomposed in the evolved copies of $S$, the copies are not connected except in the sense of persistence (in the sense we can define  $S_2=D S_1 \in S_1$) and immersion, in the analogy therefore the motion of the observer corresponds to a change of structure  of the body,  seen as
 conformal transformation through the equivalent graph Doppler effect (relevant also for the higher order symmetries is that acceleration
ordinary in the resting system is an invariant even in this model).


\begin{thebibliography}{99}
 \bibitem{start01} \emph{To be submitted}
 \bibitem{Konopka:2008hp}
  T.~Konopka, F.~Markopoulou, S.~Severini,
  Phys.\ Rev.\ D {\bf 77} 104029 (2008).

  \bibitem{Konopka:2006hu}
  T.~Konopka, F.~Markopoulou, L.~Smolin,
  hep-th/0611197.
\bibitem{Caravelli:2010xx}
  F.~Caravelli, F.~Markopoulou,
  Phys.\ Rev.\ D {\bf 84} 024002 (2011).
\bibitem{Oriti} D. Oriti, (Ed.).  \emph{Approaches to Quantum Gravity: Toward a New Understanding of Space, Time and Matter}. Cambridge: Cambridge University Press. 
       (2009).
 \bibitem{Levin} M. Levin and X. G. Wen,  Phys.
Rev.\textbf{ B 67,} 245316 (2003);  M. A. Levin and X. G. Wen, Phys. Rev. \textbf{B 71}, 045110 (2005).

  \bibitem{WH}
J.  Wheeler. Law without Law in J. Wheeler, \& W. Zurek (Eds.),\emph{ Quantum Theory and Measurement.} Princeton, NJ: Princeton University Press,  (1983). Quote <<\emph{We are no longer satisfied with insights only into particles, or fields of force, or geometry, or even space and time.  Today we demand of physics some understanding of existence itself.}>>.

\bibitem{STARTRECK}
{Spock}: <<Change is the essential process of all existence.>>
{Star Trek OS S03E15}.
<<
Spock: Incredible, Captain.
Kirk: What was that?
Spock: What my instruments read is totally unbelievable, Captain. Twice, for a split second each time, everything within range of our instruments seemed on the verge of winking out.
Kirk: I want facts, not poetry.
Spock: I have given you the facts, Captain. The entire magnetic field in this solar system simply blinked. The planet below, the mass of which we're measuring, attained zero gravity.
Kirk: That's impossible. What you're describing
Spock: Is non-existence.>> {Star Trek OS S01E27} \emph{The Alternative Factor}.
  \bibitem{P1}
    R. Penrose, \emph{Angular momentum: an approach to combinatorial space-time}, in Quantum Theory and Beyond, ed. T. Bastin, Cambridge University Press, Cambridge, 1971, pp. 151-180.


\bibitem{P2}
R. Penrose, \emph{On the nature of quantum geometry}, in Magic Without Magic, ed. J. Klauder, Freeman, San Francisco, 1972, pp. 333-354.



\bibitem{P3}
R. Penrose,
\emph{Combinatorial quantum theory and quantized directions}, in Advances in Twistor Theory, edited by L. P. Hughston and R. J. Ward, Research Notes in Mathematics, Vol. 37 (Pitman, San Francisco, 1979), pp. 301-307.




\bibitem{Padma}
T. Padmanabhan,
Classical and Quantum Gravity,  4,  4 (1987).
\bibitem{Garay:1994en}
  L.~J.~Garay,
  Int.\ J.\ Mod.\ Phys.\ A {\bf 10} 145 (1995) .

\bibitem{Science}
G. Rubino, L. A. Rozema, A. Feix, M. Araújo, J. M. Zeuner, L. M. Procopio, C. Brukner, P. Walther,
Science Advances ,  {3},
	n {3}: e1602589 (2017).
\bibitem{2015NJPh...17j2001A} Ara{\'u}jo, M., Branciard, C., Costa, F., et al.\ , New Journal of Physics, 17, 102001, (2015).

\bibitem{Sorkingeven}
R. Sorkin.\emph{ Logic is to the quantum as geometry is to gravity}. In Jeff Murugan, Amanda
Weltman, and George Ellis, editors, \emph{Foundations of Space and Time: Reflections on Quantum
Gravity.} Cambridge University Press, . arXiv:1004.1226 (2012).

\bibitem{DeHaro:2018rvw}
  S.~De Haro and H.~W.~De Regt,
  arXiv:1803.06963 [physics.hist-ph].


\bibitem{Nielsen}
M. A. Nielsen, M. Dowling, M. Gu, A. Doherty, Science 311, 1133 (2006).

\end{thebibliography}
\end{document}